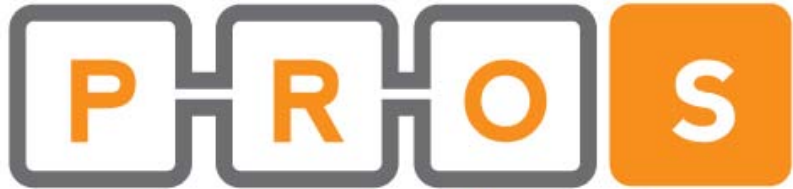

Research Centre in Software Production Methods

Universidad Politécnica de Valencia

Camino de Vera s/n  46022  Valencia (SPAIN)

# Informe Técnico / Technical Report

---

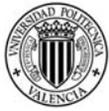
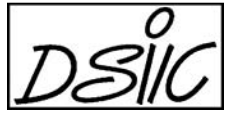
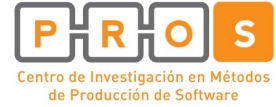

# INTEGRATION OF
# COMMUNICATION ANALYSIS
# AND THE OO-METHOD:

# MANUAL DERIVATION OF THE CONCEPTUAL MODEL

# THE SUPERSTATIONERY CO. LAB DEMO

**Authors (in alphabetical order):**

**Sergio España, Arturo González, Óscar Pastor, Marcela Ruiz**

# TABLE OF CONTENTS

# 1. INTRODUCTION

## 1.1. Motivation of this work

This document presents a lab demo[1] that exemplifies the manual derivation of an OO-Method conceptual model, taking as input a Communication Analysis requirements model. In addition, it is described how the conceptual model is created in the *OLIVANOVA* Modeler tool.

The lab demo corresponds to part of the business processes of a fictional small and medium enterprise named SuperStationery Co. This company[2] provides stationery and office material to its clients. The company acts as a as intermediary: the company has a catalogue of products that are bought from suppliers and sold to clients.

This lab demo, besides illustrating the derivation technique, demonstrates that the technique is feasible in practice. Also, the results of this lab demo provide a valuable feedback in order to improve the derivation technique.

## 1.2. Scope of this document

This document does not describe Communication Analysis and the OO-Method in full detail. For further information about these methods please refer to [González, España et al. 2008; España, González et al. 2009; España, Condori-Fernández et al. 2010; González, Ruiz et al. 2011] and [Pastor and Molina 2007], respectively.

---

[1] In a lab demo, a technique is used by researchers or method designers in an artificial environment (no conditions of practice, but context is controlled), to solve a realistic problem, in order to gain knowledge and to show that the technique could work in practice [Wieringa 2008].

[2] SuperStationery Co. is a fictional company but the work practice described herein intends to be realistic and useful for the purpose of illustrating Communication Analysis requirements modelling.

# 2. TEXTUAL CASE DESCRIPTION

## 2.1. Workpractice

SuperStationery Co. is a company that provides stationery and office material to its clients. The company acts as an intermediary: the company has a catalogue of products that are bought from suppliers and sold to clients. Most clients call the Sales Department, where they are attended by a salesman. Then the client requests one or several products that are to be sent to one or many destinations. The salesman takes note of the order (see the Order form in page 4). Other clients place orders by email or by fax. Then the Sales Manager reviews the order and assigns it to one of the many suppliers that work with the company, using his own judgement (the Sales Manager notes down the Supplier section of the Order form; additionally, the company wants to record the assignment date). An order form is sent by fax to the supplier. The supplier receives the order form and checks whether they have enough stock or not. In case they have enough stock of all the products requested in the client order, they accept the order (the supplier indicates the planned delivery date –that is, the date at which the supplier commits to deliver the order- and the salesman also notes down the response date); otherwise, they reject it. In case the order is rejected, the Sales Manager assigns it to a different supplier (this can happen many times until the order is accepted). Once the order is accepted, the salesman sends a copy of the order to the Transport Department and the Insurance Department. In the Transport Department, the Transport Manager arranges how the goods will be carried to the destinations; this implies selecting one of the truck drivers hired by the company and deciding the order in which the truck will visit each of the client destinations. The Transport Manager prefers to work in paper and pencil; then he gives the logistics information to his assistant, the assistant fills the logistics form (see page 7) and sends it to both the client and the supplier. In the Insurance Department, the clerk specifies the insurance clauses, stapling them to the order form. SuperStationery has contracted an insurance policy with an insurance company. The policy has a set of generic insurance clauses (see the Insurance Policy record in page 8). For each order, the Insurance Department clerk can specify additional clauses that extend or restrict the coverage (see page Insurance Clauses form in page 9). The clerk sends the order form and the insurance information back to the Sales Department, where the salesman faxes the insurance information to the client. When the transportation vehicle (usually a truck, but sometimes a van) picks up the goods from the supplier's warehouse, the supplier phones the company to report that the shipments are on their way to their destinations (a timestamp is recorded).

As the company prospers, the amount of orders increases and, thus, the company needs more truck drivers to deliver the goods in time. Therefore, from time to time the transport manager hires a new truck driver. Truck drivers have their own truck (see the Truck Driver record in page 8).

## 2.2. Comments

To keep the case simple, the following assumptions are made.

The catalogue was defined in the first place by the company director and it is him who decides how to update it, but it is the Sales Department manager who actually updates the catalogue table (see an example in page 11).

The company keeps information about its clients. Whenever a client makes an order for the first time, the salesman creates a new client record (see page 6).

It is a local company so it only serves products to clients within a certain range.

The company works with a set of suppliers (the salesman creates a Supplier record on request of the supplier company; see page 10). It is assumed that all providers provide all the products of the catalogue and that all of them provide each product at the same price. However, the supply chain is out of scope of the current development project. SuperStationery keeps no stock of products.

Orders are never assigned partially (i.e. half an order to one supplier and the other half to another supplier). It always happens that sooner or later one supplier has enough stock of all the products in an order and, therefore, it accepts the order.

An order is assigned to one supplier. One supplier can be assigned many orders, even if they are still not delivered.

## 2.1. Business forms

### 2.1.1. Order form

| ORDER | | | | | |
|---|---|---|---|---|---|

**Order number**: 10352        **Request date**:       31-08-2009
**Payment type**: ☒ Cash ☐ Credit ☐ Cheque  **Planned delivery date**: 05-09-2009

**Client**
        **VAT number**: 56746163-R
        **Name**: John Papiro Jr.
        **Telephone**: 030 81 48 31

**Supplier**
        **Code**: OFFIRAP
        **Name**: Office Rapid Ltd.
        **Address**: Brandenburgen street, 46, 2983 Millhaven

**Destination**: Blvd. Blue mountain, 35-14A, 2363 Toontown

**Person in charge**: Brayden Hitchcock

| # | Code | Product name | Price | Q | Amount |
|---|---|---|---|---|---|
| 1 | ST39455 | Rounded scissors (cebra) box-100 | 25,40 € | 35 | 889,00 € |
| 2 | ST65399 | Staples cooper 26-22 blister 500 | 5,60 € | 60 | 336,00 € |
| 3 | CA479-9 | Stereofoam cups box-50 (pack 120) | 18,75 € | 10 | 187,50 € |
| | | | | | 1412,50 € |

**Destination**: Greenhouse street, 23, 2989 Millhaven

**Person in charge**: Luke Padbury

| # | Code | Product name | Price | Q | Amount |
|---|---|---|---|---|---|
| 1 | ST65399 | Staples cooper 26-22 blister 500 | 5,60 € | 30 | 444,50 € |
| 2 | CA746-3 | Sugar lumps 1kg | 2,30 € | 3 | 6,90 € |
| | | | | | 451,40 € |

                                                **Total**  1863,90 €

Form 1. Example of an order form

| Field | Description |
|---|---|
| Order number | A sequential number that identifies the order. |
| Request date | The date in which the client places the order. |
| Payment type | Information about the payment type. Its value is normally either Cash, Credit or Cheque, but the salesman can freely indicate any other information here. |
| Client VAT number | The tax number of the client. |
| Client name | The name and surname of the client. This information appears in the client record. |
| Client telephone | Telephone number to contact the client at work. This information appears in the client record. |

| Supplier code | The mnemonic code of the supplier that has been assigned (and has accepted) the order. |
|---|---|
| Supplier name | The name of the supplier. This information appears in the supplier record. |
| Supplier address | An address of the supplier (e.g. street, number). This information appears in the supplier record. |
| Destination | A client destination at which the products have to be delivered. |
| Product code | The code of a product that is requested by the client. |
| Product name | The name of the product. This information appears in the catalogue. |
| Price | The price of the requested product. The product price in the line takes its value from the current price of the product in the catalogue. |
| Quantity | The amount of items of the product that the client requests at a specific destination. |
| Amount | The value that results of multiplying the price of the product by the quantity. |

One order can have many destinations. One destination can have many lines.

## 2.1.2. Client record

| CLIENT RECORD |
|---|
| **Personal data** |
| **VAT number**: 56746163-R<br>**Client name**: John Papiro Jr.<br>**Telephone**: 030 81 48 31 |
| **Addresses** |
| **Address**: Blvd. Blue mountain, 35-14A<br>**Post code**: 2363<br>**City**: Toontown |
| **Address**: Greenhouse street, 23<br>**Post code**: 2989<br>**City**: Millhaven |
| **Address**: Peggintose street, 345<br>**Post code**: 2697<br>**City**: Groovantia |
| **Address**:<br>**Post** code:<br>**City**: |
| **Registration date**: 29/03/2003 |

Form 2. Example of a client record

| Field | Description |
|---|---|
| VAT number | The tax number of the client. Clients are identified by means of their VAT number. |
| Client name | The name and surname of the client |
| Telephone | Telephone number to contact the client at work |
| Address | An address of the client (e.g. street, number).<br>A client can have several addresses. One address corresponds to only one client. |
| Post code | Post code that corresponds to the client address |
| City | City that corresponds to the client address |
| Registration date | The date in which the record is created |

### 2.1.3. Logistics form

| LOGISTICS | |
|---|---|
| **Order number**: 10352 | **Planned delivery date**:  05-09-2009 |

**Logistics**

 **Truck**: V-5568-FN
 **Driver**: Leonard Kothrapali   **Telephone**: 0666 657 889

**Origin**: Brandenburgen street, 46, 2983 Millhaven

**Itinerary**

| 1 | Greenhouse street, 23, 2989 Millhaven |
|---|---|
| 2 | Blvd. Blue mountain, 35-14A, 2363 Toontown |

**Comments**:
The stereofoam cups have to be packaged carefully or they can break.

Form 3.  Example of a logistics form

| Field | Description |
|---|---|
| Order number | The number of the client order for which the logistics are being arranged |
| Planned delivery date | The date at which the supplier commits to deliver the order. |
| Driver | The truck driver that will deliver the goods requested by the client |
| Truck | The plate number of the truck owned by the driver |
| Telephone | A mobile phone number of the driver |
| Origin | The address of the supplier of the order |
| Destination | The itinerary is an ordered list of the destinations of the order (where the truck has to stop). That is, for each destination, the transport manager establishes the stop order; that is the order in which the truck driver visits that specific destination. |
| Comments | Any additional comments that the transport manager wants to make |

Each order can only be delivered by one truck. A truck can deliver many orders.

### 2.1.4. Truck driver record

| TRUCK DRIVER |
|---|
| **VAT number**: 37236235T<br>**Driver name**: Leonard Kothrapali<br>**Telephone**: 0666 657 889<br>**Plate number**: V-5568-FN<br>**Maximum load (kg)**: 18000 |

Form 4. Example of a truck driver record

| Field | Description |
|---|---|
| VAT number | The VAT number of the truck driver. Truck drivers are identified by means of the VAT number. |
| Driver name | The name of the driver |
| Telephone | A mobile phone number of the driver |
| Plate number | The plate number of the truck owned by the driver |
| Maximum load (kg) | The maximum load permitted to be shipped in the truck |

### 2.1.5. Insurance policy record

| INSURANCE POLICY |
|---|
| **Insurance policy** |
| **Company**: Zuritsz          **Policy number**: 000663979877 |
| **Policy clauses** |
| It requires the assured to act with 'reasonable despatch'---how the owner would act in case the goods are not insured---in all circumstances within his/her control. |
| It gives the insurer the right to pay the assured for the total loss when the goods are so damaged that the cost of recovering and reconditioning would exceed their original value. |
| It excludes the loss or damage<br>   * caused by strikers, or persons taking part in labour disturbances, riots or civil commotions;<br>   * resulting from strikes, lock-outs, disturbances, riots or civil commotions. |
| It provides for reimbursing the assured for the expenses to protect the interest insured from further loss or damage. |
| It covers deliberate damage to or deliberate destruction of the property insured or any part thereof by the wrongful act of any person(s). |

Form 5. Example of an insurance policy record

| Field | Description |
|---|---|
| Company | The insurance policy with which SuperStationery has contracted the insurance policy. |
| Policy number | The policy number refers to a specific policy contracted with the insurance company. Sometimes it includes letters. The company and the policy number together allow identifying an insurance policy. |
| Policy clauses | Clauses that have been defined in the insurance policy. |

## 2.1.6. Order insurance form

| INSURANCE CLAUSES | |
|---|---|
| **Order number**: 10352 | **Planned delivery date**: 05-09-2009 |

| **Insurance policy** | |
|---|---|
| **Company**: Zuritsz | **Policy number**: 000663979877 |

**Policy clauses**

It requires the assured to act with 'reasonable despatch'---how the owner would act in case the goods are not insured---in all circumstances within his/her control.

It gives the insurer the right to pay the assured for the total loss when the goods are so damaged that the cost of recovering and reconditioning would exceed their original value.

It excludes the loss or damage
   * caused by strikers, or persons taking part in labour disturbances, riots or civil commotions;
   * resulting from strikes, lock-outs, disturbances, riots or civil commotions.

It provides for reimbursing the assured for the expenses to protect the interest insured from further loss or damage.

It covers deliberate damage to or deliberate destruction of the property insured or any part thereof by the wrongful act of any person(s).

**Extraordinary clauses**

The shipment is insured until the truck arrives at the destination dock.

Form 6. Example of an order insurance form

| Field | Description |
|---|---|
| Order number | The client order to which the insurance information is attached. |
| Planned delivery date | The date at which the supplier commited to deliver the order. |
| Company | The insurance policy with which SuperStationery has contracted the insurance policy. |
| Policy number | The number of the insurance policy that applies to the shipping of the order. |
| Policy clauses | The clauses specified by the insurance policy. |
| Extraordinary clauses | Clauses that are specifically defined for this order shipping. |

## 2.1.7. Supplier record

| SUPPLIER RECORD |
|---|
| **Code**: OFFIRAP |
| **Supplier name**: Office Rapid Ltd. |
| **VAT number**: 73658762H |
| |
| **Telephone**: 053 73 63 88    **Address**: Brandenburgen street, 46 |
| **Post code**: 2983    **City**: Millhaven |
| |
| **Registration date**:  21/11/2000 |

Form 7.  Example of a supplier record

| Field | Description |
|---|---|
| Code | A mnemonic code that is assigned to the supplier |
| Supplier name | The name and surname of the supplier |
| VAT number | The tax number of the supplier |
| Telephone | Telephone number to contact the supplier at work |
| Address | An address of the supplier (e.g. street, number) |
| Post code | Post code that corresponds to the supplier address |
| City | City that corresponds to the supplier address |
| Registration date | The date in which the record is created |

Suppliers are identified by means of the mnemonic code.

## 2.1.8. Company catalogue

| CATALOGUE |
|---|
| **Code**: ST39450 **Price**: 1,80 € **Product name**: GBooo stock text stamp - Draft<br>**Comments**: Premade text stamp with the word Draft |
| **Code**: ST39451 **Price**: 1,40 € **Product name**: Address labels 100x50mm (150)<br>**Comments**: |
| **Code**: ST39452 **Price**: 1,80 € **Product name**: Address labels 75x38mm (200)<br>**Comments**: |
| **Code**: ST39453 **Price**: 14,50 € **Product name**: Rounded scissors (plain) box-100<br>**Comments**: Each pair of scissors has its own plastic cover |
| **Code**: ST39454 **Price**: 30,40 € **Product name**: Sharp scissors alum. box-40<br>**Comments**: Rustproof and handy office scissors |
| **Code**: ST39455 **Price**: 17,20 € **Product name**: Rounded scissors (cebra) box-100<br>**Comments**: Each pair of scissors has its own plastic cover |
| **Code**: ST39456 **Price**: 0,30 € **Product name**: Stick'n'tape clear tape 12mm x 33m<br>**Comments**: |
| **Code**: ST39457 **Price**: 0,40 € **Product name**: Stick'n'tape clear tape 12mm x 66m<br>**Comments**: |
| **Code**: ST39458 **Price**: 0,45 € **Product name**: Stick'n'tape clear tape 18mm x 66m<br>**Comments**: |
| **Code**: ST39459 **Price**: 0,60 € **Product name**: Stick'n'tape clear tape 24mm x 66m<br>**Comments**: |

Form 8. Example of a page of the catalogue

| Field | Description |
|---|---|
| Code | A mnemonic code that is assigned to a product |
| Price | The price of the product at which is sold to clients |
| Product name | The name of the product |
| Comments | An optional note about the product |

Products are identified by means of a mnemonic code.

# 3. COMMUNICATION ANALYSIS REQUIREMENTS MODEL

## 3.1. Introduction

SuperStationery Co. is a company that provides stationery and office material to its clients. The company acts as a as intermediary: the company has a catalogue of products that are bought from suppliers and sold to clients.

The company has a director and is divided into four departments. Figure 1 shows the organisation chart.

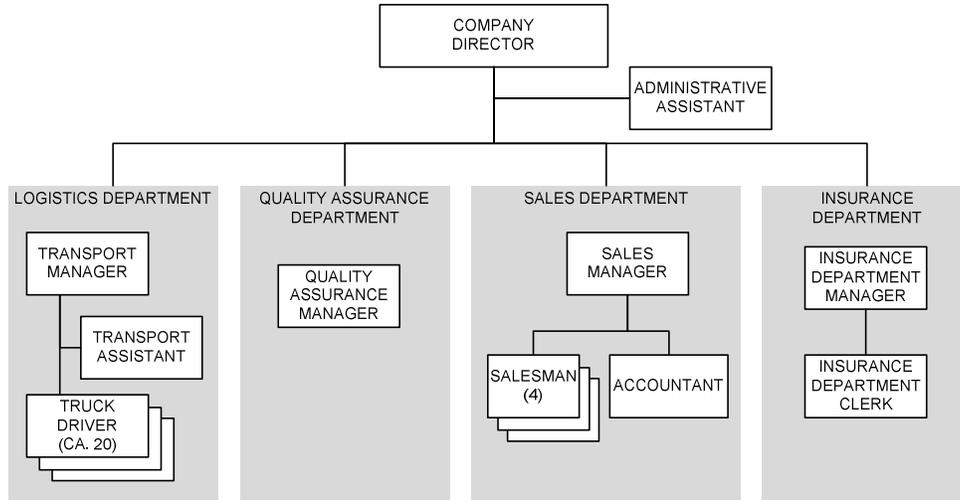

Figure 1. SuperStationery Co. organisation chart

The company work practice has been decomposed into business processes. The following table relates the business processes and the departments that participate in them.

| Business process | | Departments | | | | |
|---|---|---|---|---|---|---|
| Acronym | Name | Director | Logistics | Quality A. | Sales | Insurance |
| CLIE | Client management | X | | | X | |
| PROD | Product management | X | | X | | |
| LOGI | Logistics | | X | | | |
| SALE | Sales management | | X | | X | X |
| RISK | Risk management | | | | | X |
| ACCO | Accounting | | | | X | |
| SUPP | Supplier management | X | | | X | |

Some business processes fall entirely within the scope of a single department (e.g. *Logistics* business process), while others crosscut several departments (e.g. *Sales management* business process).

## 3.2. Communicative Event Diagram

Figure 2 presents part of the communicative event diagram of the *Sales management* business process[3].

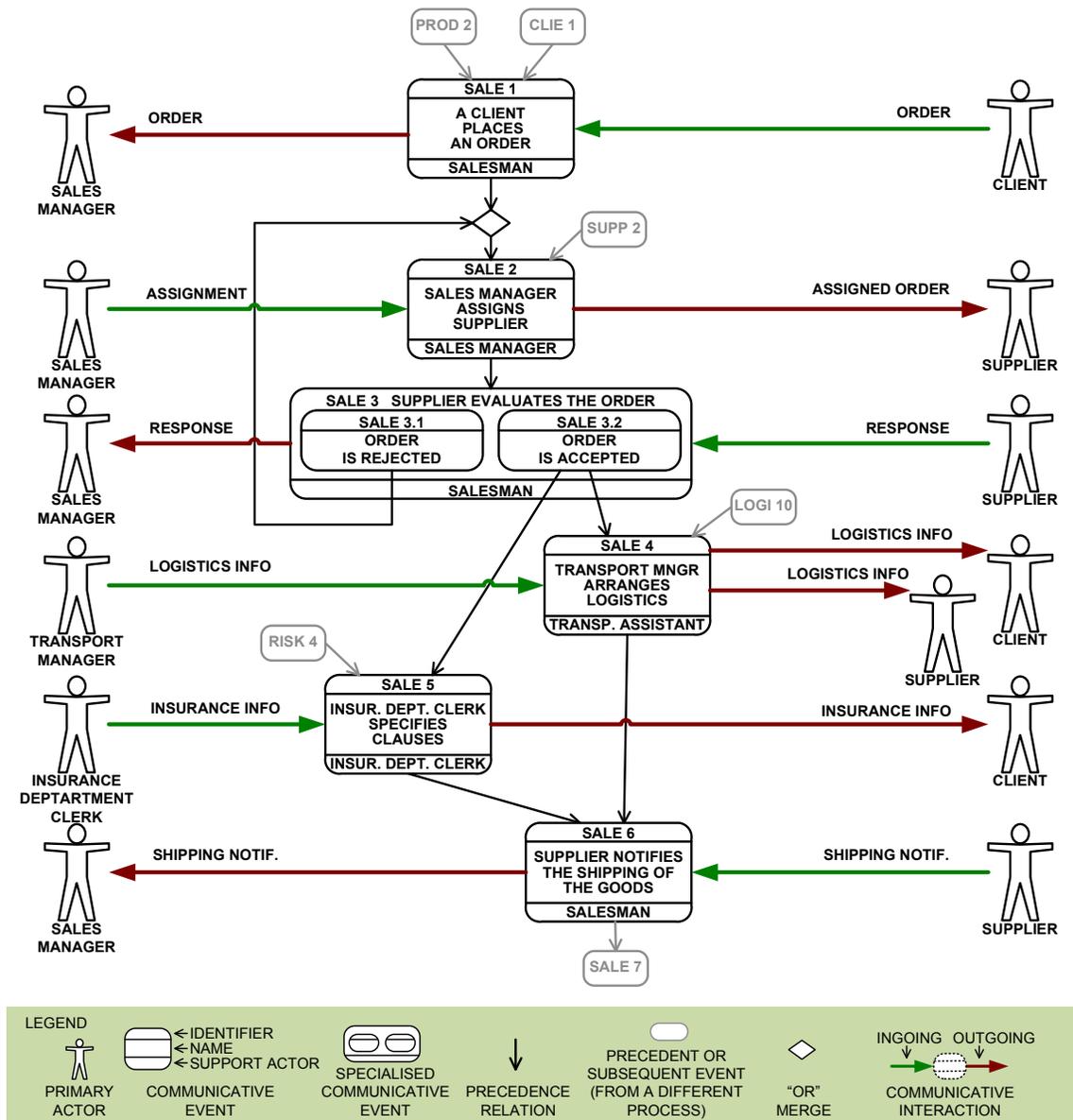

Figure 2.  Communicative event diagram of SuperStationery Co. *Sales management* business process (SALE)

Note that some communicative events that appear in Figure 2 have precedent events that belong to other business processes. For instance, the precedent communicative events of SALE 1 are PROD 2 and CLIE 1 and they belong to *Product management* and *Client management* business processes, respectively. These business processes are specified separately; i.e. they have their own communicative event diagrams and event specification templates. For instance, Figure 3 presents the communicative event diagram of the *Client management* business process.

---

[3] Some communicative events are left out of the scope of this lab demo (e.g. SALE 7. Truck returns to base).

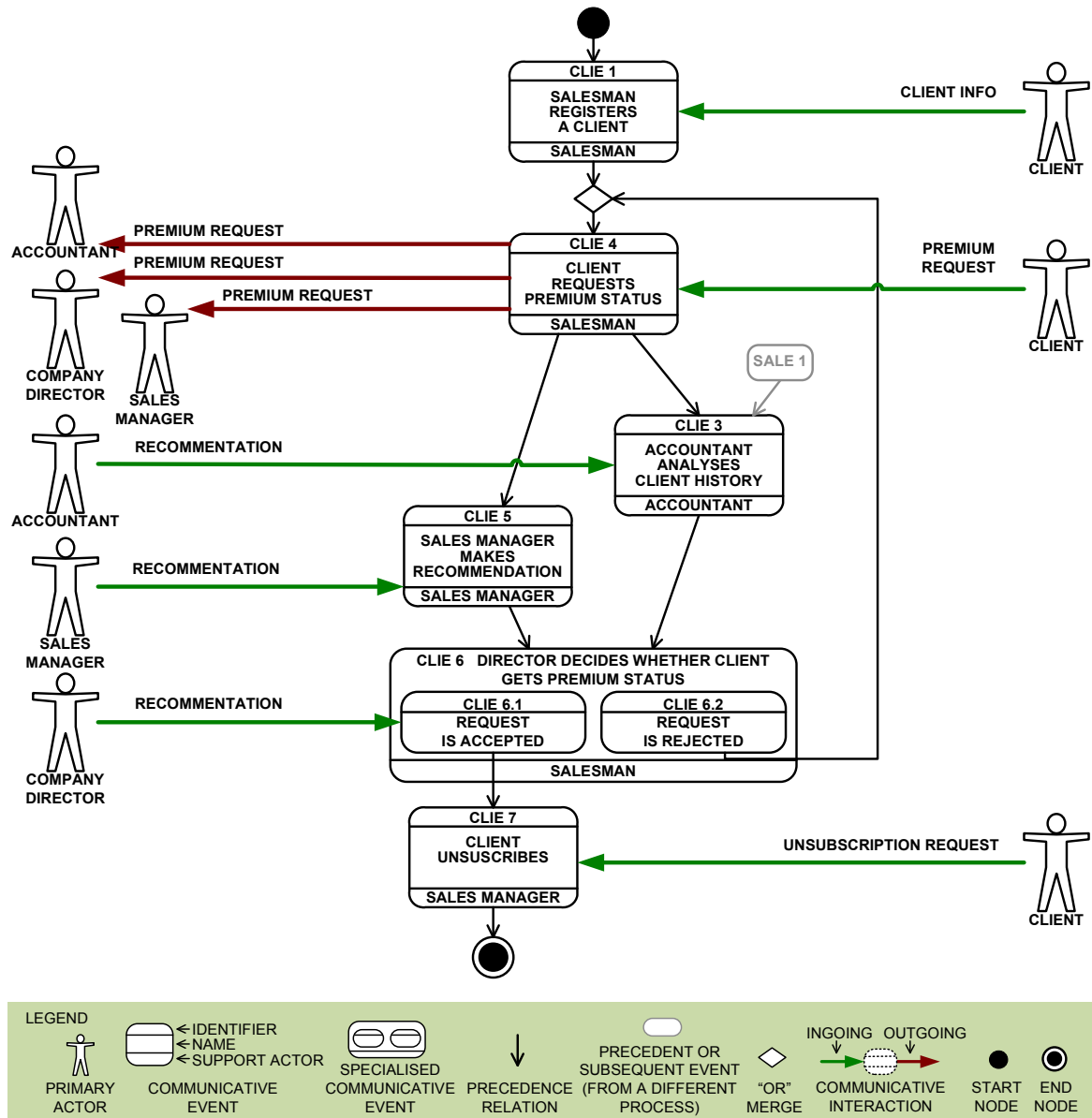

Figure 3. Communicative event diagram of SuperStationery Co. *Client management* business process (Clie)

It is important to remark that the numbering of the communicative events in a Communicative Event Diagram does not imply temporal precedence; Sale 1, Sale 2, Clie 1 and Clie 2 are simply identifiers. In the communicative event diagram of the Sale business process, the order of the numbering and the temporal order of the events coincide. The diagram specifies that Sale 1 precedes Sale 2 by means of a precedence relationship. In contrast, in the communicative event diagram of the Clie business process, order of the numbering and the temporal order of the events does not coincide (e.g. despite the numbering, Clie 4 precedes Clie 3). Moreover, there is an identifier that is missing; namely Clie 2. This does not imply incompleteness of the specification. It may simply be due to a business process reengineering that has resulted in the elimination of a communicative event, or it may have happened inadvertently while the analyst numbered the events.

It is also noteworthy that the communicative event diagram of the Sale business process does not include a start node, while Clie does. A communicative event is preceded by the start node when it is not preceded by any communicative event. In the Sale business process each of its communicative events has at least one precedent event, either from the Sale business process itself (e.g. Sale 2 has Sale 1 as precedent) or from a different business process (e.g. Sale 1 has Clie 1 and Prod 2 as precedents). In contrast, in the Clie business process no communicative event precedes Clie 1; therefore Clie 1 is preceded by the start node.

A similar line of reasoning is done for the end node. A communicative event is followed by an end node when it does not precede any other communicative event. In the Sale business process each of its communicative events precedes another event (even Sale 6 precedes Sale 7, although it is out of the scope of the lab demo). In contrast, in the Clie business process Clie 7 does not precede any other communicative event; therefore, Clie 7 is followed by an end node.

## 3.3. Event description templates

| **SALE 1. A client places an order** |
| --- |

### 1    General information

**Goals**

The objective of the organisation is to attend the clients when they request goods.

From the point of view of the information system, the objective of this event is to record the order that the client places, and to let the Sales Manager know that a new order has arrived.

**Description**

Most clients call the Sales Department, where they are attended by a salesman. Then the client requests one or several products that are to be sent to one or many destinations. The salesman takes note of the order. Other clients place orders by email or by fax.

### 2    Contact requirements

**Actor responsibilities**
- **Primary actor**: Client
- **Communication channel**: In person, by phone, by fax
- **Support actor**: Salesman

**Temporal requirements**
- **Occurrence temporal restrictions**: Only working days during reception hours (09:00-18:00)
- **Frequency of occurrence**: 500 orders per week

**Business forms**

| ORDER | | | | | |
|---|---|---|---|---|---|

**Order number**: 10352      **Request date**:      31-08-2009
**Payment type**: ☒ Cash ☐ Credit ☐ Cheque      **Planned delivery date**: 05-09-2009

**Client**
     **VAT number**: 56746163-R
     **Name**: John Papiro Jr.
     **Telephone**: 030 81 48 31

**Supplier**
     **Code**: OFFIRAP
     **Name**: Office Rapid Ltd.
     **Address**: Brandenburgen street, 46, 2983 Millhaven

**Destination**: Blvd. Blue mountain, 35-14A, 2363 Toontown

**Person in charge**: Brayden Hitchcock

| # | Code | Product name | Price | Q | Amount |
|---|---|---|---|---|---|
| 1 | ST39455 | Rounded scissors (cebra) box-100 | 25,40 € | 35 | 889,00 € |
| 2 | ST65399 | Staples cooper 26-22 blister 500 | 5,60 € | 60 | 336,00 € |
| 3 | CA479-9 | Stereofoam cups box-50 (pack 120) | 18,75 € | 10 | 187,50 € |
| | | | | | 1412,50 € |

**Destination**: Greenhouse street, 23, 2989 Millhaven

**Person in charge**: Luke Padbury

| # | Code | Product name | Price | Q | Amount |
|---|---|---|---|---|---|
| 1 | ST65399 | Staples cooper 26-22 blister 500 | 5,60 € | 30 | 444,50 € |
| 2 | CA746-3 | Sugar lumps 1kg | 2,30 € | 3 | 6,90 € |
| | | | | | 451,40 € |
| | | | | **Total** | 1863,90 € |

Form 9.  Example of an order form

Some parts of the form are not yet filled in this event.

## 3   Communication requirements

**Message structure**

| FIELD | OP | DOMAIN | EXAMPLE VALUE |
|---|---|---|---|
| ORDER = | | | |
| < Order number + | g | number | 10352 |
|    Request date + | i | date | 31-08-2009 |
|    Payment type + | i | text | Cash |
|    Client + | i | Client | 56746163-R, John Papiro Jr. |
|    DESTINATIONS = | | | |
|    { DESTINATION = | | | |
|     < Address + | i | Client address | Blvd. Blue mountain, 35-14A, 2363 Toontown |
|      Person in charge + | i | text | Brayden Hitchcock |
|     LINES = | | | |
|     { LINE = | | | |
|      < Product + | i | Product | ST39455, Rounded scissors (cebra) box-100 |
|       Price + | i | money | 25,40 € |
|       Quantity > | i | number | 35 |
|     } | | | |
|     > | | | |
|    } | | | |
| > | | | |

| Field | Description |
|---|---|
| Order number | A sequential number that identifies the order. |
| Request date | The date in which the client places the order. |
| Payment type | Information about the payment type. Its value is normally either Cash, Credit or Cheque, but the salesman can freely indicate any other information here. |
| Client | The client that places the order. |
| Address | A client destination at which the products have to be delivered. |
| Person in charge | The name of the person that will receive the order at the destination. From one order to another, this person can be different. |
| Product | A product that is requested by the client. |
| Price | The price of the requested product. |
| Quantity | The amount of items of the product that the client requests at a specific destination. |

**Structural restrictions**

One order can have many destinations.

One destination can have many lines.

**Contextual restrictions**

Orders are identified by Order number.

The product price in the line takes its value from the current price of the product in the catalogue.

## 4   Reaction requirements

**Treatments**

The order is recorded.

**Linked communications**

The Sales Manager is informed of the order placement.

| SALE 2. Sales Manager assigns supplier |
|---|

## 1 General information

**Goals**

The objective of the organisation is to select the most appropriate supplier so that they serve the goods that the customer has ordered.

From the point of view of the information system, the objective is to allow the Sales Manager take an informed decision when choosing the supplier.

**Description**

The Sales Manager reviews the order and assigns it to one of the many suppliers that work with the company (the Sales Manager notes down the Supplier section of the order form; additionally, the company wants to record the assignment date). An order form is sent by fax to the supplier.

## 2 Contact requirements

**Actor responsibilities**
- **Primary actor**: Sales Manager
- **Communication channel**: In person
- **Support actor**: Sales Manager

**Temporal requirements**
- **Frequency of occurrence**: 500 orders per week

**Business forms**

Order form, see event SALE 1 (page 15).

## 3 Communication requirements

**Message structure**

| FIELD | OP | DOMAIN | EXAMPLE VALUE |
|---|---|---|---|
| ASSIGNMENT = | | | |
| < Order + | i | Client order | 2008-00352 |
| Supplier + | i | Supplier | OFFIRAP, Office Rapid Ltd. |
| Assignment date | i | date | 01-09-2009 |
| > | | | |

| Field | Description |
|---|---|
| Order | The order that is being assigned |
| Supplier | The supplier to which the order is assigned |
| Assignment date | The date in which the assignment is done |

**Structural restrictions**

An order is assigned to one supplier.

One supplier can be assigned many orders, even if they are still not delivered.

## 4 Reaction requirements

**Treatments**

The order is updated.

**Linked communications**

The order form is sent to the supplier.

---

## SALE 3. Supplier evaluates the order

## 1    General information

**Goals**

The objective of the organisation is to know whether the selected supplier will serve the order or not in order to act accordingly (e.g. to select a different supplier in case of rejection). From the point of view of the information system, the objective is to let the manager know the supplier's response.

**Description**

The supplier receives the order form and checks whether they have enough stock or not. In case they have enough stock, they accept the order; otherwise, they reject it. In case the order is rejected, the Sales Manager assigns it to a different supplier (this can happen many times until the order is accepted). Once the order is accepted, the salesman sends a copy of the order to the Transport Department and the Insurance Department.

## 2    Contact requirements

**Actor responsibilities**
- **Primary actor**: Supplier.
- **Communication channel**: Mainly by fax; sometimes phone or email.
- **Support actor**: Salesman

**Temporal requirements**
- **Frequency of occurrence**: 500 orders per week

**Required supports**

Order form, see event SALE 1 (page 15).

## 3    Communication requirements

**Message structure**

| FIELD | OP | DOMAIN | EXAMPLE VALUE |
|---|---|---|---|
| RESPONSE = | | | |
| < Order + | i | Client order | 10352 |
| Decision + | i | [accepted\|rejected] | accepted |
| Planned delivery date + | i | date | 5-09-2009 |
| Response date | i | date | 1-09-2009 |
| > | | | |

| Field | Description |
|---|---|
| Order | The client order for which the supplier is giving a response. |
| Decision | The decision of the supplier; i.e. whether the supplier accepts or rejects supplying the order (this mainly depends on the available stock). |
| Planned delivery date | The date at which the supplier commits to deliver the order. |
| Response date | The date at which the supplier responds. |

## 4    Reaction requirements

**Treatments**

In case the order is rejected, the supplier information is kept but the order is marked as rejected so the Sales Manager can reassign it to a different supplier.

**Linked behaviour**

In case the order is rejected, the Sales Manager has to reassign it to a different supplier.

**Linked communications**

Once the order is accepted, the salesman sends a copy of the order to the Transport Department and the Insurance Department.

## SALE 4. Transport manager arranges logistics

### 1    General information

**Goals**
The objective of the organisation is to arrange how the goods will be shipped to the customer.

**Description**
In the Transport Department, the Transport Manager arranges how the goods will be carried to the destinations; this implies selecting one of the trucks owned by the company and deciding the order in which the truck will visit each of the client destinations. The Transport Department prefers to work in paper and pencil; then he gives the logistics information to his assistant, the assistant fills the logistics form and sends it to both the client and the supplier.

### 2    Contact requirements

**Actor responsibilities**
- **Primary actor**: Transport Manager.
- **Communication channel**: In person.
- **Support actor**: Assistant

**Temporal requirements**
- **Frequency of occurrence**: 500 orders per week

**Business forms**

| LOGISTICS | |
|---|---|
| **Order number**: 10352 | **Planned delivery date**:   05-09-2009 |

| Logistics |
|---|
| **Truck**: V-5568-FN |
| **Driver**: Leonard Kothrapali            **Telephone**: 0666 657 889 |

| **Origin**: Brandenburgen street, 46, 2983 Millhaven |
|---|

| Itinerary | |
|---|---|
| 1 | Greenhouse street, 23, 2989 Millhaven |
| 2 | Blvd. Blue mountain, 35-14A, 2363 Toontown |

| **Comments**: The stereofoam cups have to be packaged carefully or they can break. |
|---|

Form 10.  Example of a logistics form

## 3 Communication requirements

**Message structure**

| FIELD | OP | DOMAIN | EXAMPLE VALUE |
|---|---|---|---|
| LOGISTICS INFO = | | | 10352 |
| < Order + | i | Client order | |
|   Driver + | i | Truck driver | V-5568-FN  Leonard Kothrapali    0666 657 889 |
|   Logistic comments  + | i | Text | Stereofoam cups have to be packaged carefully or they can break. |
| | | | |
|   ITINERARY = | | | |
|   { Destination + | i | { Order.Destinations } | Blvd. Blue mountain, 35-14A, 2363 Toontown |
|     Stop order } | i | number | 1 |
| > | | | |

| Field | Description |
|---|---|
| Order | The client order for which the logistics are being arranged |
| Driver | The truck driver that will deliver the goods requested by the client |
| Logistic comments | Any additional comments that the transport manager wants to make |
| Destination | The itinerary is an ordered list of the destinations of the order (where the truck has to stop). |
| Stop order | That is, for each destination, the transport manager establishes the stop order; that is the order in which the truck driver visits that specific destination. |

**Structural restrictions**

Each order can only be delivered by one truck.

A truck can deliver many orders.

The itinerary can be composed of several destinations.

Each destination can only appear in one itinerary.

**Contextual restrictions**

Itinerary consists of an ordered list of all the destinations in the order.

## 4 Reaction requirements

**Treatments**

The logistics information is stored.

**Linked communications**

Logistics information is sent to both the client and the supplier.

## SALE 5. Insurance department clerk specifies clauses

### 1    General information

**Goals**

The objective of the organisation is to specify which insurance policy covers the shipment.

**Description**

In the Insurance Department, the clerk specifies the insurance clauses, stapling them to the order form. SuperStationery has contracted an insurance policy with an insurance company. The policy has a set of generic insurance clauses. For each order, the Insurance Department clerk can specify additional clauses that extend or restrict the coverage. The clerk sends the order form and the insurance information back to the Sales Department, where the salesman faxes the insurance information to the client.

### 2    Contact requirements

**Actor responsibilities**
- **Primary actor**: Insurance Department clerk.
- **Communication channel**: In person.
- **Support actor**: Salesman

**Temporal requirements**
- **Frequency of occurrence**: 500 orders per week

**Business forms**

| INSURANCE CLAUSES |
|---|
| **Order number**: 10352                    **Planned delivery date**:    05-09-2009 |
| **Insurance policy** <br>                    **Company**: Zuritsz                    **Policy number**: 000663979877 |
| **Policy clauses** |
| It requires the assured to act with 'reasonable despatch'---how the owner would act in case the goods are not insured---in all circumstances within his/her control. |
| It gives the insurer the right to pay the assured for the total loss when the goods are so damaged that the cost of recovering and reconditioning would exceed their original value. <br> It excludes the loss or damage <br>    * caused by strikers, or persons taking part in labour disturbances, riots or civil commotions; <br>    * resulting from strikes, lock-outs, disturbances, riots or civil commotions. |
| It provides for reimbursing the assured for the expenses to protect the interest insured from further loss or damage. |
| It covers deliberate damage to or deliberate destruction of the property insured or any part thereof by the wrongful act of any person(s). |
| **Extraordinary clauses** |
| The shipment is insured until the truck arrives at the destination dock. |

Form 11.  Example of an order insurance form

## 3   Communication requirements

**Message structure**

| FIELD | OP | DOMAIN | EXAMPLE VALUE |
|---|---|---|---|
| INSURANCE INFO = | | | |
| < Order + | i | Client order | 10352 |
| Policy + | i | Insurance policy | Zuritsz p.n. 000663979877 |
| Extra clauses | i | Text | The shipment is insured until the truck arrives at the destination dock |
| > | | | |

| Field | Description |
|---|---|
| Order | The client order to which the insurance information is attached. |
| Policy | The insurance policy that applies to the shipping of the order. The insurance policy has contracted by SuperStationery with an insurance company. |
| Extra clauses | Clauses that are specifically defined for this order shipping. |

**Structural restrictions**

Each order can be added several extraordinary clauses, but they can be treated as a mere textual clarification.

## 4   Reaction requirements

**Treatments**

The insurance information is stored.

**Linked communications**

Insurance information is sent to the client.

## SALE 6. Supplier notifies the shipping of the goods

### 1    General information

**Goals**

The objective of the organisation is to know ensure that the delivery is made on time.

**Description**

When the transportation vehicle (usually a truck, but sometimes a van) picks up the goods from the supplier's warehouse, the supplier phones the company to report that the shipments are on their way to their destinations. Salesman usually picks up the phone and reports the information to the Sales Manager.

### 2    Contact requirements

**Actor responsibilities**
- **Primary actor**: Supplier
- **Communication channel**: In person
- **Support actor**: Salesman

**Temporal requirements**
- **Frequency of occurrence**: 500 orders per week

### 3    Communication requirements

**Message structure**

| FIELD | OP | DOMAIN | EXAMPLE VALUE |
|---|---|---|---|
| DELIVERY NOTIFIC = | | | |
| < Order + | i | Client order | 10352 |
| Shipping timestamp | i | date | 05-09-2009 12:34 |
| > | | | |

| Field | Description |
|---|---|
| Order | The client order that has been shipped |
| Shipping timestamp | The date and time in which the order has been shipped in the truck |

### 4    Reaction requirements

**Treatments**

The order is marked as "Delivered".

**Linked communications**

Sales Manager has to know about this occurrence.

## 3.4. Event description templates (from other business processes)

---
**CLIE 1. Salesman registers a client**
---

### 1 General information

**Goals**

The objective of the organisation is having the more clients the better.

From the point of view of the information system, the objective of this event is to keep a registry of clients.

**Description**

Whenever a client places an order for the first time, the salesman creates a client record.

### 2 Contact requirements

**Actor responsibilities**
- **Primary actor**: Client
- **Communication channel**: In person, by phone, by fax
- **Support actor**: Salesman

**Temporal requirements**
- **Occurrence temporal restrictions**: Only attended on working days during reception hours (09:00-18:00)
- **Frequency of occurrence**: 10 new clients per month.

**Business forms**

| CLIENT RECORD |
| --- |
| **Personal data** |
| **VAT number**: 56746163-R<br>**Client name**: John Papiro Jr.<br>**Telephone**: 030 81 48 31 |
| **Addresses** |
| **Address**: Blvd. Blue mountain, 35-14A<br>**Post code**: 2363<br>**City**: Toontown |
| **Address**: Greenhouse street, 23<br>**Post code**: 2989<br>**City**: Millhaven |
| **Address**: Peggintose street, 345<br>**Post code**: 2697<br>**City**: Groovantia |
| **Address**:<br>**Post** code:<br>**City**: |
| **Registration date**:  29/03/2003 |

Form 12.  Example of a client record

## 3    Communication requirements

**Message structure**

| FIELD | OP | DOMAIN | EXAMPLE VALUE |
| --- | --- | --- | --- |
| CLIENT INFO = | | | |
| < VAT number + | i | text | 56746163-R |
|     Client name + | i | text | John Papiro Jr. |
|     Telephone + | i | text | 030 81 48 31 |
|     Registration date + | i | date | 29/03/2003 |
|     ADDRESSES = | | | |
|     { CLIENT ADDRESS = | | | |
|       < Address + | i | text | Blvd. Blue mountain, 35-14A |
|           Post code + | i | text | 2363 |
|           City> | i | text | Toontown |
|     } | | | |
| > | | | |

| Field | Description |
|---|---|
| VAT number | The tax number of the client |
| Client name | The name and surname of the client |
| Telephone | Telephone number to contact the client at work |
| Registration date | The date in which the record is created |
| Address | An address of the client (e.g. street, number) |
| Post code | Post code that corresponds to the client address |
| City | City that corresponds to the client address |

**Structural restrictions**

A client can have several addresses.

One address corresponds to only one client.

**Contextual restrictions**

Clients are identified by means of their VAT number.

## 4   Reaction requirements

**Treatments**

The client information is recorded.

## SUPP 2. Salesman registers a supplier

## 1    General information

**Goals**

The objective of the organisation is having suppliers that can provide the goods that the clients order.

From the point of view of the information system, the objective of this event is to keep a registry of suppliers.

**Description**

Salesmen are contacted by suppliers that want to work with the company. The salesman then creates a supplier record to keep the information of the supplier. Suppliers also indicate the products they can provide the company.

## 2    Contact requirements

**Actor responsibilities**
- **Primary actor**: Supplier
- **Communication channel**: In person, by phone, by fax
- **Support actor**: Salesman

**Temporal requirements**
- **Occurrence temporal restrictions**: Only attended on working days during reception hours (09:00-18:00)
- **Frequency of occurrence**: 2 new suppliers per month.

**Business forms**

| SUPPLIER RECORD |
| --- |
| **Code**: OFFIRAP |
| **Supplier name**: Office Rapid Ltd. |
| **VAT number**: 73658762H |
| |
| **Telephone**: 053 73 63 88    **Address**: Brandenburgen street, 46 |
| **Post code**: 2983    **City**: Millhaven |
| |
| **Registration date**:  21/11/2000 |

Form 13.  Example of a supplier record

## 3 Communication requirements

**Message structure**

| FIELD | OP | DOMAIN | EXAMPLE VALUE |
|-------|----|--------|--------------|
| SUPPLIER INFO = | | | |
| < Code + | i | text | OFFIRAP |
|    Supplier name + | i | text | Office Rapid Ltd. |
|    VAT number + | i | text | 73658762H |
|    Telephone + | i | text | 053 73 63 88 |
|    Address + | i | text | Brandenburgen street, 46 |
|    Post code + | i | text | 2983 |
|    City + | i | text | Millhaven |
|    Registration date | i | date | 21/11/2000 |
| > | | | |

| Field | Description |
|-------|-------------|
| Code | A mnemonic code that is assigned to the supplier |
| Supplier name | The name and surname of the supplier |
| VAT number | The tax number of the supplier |
| Telephone | Telephone number to contact the supplier at work |
| Address | An address of the supplier (e.g. street, number) |
| Post code | Post code that corresponds to the supplier address |
| City | City that corresponds to the supplier address |
| Registration date | The date in which the record is created |

**Contextual restrictions**

Suppliers are identified by means of a mnemonic code.

## 4 Reaction requirements

**Treatments**

The supplier information is recorded.

## PROD 2. Company director defines catalogue

### 1    General information

**Goals**
The objective of the organisation is defining a catalogue of products that are offered to clients.

**Description**
The general director decides which products belong to the company catalogue. He sets the price at which each product is sold to the clients.

### 2    Contact requirements

**Actor responsibilities**
- **Primary actor**: Company director
- **Support actor**: Sales Department manager

**Temporal requirements**
- **Frequency of occurrence**: Once or twice a year.

**Business forms**

| CATALOGUE |
|---|
| **Code**: ST39450    **Price**:    1,80 €    **Product name**: GBooo stock text stamp - Draft<br>**Comments**: Premade text stamp with the word Draft |
| **Code**: ST39451    **Price**:    1,40 €    **Product name**: Address labels 100x50mm (150)<br>**Comments**: |
| **Code**: ST39452    **Price**:    1,80 €    **Product name**: Address labels 75x38mm (200)<br>**Comments**: |
| **Code**: ST39453    **Price**:    14,50 €    **Product name**: Rounded scissors (plain) box-100<br>**Comments**: Each pair of scissors has its own plastic cover |
| **Code**: ST39454    **Price**:    30,40 €    **Product name**: Sharp scissors alum. box-40<br>**Comments**: Rustproof and handy office scissors |
| **Code**: ST39455    **Price**:    17,20 €    **Product name**: Rounded scissors (cebra) box-100<br>**Comments**: Each pair of scissors has its own plastic cover |
| **Code**: ST39456    **Price**:    0,30 €    **Product name**: Stick'n'tape clear tape 12mm x 33m<br>**Comments**: |
| **Code**: ST39457    **Price**:    0,40 €    **Product name**: Stick'n'tape clear tape 12mm x 66m<br>**Comments**: |
| **Code**: ST39458    **Price**:    0,45 €    **Product name**: Stick'n'tape clear tape 18mm x 66m<br>**Comments**: |
| **Code**: ST39459    **Price**:    0,60 €    **Product name**: Stick'n'tape clear tape 24mm x 66m<br>**Comments**: |

Form 14.  Example of a page of the catalogue

## 3    Communication requirements

**Message structure**

| FIELD | OP | DOMAIN | EXAMPLE VALUE |
|---|---|---|---|
| CATALOGUE = <br> { PRODUCT = <br>    < Product code + <br>      Product name + <br>      Price + <br>      Comments <br>   > <br> } | <br><br> i <br> i <br> i <br> i | <br><br> text <br> text <br> money <br> text | <br><br> ST39455 <br> Rounded scissors (cebra) box-100 <br> 17,20 € <br> Each pair of scissors has its own plastic cover |

| Field | Description |
|---|---|
| Product code | A mnemonic code that is assigned to the product |
| Product name | The name of the product |
| Price | The price of the product at which is sold to clients |
| Comments | An optional note about the product |

**Contextual restrictions**

Products are identified by means of a mnemonic code.

## 4    Reaction requirements

**Treatments**

The product information is recorded.

## LOGI 10. *Transport manager hires truck driver*

### 1    General information

**Goals**

The objective of the organisation is to have a small fleet of trucks that can deliver the orders.

From the point of view of the IS, the objective is to keep record of the available truck drivers.

**Description**

As the company prospers, the amount of orders increases and, thus, the company needs more truck drivers to deliver the goods in time. Therefore, from time to time the transport manager hires a new truck driver. Truck drivers have their own truck.

### 2    Contact requirements

**Actor responsibilities**
- **Primary actor**: Truck driver
- **Support actor**: Transport manager

**Temporal requirements**
- **Frequency of occurrence**: 3 or 4 times a year.

**Business forms**

| TRUCK DRIVER |
|---|
| **VAT number**: 37236235T<br>**Driver name**: Leonard Kothrapali<br>**Telephone**: 0666 657 889<br>**Plate number**: V-5568-FN<br>**Maximum load (kg)**: 18000 |

Form 15.  Example of a truck driver record

## 3 Communication requirements

**Message structure**

| FIELD | OP | DOMAIN | EXAMPLE VALUE |
|---|---|---|---|
| TRUCK DRIVER = | | | |
| < VAT_number + | i | text | 37236235T |
|   Driver name + | i | text | Leonard Kothrapali |
|   Telephone + | i | text | 0666 657 889 |
|   Plate number + | i | text | V-5568-FN |
|   Maximum load (kg) | i | number | 18000 |
| > | | | |

| Field | Description |
|---|---|
| VAT number | The VAT number of the truck driver |
| Driver name | The name of the driver |
| Telephone | A mobile phone number of the driver |
| Plate number | The plate number of the truck owned by the driver |
| Maximum load (kg) | The maximum load permitted to be shipped in the truck |

**Contextual restrictions**

Truck drivers are identified by means of the VAT number.

## 4 Reaction requirements

**Treatments**

The information of the truck driver is recorded.

## RISK 4. Insurance department clerk contracts insurance policy

## 1  General information

**Goals**

The objective of the organisation is to have the shipping of each delivery covered by an insurance policy.

From the point of view of the information system, the objective of this event is to keep record of the contracted insurance policies.

**Description**

The insurance department clerk is always looking for insurance policies that can be contracted to cover the shipping during deliveries. When he finds a suitable insurance policy then he contracts it and makes an insurance policy record.

## 2  Contact requirements

**Actor responsibilities**
- **Primary actor**: Truck driver
- **Support actor**: Transport manager

**Temporal requirements**
- **Frequency of occurrence**: 3 or 4 times a year.

**Business forms**

| INSURANCE POLICY | |
|---|---|
| **Insurance policy** | |
| **Company**: Zuritsz | **Policy number**: 000663979877 |
| **Policy clauses** | |
| It requires the assured to act with 'reasonable despatch'---how the owner would act in case the goods are not insured---in all circumstances within his/her control. | |
| It gives the insurer the right to pay the assured for the total loss when the goods are so damaged that the cost of recovering and reconditioning would exceed their original value. | |
| It excludes the loss or damage<br> * caused by strikers, or persons taking part in labour disturbances, riots or civil commotions;<br> * resulting from strikes, lock-outs, disturbances, riots or civil commotions.<br>It provides for reimbursing the assured for the expenses to protect the interest insured from further loss or damage. | |
| It covers deliberate damage to or deliberate destruction of the property insured or any part thereof by the wrongful act of any person(s). | |

Form 16.  Example of an insurance policy record

## 3 Communication requirements

**Message structure**

| FIELD | OP | DOMAIN | EXAMPLE VALUE |
|---|---|---|---|
| INSURANCE POLICY = <br>< Company + <br>  Policy number + <br>  Clauses + <br>> | i <br>i <br>i | Text <br>Text <br>Text | Zuritsz <br>000663979877 <br>It requires the assured to act with 'reasonable despatch'... |

| Field | Description |
|---|---|
| Company | The insurance company |
| Policy number | The policy number refers to a specific policy contracted with the insurance company. Sometimes it includes letters. |
| Clauses | Clauses that have been defined in the insurance policy . |

**Contextual restrictions**

The company and the policy number together allow identifying an insurance policy.

## 4 Reaction requirements

**Treatments**

The information of the insurance policy is recorded.

# 4. OO-METHOD CONCEPTUAL MODEL DERIVATION

We first present an overview of the derivation strategy. Then, we describe how the SuperStationery Co. requirements model is processed in order to derive the corresponding conceptual model.

## 4.1. Overview of the derivation strategy

As shown by Figure 4, the Object Model is derived from the elements that appear in the Communicative Event Diagram and the Event Specification Templates. The Dynamic Model is initially derived from the Communicative Event Diagram; then some transitions are added when processing the Event Specification Templates. The Functional Model is derived from the Event Specification Templates. Although the Presentation Model can also be reasoned from the information contained in the Requirements Model (preliminary work can be found in [España 2005]), this report focuses on the derivation of the Object Model (i.e. the reasoning of the class diagram). However, some guidelines for the derivation of the Dynamic Model are also proposed.

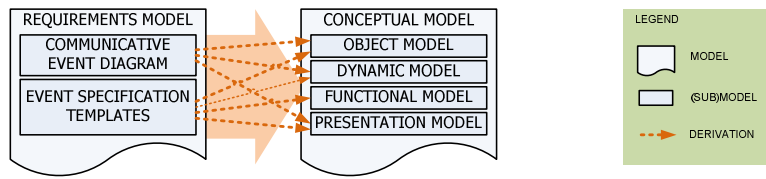

Figure 4. Strategy for the derivation of OO-Method conceptual models from Communication Analysis requirements models

### 4.1.1. Outline of the derivation of the Object Model

The Object Model specifies the memory of the IS by means of an extended UML class diagram. In order to obtain the Object Model, the communicative events of the Communicative Event Diagram are processed. Requirements related to communicative events are specified in detail in the Event Specification Templates; therefore these templates contain relevant information for the derivation procedure. Specifically, for each communicative event, the message structure is processed in order to derive a class diagram view. We refer as **class diagram view** to a portion of the class diagram that includes one or several portions of interrelated classes. By portion of a class we mean some or all of the attributes and services of the class. The concept of class diagram view with regard to conceptual models is analogous to the concept of relational view with regard to relational database schemas[4]; relational views allow displaying different perspectives of the same database.

---

[4] In some works, a relational (or database) view is considered to be a virtual relation derived from base relations using a set of view defining operations such as projection, join, and others [Codd 1974]. However, other works consider views as relational schemas (or even conceptual models or data-oriented requirements specifications) that correspond to part of a complex domain [Batini, Lenzerini et al. 1986]. Our definition of class diagram view, at the point of conceptual model derivation, is closer to the latter notion.

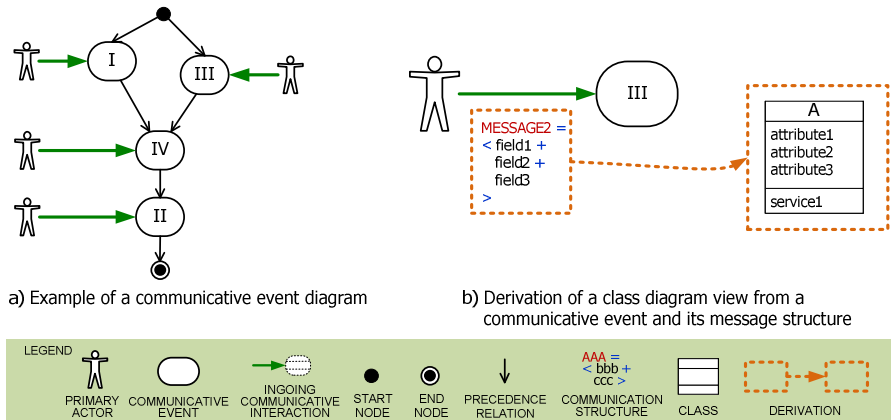

a) Example of a communicative event diagram

b) Derivation of a class diagram view from a communicative event and its message structure

Figure 5. Simplified example of the derivation of a class diagram view from a communicative event

Figure 5.a presents a simple example of a communicative event diagram that is used for illustration purposes[5]. Figure 5.b informally represents how a class diagram view is derived from a communicative event. Two dashed-line rectangles and an arrow represent that a class diagram view (on the right) is obtained from processing the message structure (on the left) that specifies the message conveyed by the primary actor of the communicative event III.

Each communicative event derives a different class diagram view. The derived class diagram views can be integrated to obtain the complete class diagram, the same way that different relational views are integrated to obtain a single database schema [Batini, Lenzerini et al. 1986]. This integration can be approached in two ways; we opt for the second:

a) Post-process view integration. First, the communicative events are processed so as to obtain their corresponding class diagram views. The events can be processed in any order. Then, all the class diagram views are integrated into a single class diagram. Figure 6.a exemplifies this approach.

b) Incremental view integration. First, the communicative events are sorted (see Section 4.2.2). Then, communicative events are processed in order, one by one (see Sections 4.2.3 to 4.2.13). The class diagram view that results from processing each communicative event is integrated into the class diagram under construction. This way, new classes can be added to the class diagram, class attributes and services can be added to existing classes, and structural relationships[6] among existing classes can be added to the class diagram. Figure 6.b exemplifies this approach.

---

[5] Some details of the communicative event diagram have been omitted for the sake of simplicity.

[6] We refer as structural relationships to semantic relationships between classes as defined by Pastor and Molina [2007, pp. 80-95]. Although the OO-Method defines several types of structural relationships (namely, association, aggregation and composition), only associations are derived for the moment.

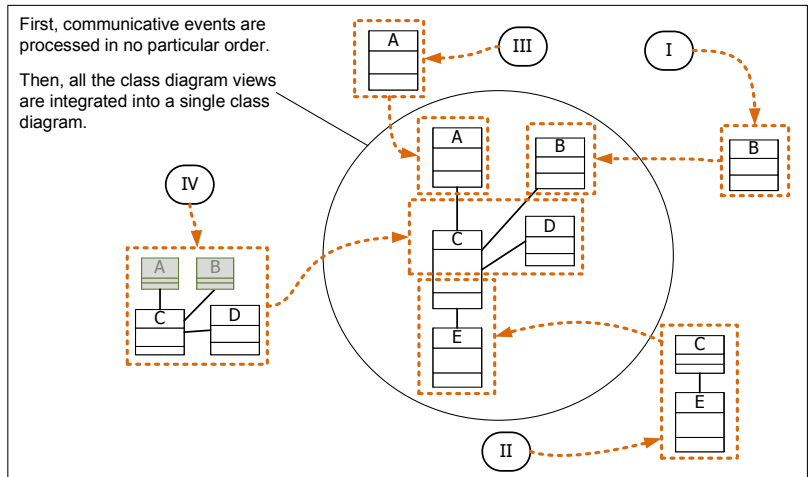

a) Construction of the class diagram by means of post-process view integration

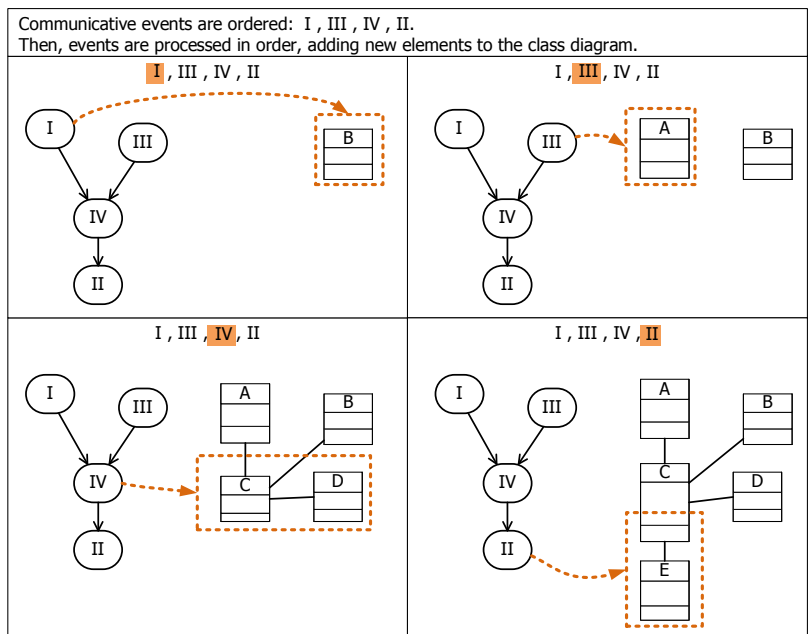

b) Construction of the class diagram by means of incremental view integration

Figure 6.   Approaches to class diagram construction

By performing an incremental view integration and by processing the communicative events in a predefined order, we intend to prevent a situation in which a newly-derived class diagram view refers to a class diagram element that has not yet been derived (e.g. when trying to add an attribute to an inexistent class). In any case, if the requirements model suffers incompleteness, this situation cannot be fully prevented.

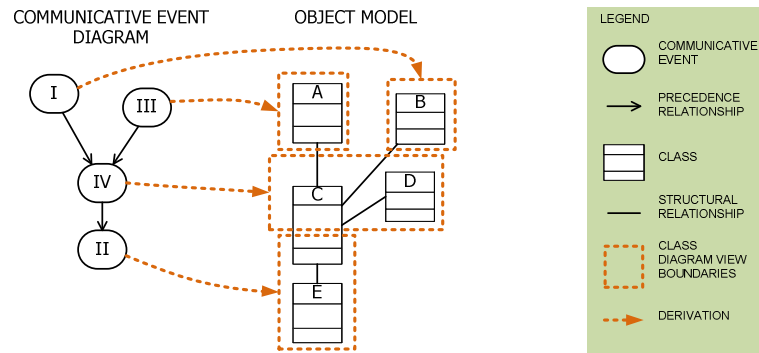

Figure 7.   Simplified example of communicative events and their corresponding class diagram views

Figure 7 informally represents the correspondences between the communicative events of a business process and the class diagram views. Note that the relation among communicative events and classes has a *many-to-many* cardinality:

- One communicative event can derive one class (e.g. communicative events I and II). It is the case of communicative events with a simple message structure, such as Supp 2 (see Section 4.2.3).
- One communicative event can affect several classes (e.g. communicative event III). Communicative events with more complex message structures result in class diagram views that contain several classes and relationships among them, as in Sale 1 (see Section 4.2.6).
- One class can be affected[7] by several communicative events (e.g. class C is affected by events IV and II). This is normally due to the fact that the occurrence of one communicative event creates a business object that is later updated by the occurrence of another communicative event, as it happens with events Sale 1 and Sale 2 (see Sections 4.2.6 and 4.2.7, respectively).

### 4.1.2. Outline of the derivation of the Dynamic Model

The Dynamic Model specifies the valid lifecycles of objects and the interaction between them by means of a collection of state-transition diagrams (that are compliant with the UML State Machine Diagram). Each class of an Object Model has its own state-transition diagram in the Dynamic Model. The Dynamic Model can be obtained mainly by processing the Communicative Event Diagram. However, some elements can be added to the Dynamic Model as a result of processing the Event Specification Templates.

Two types of state-transition diagrams can be derived for a class, depending how many communicative events affect the class:

- A class is affected by only one communicative event. For each class in the conceptual model that is affected by only one communicative event, a basic state-transition diagram is derived by default. The basic state-transition diagram represents the essential valid lifetimes for the instances of a class. The diagram includes three states; namely a pre-creation state, an intermediate state and a destruction state. The diagram also includes one transition for each creation service, that departs from the pre-creation state and arrives to the intermediate state; one transition for each destruction service, that departs from the intermediate state and arrives to the destruction state; one transition for each one of the rest of services of the class, that departs from and arrives to the intermediate state departs from the pre-creation state and arrives at the intermediate state [Pastor and Molina 2007, pp. 120-121]. To derive the basic state-transition diagram, only the Object Model is needed.

---

[7] When processing a communicative event for derivation purposes, a class can be created for the first time or it can extended with new attributes or services. In either case, the class is said to be part of the class diagram view of that particular communicative event. Also, the class is said to be *affected* by that communicative event.

- A class is affected by more than one communicative event. For each class in the conceptual model that is affected by more than one communicative event, a more complex state-transition diagram is derived (see Figure 8). The main derivation guideline is the following. Communicative events are converted into transitions (e.g. IV, II) and precedence relationships are converted into states (e.g. IVed, IIed[8]). Additional transitions can be added to the state transition diagram; for instance, transitions that correspond to edition and destruction atomic services (e.g. edit and destroy, respectively), as well as transitions that correspond to atomic services that take part in a transaction (these transitions appear as a result of processing the event specification templates).

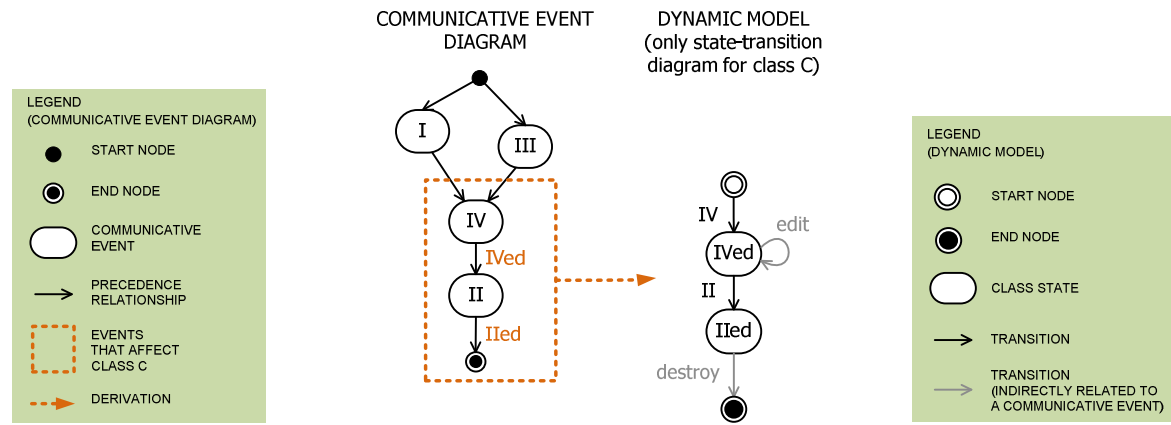

Figure 8.  Simplified example of the derivation of a (non-default) state-transition diagram from a communicative event diagram

A more illustrative example is shown in Section 4.3.

## 4.2. Derivation of the Object Model

Before applying the transformation guidelines, the communicative event diagram needs to be pre-processed. For this purpose, the following actions are taken:

1. The communicative event diagram is extended with communicative events from other processes.

2. The extended communicative event diagram is removed loopback precedence relationships (as well as other elements) until a partially ordered set of communicative events is obtained.

3. The resulting rooted directed tree is processed to obtain an ordered list of communicative events.

The events are then processed in the defined order, applying the derivation guidelines.

---

[8] To facilitate the understandability of the example, precedence relationships and transitions have been labelled with a name that is derived from the name of the precedent communicative event. The English participle suffix *–ed* is added to the event name. This intuitively represents the state in which an object remains after an event. For instance, after an occurrence of the event IV the affected object remains in the state IVed. It makes more sense with real words: after an occurrence of the communicative event "A customer submits a claim" an object of the class CLAIM remains in the state Submitted; after an occurrence of the communicative event "A clerk classifies the claim" the object remains in the state Classified.

### 4.2.1. Extension of the communicative event diagram

The communicative event diagram is extended with all the precedent communicative events so as to obtain a rooted directed graph[9], where the start node is designated as the root. For this purpose, we need to include any communicative event that is not included in the diagram but precedes a communicative event which is included in the diagram. This needs to be repeated until no more communicative events are included. The resulting diagram is referred to *as extended communicative event diagram*.

Figure 9 presents the extended communicative event diagram that corresponds to the *Sales management* (Sale) business process. It extends the original diagram in Figure 2 with the communicative events that are needed to obtain a rooted directed graph. The communicative event Prod 2 belongs to the *Product management* business process, Clie 1 belongs to *Client management*, Supp 2 belongs to *Supplier management*, Logi 10 belongs to *Logistics*, and Risk 4 belongs to *Risk management*.

Despite the guidelines for including start and end nodes in communicative event diagrams (see page 14), the start note and end nodes can be omitted for the sake of simplicity (i.e. to avoid many line crossings). The extended communicative diagram in Figure 10 omits the start node[10].

---

[9] A *rooted directed graph* is a directed graph in which there is a node designated as the root, from which there is a path to every other node.

[10] In any case, note that the start node is implicit in the diagram in Figure 10. The precedence relationships between the start node and those communicative events that do not have a precedent communicative event are also implicit. Therefore, it can still be considered a rooted directed graph.

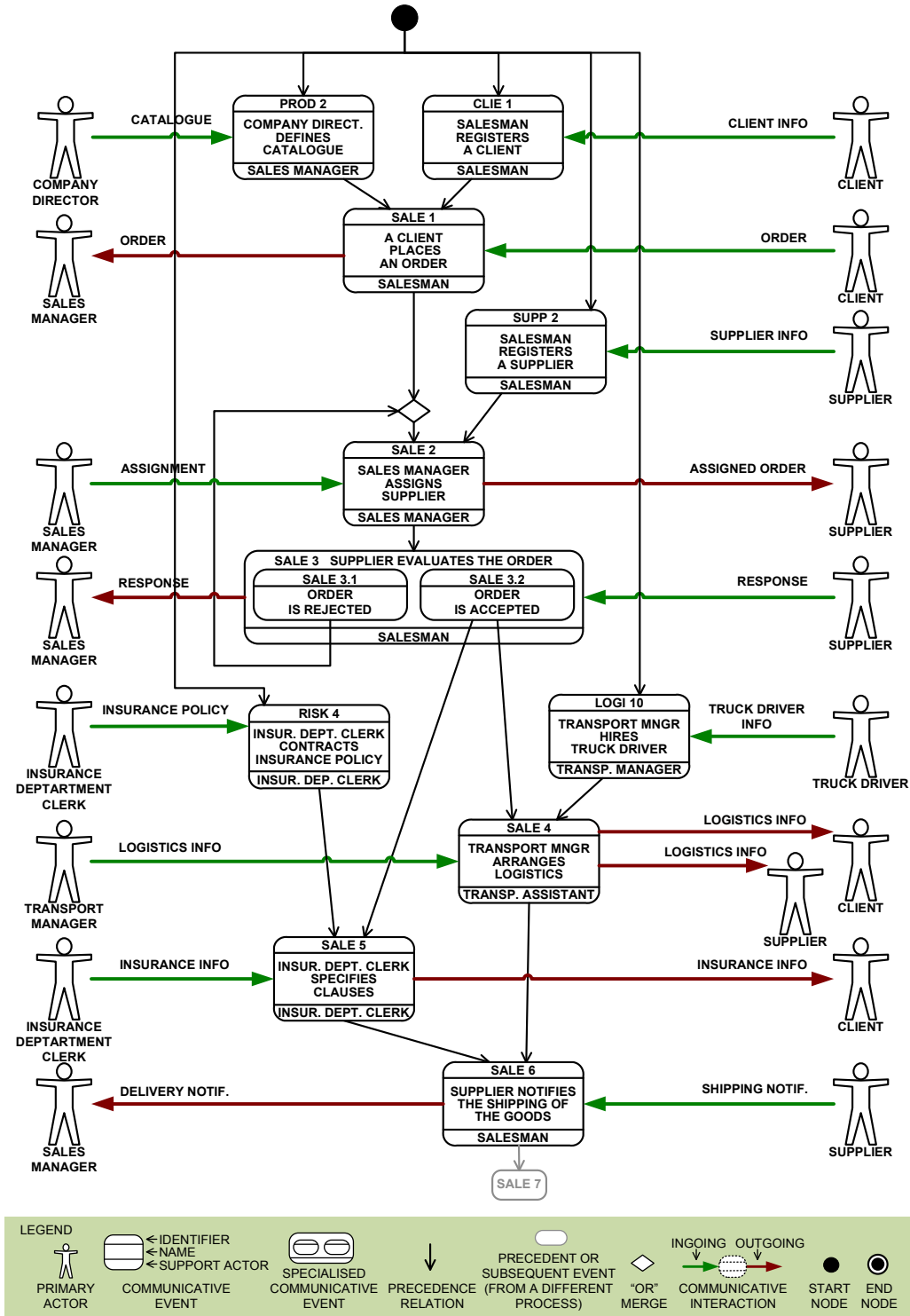

Figure 9. Extended diagram that includes communicative events from other business processes

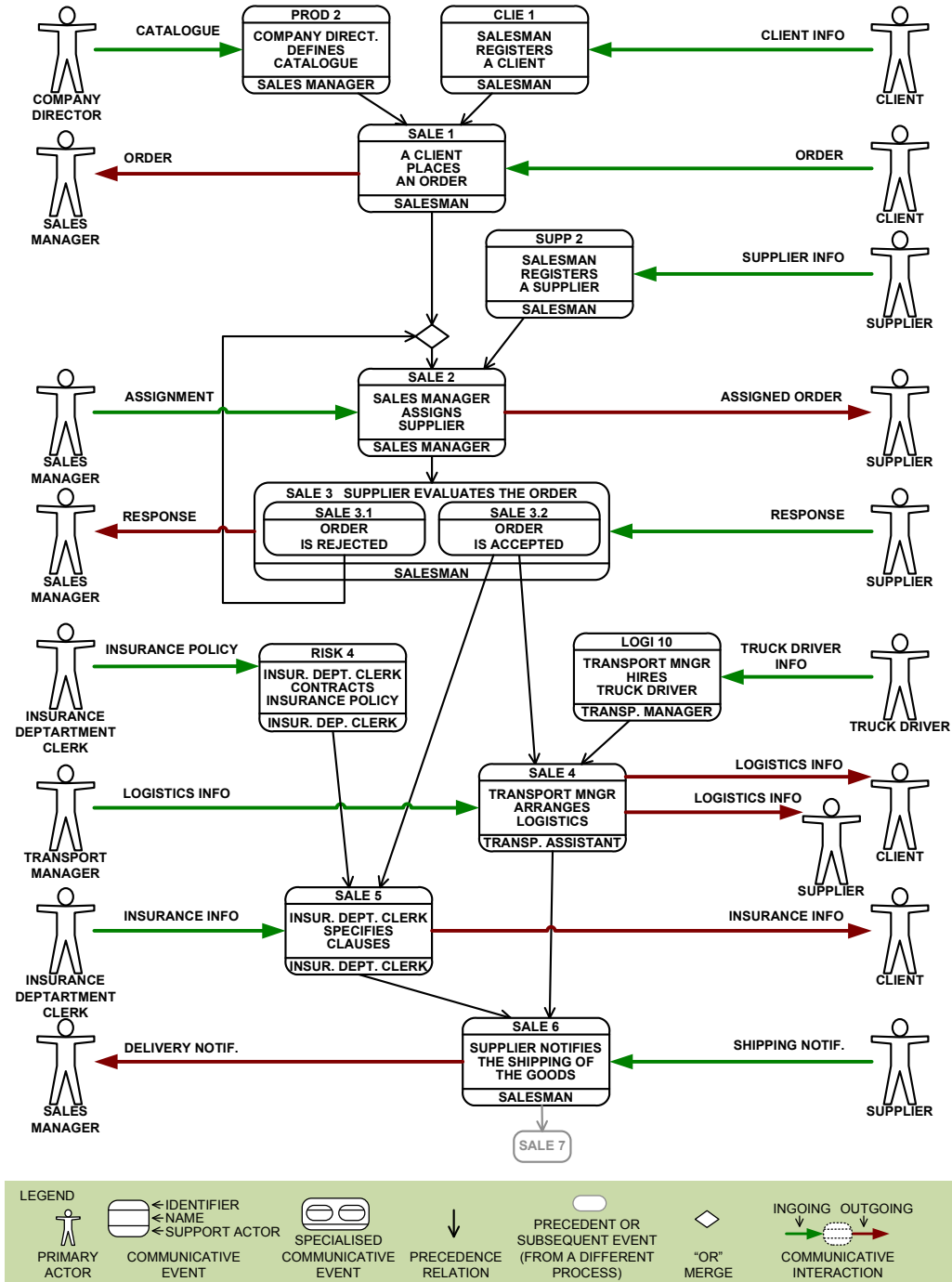

Figure 10.  Extended diagram that omits the start node for the sake of clarity

## 4.2.2. Event ordering

The communicative events depicted in the extended communicative event diagram are ordered according to the precedence relationships. For this purpose, a *partially ordered set of communicative events* needs to be obtained from the extended communicative event diagram. To obtain the partially ordered set (a.k.a. poset)[11], the loopbacks that might appear in an extended communicative event diagram are removed and the diagram is simplified in order to remove every other model element except for communicative events and precedence relationships. The remaining precedence relationships define a strict partial order among the communicative events. Figure 11.a shows the poset of communicative events of the diagram in Figure 10; note that the loopback from event SALE 3 to SALE 2 has been removed.

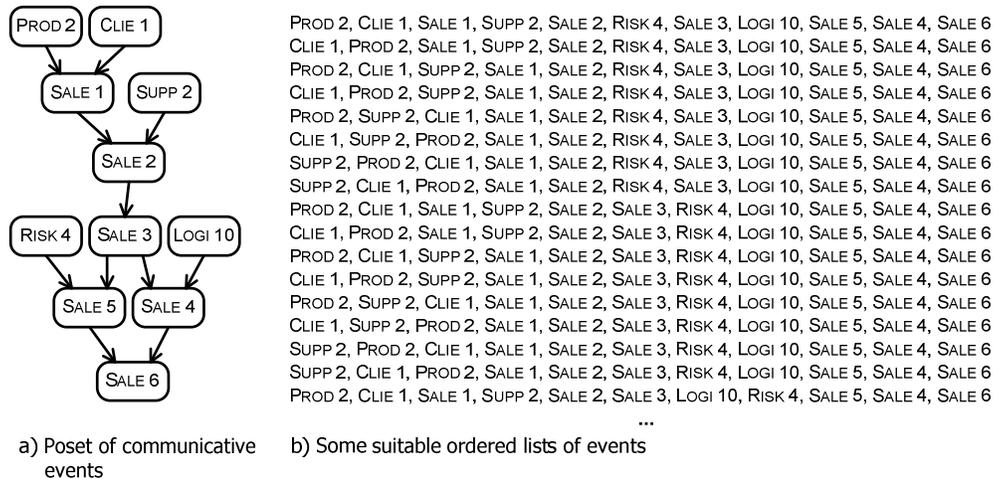

a) Poset of communicative events

b) Some suitable ordered lists of events

Figure 11.  Obtaining ordered lists of events from the communicative event diagram

Any ordered list of events that is compatible with the partial order is suitable for the conceptual model derivation (see Figure 11.b). Any known procedure for topological sorting can be used for obtaining the ordered list of events; for instance, the algorithm proposed by Kahn [1962]. The following ordered list of events has been chosen for pedagogical purposes:

SUPP 2, PROD 2, CLIE 1, SALE 1, SALE 2, SALE 3, LOGI 10, SALE 4, RISK 4, SALE 5, SALE 6

In the following, the communicative events are processed following this order.

### 4.2.3.  SUPP 2. Salesman registers a supplier

In order to derive the class diagram view that corresponds to communicative event SUPP 2, the message structure is processed. SUPPLIER INFO is a quite simple message structure; it is an aggregation of fields (i.e. a substructure of the form < + >). Therefore, a class named SUPPLIER is derived. The name of the corresponding substructure can be used to name the class (that is, the class could have been named SUPPLIERINFO); however, the analyst can decide to give the class a different name.

For each data field of the aggregation substructure, class SUPPLIER is added an attribute (e.g. the data field Supplier code leads to the derivation of the attribute supplier_code). Figure 12 depicts this derivation.

---

[11] A poset consists of a set together with a binary relation that indicates that, for certain pairs of elements in the set, one of the elements precedes the other. In our case, the binary relation is defined by the precedence relationships. If a communicative event A precedes another communicative event B then we can consider that A<B. The start node (even if it is implicit) is the *least element* of the poset.

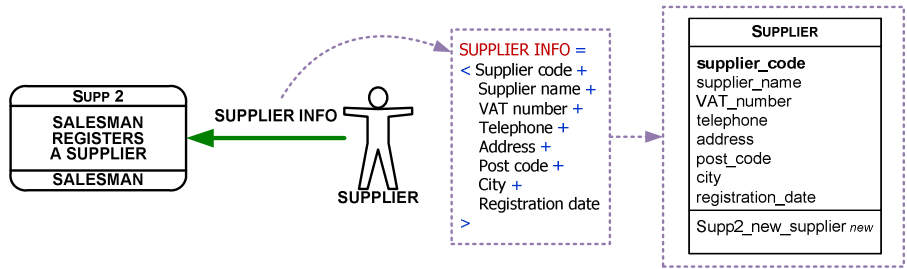

Figure 12.   Class diagram view of Supp 2

The attribute definitions for class Supplier are shown in Table 1.

Table 1.   Specification of attributes of class Supplier

| Attribute name | Id | Attribute type | Data type | Size | Requested | Null allowed |
|---|---|---|---|---|---|---|
| supplier_code | yes | Constant | Autonumeric | | yes | no |
| supplier_name | no | Variable | String | 120 | yes | no |
| VAT_number | no | Variable | Nat | | yes | no |
| telephone | no | Variable | String | 12 | yes | yes |
| address | no | Variable | String | 300 | yes | yes |
| post_code | no | Variable | String | 12 | yes | yes |
| city | no | Variable | String | 100 | yes | yes |
| registration_date | no | Variable | Date | | yes | yes |

The names of the attributes are derived from the names of the data fields. Lowercase letters are used for the attribute name, except for acronyms such as VAT[12], which remain unchanged.

The requirements model should specify how the organisational actors identify business objects of each kind. If available, this information is specified as a contextual restriction and it offers guidance for selecting which attributes constitute the class identifier (in the absence of such restrictions, then the analyst should ask the users or decide relying on his/her own judgement). In this case there is a contextual restriction in the requirements model that states "suppliers are identified by means of a mnemonic code", so the attribute supplier_code is designated the class identifier (column Id in Table 1).

Whether an attribute should be defined as a constant or as a variable attribute should be asked to the users (or the analyst should decide relying on his/her own judgement)[13]. However, by default, all attributes that are part of the class identifier are defined as constant; the rest are defined as variable (column Attribute type in Table 1).

Attribute data types are derived from data field domains. During requirements engineering, the domains of the data fields just serve the purpose of clarifying the meaning of the messages that are conveyed when a communicative event occurs. In contrast, during conceptual modelling, a detailed specification of data types is needed (especially if automatic code generation is intended). For this reasons, Communication Analysis prescribes a few basic domains for data fields, whereas the OO-Method offers a wider selection of data types for attributes. Table 2 defines a correspondence between Communication Analysis data field domains and OO-Method attribute data types, thus offering some guidance for the conversion. Some data field domains that can be converted to several attribute data types; in these cases, the analysts should apply their criteria. In any case, the data types that are underlined are the ones that are recommended in case of doubt.

---

[12] A value added tax (VAT) is a form of consumption tax that imposes a tax on the value that is added to a product at each stage of its manufacture or distribution.

[13] As defined by the OO-Method, once a constant attribute has been initialised, its value cannot be modified. In contrast, variable attributes can always be modified (as long as the proper class services are defined) [Pastor and Molina 2007].

Table 2.    Conversion table for attribute data types

| Communication Analysis Data field domain | OO-Method Attribute data type[14] |
|---|---|
| Number | Nat, Int, <u>Real</u>, Autonumeric |
| Text | <u>String</u>, Text |
| Datetime | Time, Date, <u>DateTime</u> |
| Money | Real |
| *not considered* | Bool, Image, Blob |

If the String data type is selected, then the size of the attribute (i.e. the maximum number of characters the attribute value can have) needs to be defined. The size depends on the expected content of the attribute (e.g. local telephone numbers can be stored in 12 characters, whereas we should provide at least 120 characters length to store company names). Analysts should ask the users or act according to their own criteria.

Since the current communicative event has led to the derivation of class SUPPLIER (i.e. class SUPPLIER appears for the first time during the processing of this event) then all the attributes are set as requested at creation time.

Whether an attribute should allow null values or not needs to be asked to the users (or the analyst should decide relying on his/her own judgement). However, by default, all attributes that are part of the class identifier are added a restriction so that they do not allow null values; the rest of the attributes are set to allow null values (see column Null values in Table 1).

Additionally, a creation service named Supp2_new_supplier is added to the class; the prefix Supp2_ stands for SUPP 2 (the identifier or the communicative event). This way it is possible to identify which service of a class triggers the IS reaction to a communicative event; that is, there may exist several services that are related to a communicative event (e.g. in complex business objects there is at least one creation service for each class that is part of the business object) but only one service actually corresponds to the IS reaction to the communicative event. Anyway, this is not the case of the event SUPP 2, since it only affects one class.

For each attribute of the class, an inbound argument is added to the creation service. The inbound arguments definitions for service Supp2_new_supplier are shown in Table 3.

Table 3.    Specification of inbound arguments of service Supp2_new_supplier

| Argument name | Data type | Size | Null allowed |
|---|---|---|---|
| p_atrsupplier_code | Autonumeric | | no |
| p_atrsupplier_name | String | 120 | no |
| p_atrVAT_number | Nat | | no |
| p_atrtelephone | String | 12 | yes |
| p_atraddress | String | 300 | yes |
| p_atrpost_code | String | 12 | yes |

---

[14] We clarify the meaning of some data types that are not straightforward: *Nat* stands for natural number, *Int* for integer, *Autonumeric* refers to a sequence of natural numbers that is incremented automatically every time an instance is created, *String* refers to an array of characters of a specified length, *Text* refers to a multi-line text, *Bool* stands for Boolean, *Blob* stands for binary large object.

| Argument name | Data type | Size | Null allowed |
|---|---|---|---|
| p_atrcity | String | 100 | yes |
| p_atrregistration_date | Date | | yes |

The argument name is derived from the attribute name. In principle, the argument can have the same name as their corresponding attribute. However, in order to conform to the *OLIVA**NOVA*** Modeler nomenclature, the following prefixes can be added to the attribute name: *p_atr*, *pt_*, *p_this*, and *p_evc*[15]. The rest of the argument properties (data type, size, and whether null values are allowed) are set to the same value as their corresponding attribute properties.

In the rest of the communicative events, the guidelines described above are used for determining the properties of the class attributes and the properties for the service arguments. For the sake of brevity, the detailed explanations about this topic are omitted, as well as the attribute and argument specification tables[16].

### 4.2.4. PROD 2. Company director defines catalogue

The message structure of communicative event PROD 2, named CATALOGUE, is an iteration of an aggregation substructure named PRODUCT. The iteration (a complex substructure of the form { }) specifies that the company director provides the details of many products at a time (which constitutes the catalogue). Thus, a class named PRODUCT is derived. For each field of the aggregation substructure, an attribute is added to the class (e.g. the data field Comments leads to the derivation of the class attribute comments). Also, a creation service named Prod2_new_product is added to the class; this service has as many inbound arguments as class attributes have been derived.

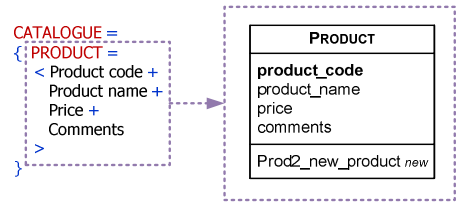

Figure 13. Class diagram view of PROD 2

### 4.2.5. CLIE 1. Salesman registers a client

The message structure of this communicative event is more complex than those of the previous events. The initial substructure of CLIENT INFO is an aggregation substructure that includes several fields (e.g. VAT number, Client name) and an iteration structure named ADDRESSES that is an iteration substructure of CLIENT ADDRESS. In turn, CLIENT ADDRESS is an aggregation substructure that includes several fields (e.g. Address).

The initial aggregation substructure CLIENT INFO leads to the derivation of a class named CLIENT. Each data field of this substructure leads to the derivation of an attribute of the class (e.g. the data field Client name leads to the derivation of the class attribute client_name).

---

[15] The guidelines for naming the arguments are not relevant, since the modelling tool already provides a name by default that follows the expected nomenclature. Therefore, these guidelines are not provided in this document.

[16] In any case, the properties of the class attributes are displayed in the *OLIVA**NOVA*** screenshots in Section 0. With regards to the properties of the service arguments, only the name is displayed in the screenshots.

The aggregation substructure CLIENT ADDRESS leads to the derivation of a class named CLIENTADDRESS. For each field of the substructure, an attribute is added to the class (e.g. Post code leads to the derivation of post_code).

Since both substructures CLIENT INFO and CLIENT ADDRESS are related by means of an iteration substructure (the complex substructure of the form { } named ADDRESSES), then a structural relationship between the corresponding classes (CLIENT and CLIENTADDRESS, respectively) is created. With regards to the cardinality[17] of the structural relationship, the existence of an iteration substructure determines a maximum cardinality of M on the side of CLIENTADDRESS (the iteration specifies that one client can have *many* addresses). The rest of the cardinalities depend on the structural restrictions that appear in the requirements specification (specifically in the event specification template of CLIE 1). For instance, the structural restriction "One address corresponds to only one client." (see page 28) implies a maximum cardinality of 1 on the side of CLIENT. In the absence of such restrictions, then the analyst should ask the users or decide relying on his/her own judgement. In this case, the cardinality is finally set as follows: CLIENT 1:1 --- 0:M CLIENTADDRESS.

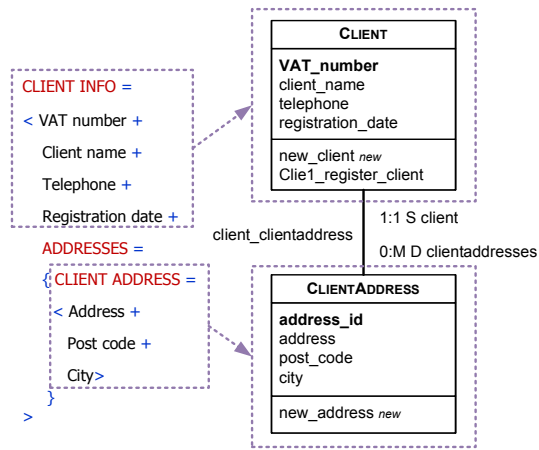

Figure 14. Class diagram view of CLIE 1

With regards to the services, for each newly-derived class a creation service is added. This way, the creation services new_client and new_address are added to classes CLIENT and CLIENTADDRESS, respectively. However, the client record is a complex business object. Therefore, apart from the creation service new_client, another service named Clie1_register_client is added to the class CLIENT. The service has an inbound argument named p_thisClient (see Table 4). This argument represents the instance of the class for which the service is invoked (the OO-Method prescribes this type of argument for every service that is not a creation service). This type of service is referred to as *end of editing* service because it is invoked by the support actor when s/he has finished entering the information of the message related to the a communicative event (in this particular case, the salesman invokes the service when s/he has finished entering the information of the client, including the client addresses). Only after this service is executed, the client information is considered to be recorded[18].

---

[17] We acknowledge that cardinality is often referred to as multiplicity since the adoption of the UML as a *de-facto* standard; however, we use the original term because the current specification of the OO-Method does so [Pastor and Molina 2007]. Also, the notation for the cardinality in *OLIVANOVA*, an OO-Method CASE tool, departs from the UML standard. The syntax for specifying cardinality is the following: CLASSA roleA $min_A$:$max_A$ --- $min_B$:$max_B$ roleB CLASSB. In the absence of ambiguity, the roles can be omitted: CLASSA $min_A$:$max_A$ --- $min_B$:$max_B$ CLASSB. This way, CLIENT 1:1 --- 0:M CLIENTADDRESS has the following implications: a given client can have from 0 to many (i.e. an indeterminate number of) addresses, and a given address corresponds to 1 and only 1 client.

Table 4.    Specification of inbound arguments of service Clie1_register_client

| Argument name | Data type | Size | Null allowed |
|---|---|---|---|
| p_thisClient | Client | | no |

### 4.2.6. SALE 1. A client places an order

The message structure of this event consists of an initial aggregation structure named ORDER that includes an iteration substructure named DESTINATIONS, which includes another iteration substructure named LINES. Both iteration substructures have an aggregation substructure inside of them (DESTINATION and LINE respectively).

The three aggregation substructures lead to the derivation of the classes CLIENTORDER, DESTINATION and LINE. For each data field of the substructures, an attribute is added to the corresponding class (e.g. the data field Order number leads to the derivation of the class attribute order_number)[19]. Figure 15 shows the derived classes and attributes; the properties of attributes of class CLIENTORDER are shown in Table 5.

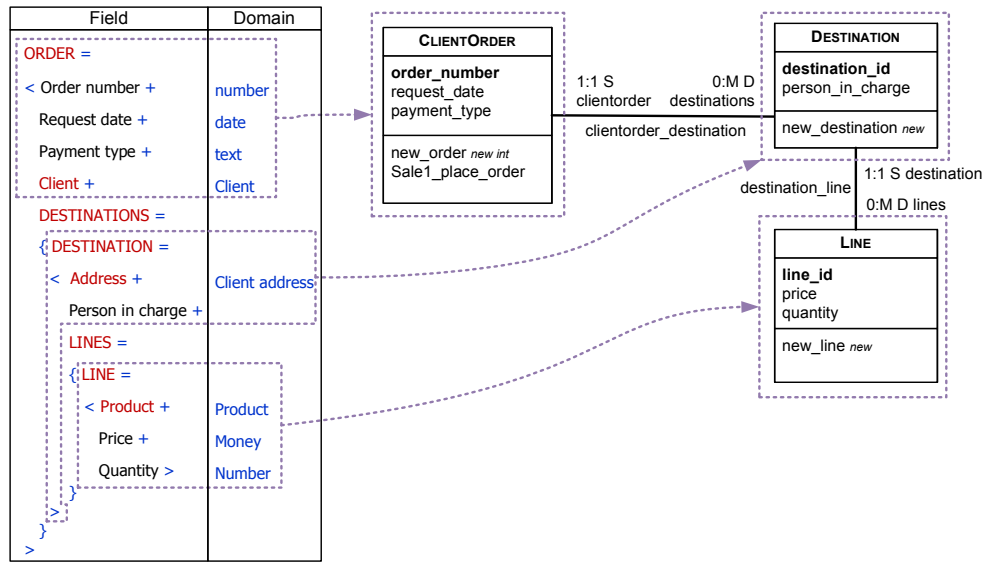

Figure 15.   Derivation of classes and structural relationships based on message substructures (SALE 1)

Since the substructure DESTINATION is part of the substructure ORDER, then the corresponding classes (CLIENTORDER and DESTINATION, respectively) are related by means of a structural relationship. Furthermore, since DESTINATIONS (the substructure that relates ORDER and DESTINATION) is an iteration substructure, then the structural relationship has maximum cardinality M on the side of class DESTINATION (the iteration specifies that one client order can have *many* destinations). As explained above, the rest of the cardinalities depend on structural restrictions that appear in the event description template of SALE 1 (in the absence of such restrictions, then the analyst should ask the users or decide relying on his/her own judgement). See the structural relationship and its cardinalities in Figure 15.

---

[18] Section 6 includes a discussion on a possible improvement for the derivation guidelines. This improvement would avoid the derivation of *end-of-edition* events for those complex business objects that do not require such event (as in the case of the client record).

[19] Reference fields (i.e. fields of the message structure that reference business objects, such as Client) do not lead to the derivation of attributes but to the derivation of structural relationships, but this is explained later on.

Similarly, since LINE is part of the substructure DESTINATION, then the corresponding classes (DESTINATION and LINE, respectively) are related by means of a structural relationship. Furthermore, since LINES (the substructure that relates DESTINATION and LINE) is an iteration substructure, then the maximum cardinality is M on the side of class LINE (the iteration specifies that one destination can have *many* lines). Again, the rest of the cardinalities depend on structural restrictions, on the user response, or on the analyst criteria. See the structural relationship and its cardinalities in Figure 15.

Table 5.    Specification of attributes of class CLIENTORDER

| Attribute name | Id | Attribute type | Data type | Size | Requested | Null allowed |
|---|---|---|---|---|---|---|
| order_number | yes | Constant | Autonumeric | | yes | no |
| request_date | no | Constant | Date | | yes | no |
| payment_type | no | Variable | String | 30 | yes | yes |

Those fields that reference business objects imply adding further structural relationships to the class diagram view.

The reference field named Client leads to adding a structural relationship between its corresponding class and the class that was derived from the CLIENT INFO aggregation substructure (see the processing of the communicative event CLIE 1. *Salesman registers a client*). Thus, a structural relationship is created between the classes CLIENTORDER and CLIENT. As part of the derivation, the maximum cardinality is 1 on the side of class CLIENT. The reasoning is as follows. Note that the meaning of the aggregation substructure ORDER is that for each order the following information is provided: an order number, the date in which the order is requested, the payment type that the client has chosen, the client that places the order and a set of destinations. Since the client is specified as a reference field, only one client can be associated to the order. The rest of the cardinalities depend on structural restrictions, on the user response, or on the analyst criteria. See the structural relationship and its cardinalities in Figure 16.

The same applies to the reference field Address, which refers to the aggregation substructure CLIENT ADDRESS (also in the communicative event CLIE 1). CLIENT ADDRESS led to the derivation of class CLIENTADDRESS. Therefore, a structural relationship is created between the classes DESTINATION and CLIENTADDRESS. Again, the maximum cardinality 1 in the side of class CLIENTADDRESS can be derived straightforwardly. The rest of the cardinalities depend on structural restrictions, on the user response, or on the analyst criteria. See the structural relationship and its cardinalities in Figure 16.

Lastly, the reference field Product refers to a product in the catalogue, which corresponds to the aggregation substructure PRODUCT (see the processing of the communicative event PROD 2. *Company director defines catalogue*). Therefore, a structural relationship is created between the classes LINE and PRODUCT.

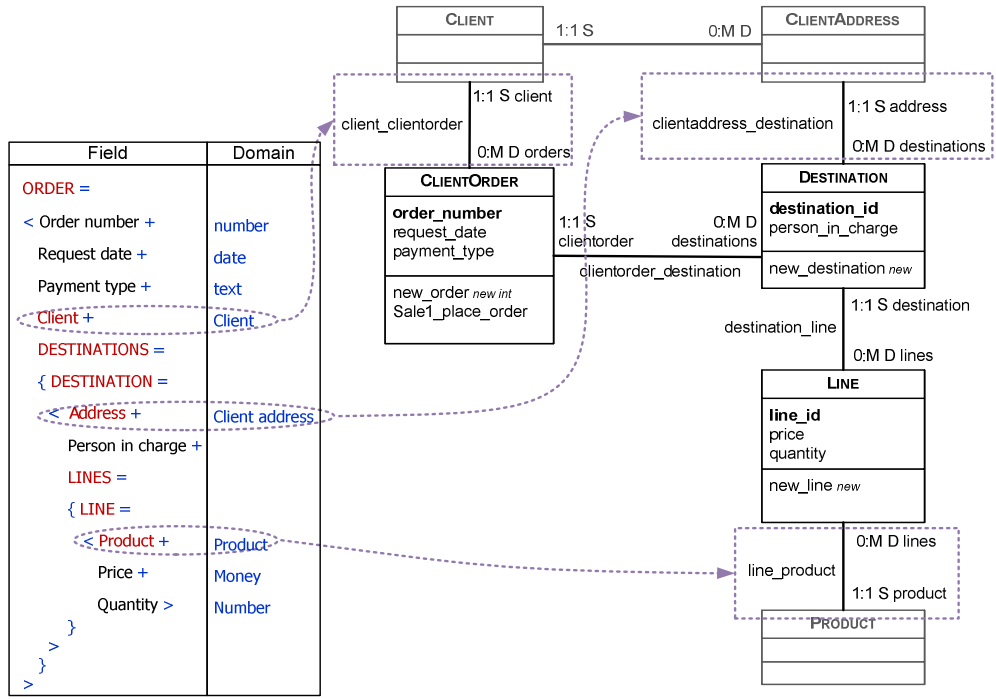

Figure 16. Derivation of structural relationships based on reference fields (SALE 1)

With regards to the names of the structural relationships and the roles, they can be derived applying the following rules:

1. The name of the structural relationship is the concatenation of the names of the related classes. For instance, the structural relationship in the example is named client_address_destination.

2. The name of the role at the side of the *referenced* class is the name of the field. For instance, the role at the side of class CLIENTADDRESS is named address after the field Address.

3. The name of the role at the side of the *referencer* class is the name of the class itself.

   a. If the maximum cardinality at this side is 1 then the name is left in singular form.

   b. If the maximum cardinality at this side is M then the name is written in the plural.

   For instance, the role at the side of the class DESTINATION is named destinations because the maximum cardinality at this side of the relationship is M.

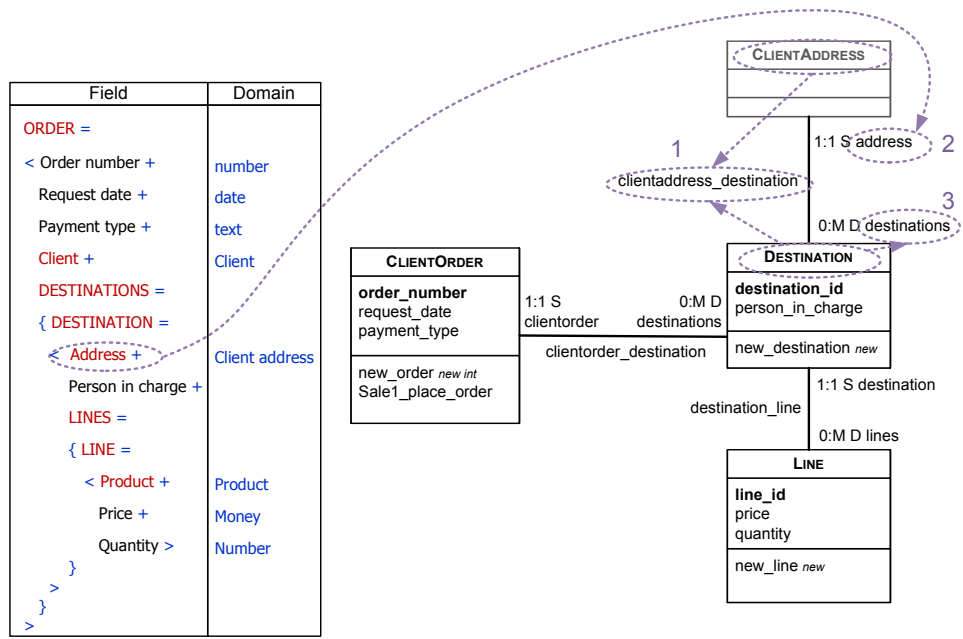

Figure 17.   Derivation of the names of structural relationships and roles (SALE 1)

With regards to the services, for each newly-derived class a creation service is added. For instance, a service named new_order is added to the class. For each attribute of the class, an inbound argument is added to the creation service (e.g. the argument p_atrorder_number corresponds to the attribute order_number). Additionally, an inbound argument named p_agrClient is added to the service; this attribute defines which client places the order. The inbound arguments definitions for service new_order are shown in Table 6.

Table 6.   Specification of inbound arguments of service new_order

| Argument name | Data type | Size | Null allowed |
|---|---|---|---|
| p_atrorder_number | Autonumeric | | no |
| p_atrrequest_date | Date | | no |
| p_atrpayment_type | String | 30 | yes |
| p_agrClient | Client | | no |

In this case, the client order is a complex business object. Therefore, apart from the creation service new_order, an *end of editing* service named S1_place_order is added to the class CLIENTORDER. This service is triggered by the salesman whenever s/he has finished introducing the information if the order; that is, after introducing the destinations and the lines. Only after this service is executed, the order is considered to be placed. Table 7 shows the details of the inbound argument (according to the OO-Method guidelines for this kind of services, it only needs to define an argument for the client order which is being placed).

Table 7.   Specification of inbound arguments of service Sale1_place_order

| Argument name | Data type | Size | Null allowed |
|---|---|---|---|
| p_thisClientOrder | ClientOrder | | no |

### 4.2.7.  SALE 2. Sales Manager assigns supplier

After an order is placed, the Sales Manager assigns the order to one of the many suppliers that work with SuperStationery. Thus, communicative event SALE 2 affects the same business object as communicative event Sale 1; namely, the client order. Note that the reference field Order in the message structure of Sale 2 indicates the business object being modified. This way, the class that corresponds to this business order is extended; namely, ClientOrder is affected by Sale 2. The data fields in the message structure of Sale 2 lead to adding new attributes to this class, whereas the reference fields in the message structure lead to adding new structural relationships between class ClientOrder and other classes that (presumably[20]) already exist in the class diagram under construction.

With regards to data fields, the field Assignment date leads to adding an attribute named assignment_date to the class ClientOrder. Figure 18 depicts this derivation and Table 8 specifies the details of the new attribute. All attributes that are added to a class as a result of a class extension have the following properties: these attributes are not part of the identification function, the attribute type is Variable, they are not requested upon creation, and they allow nulls. The data type is derived from the domain of the field (in this case, the data type of assignment_date is Date because the field domain of Assignment date is date).

Table 8.    Specification of the new attribute added to class ClientOrder

| Attribute name | Id | Attribute type | Data type | Size | Requested | Null allowed |
|---|---|---|---|---|---|---|
| assignment_date | no | Variable | Date | | no | yes |

With regards to reference fields, the field Supplier references a business object that was processed in the communicative event Supp 2. Therefore, a structural relationship is defined between the class ClientOrder and the class Supplier. The cardinality is defined as 0:1 in the side of Supplier because the orders are not assigned to suppliers when they are placed, but it occurs in a later moment in time. For the same reason, it is dynamic.

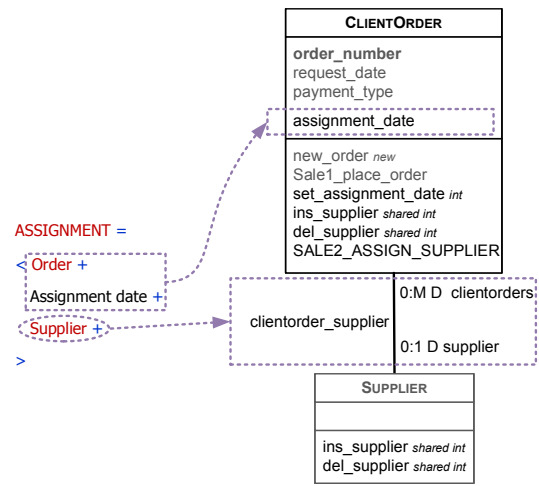

Figure 18.   Class diagram view of Sale 2

With respect to the services, several are added. An service is added to the class in order to introduce the values of this attribute; it is named set_assignment_date. Two shared services are included in both classes due to the cardinality of the structural relationship, and the fact that it is dynamic; namely an insertion shared service named ins_supplier and a deletion shared service named del_supplier. A transaction named S2_ASSIGN_SUPPLIER is defined[21] in order to execute atomically the services set_assignment_date and ins_supplier.

---

[20] As argued above, as long as the communicative event diagram has been properly extended so as to include all the precedent communicative events and the requirements model is complete, the classes to be related already exist in the class diagram. In case of incompleteness, the classes may not exist.

### 4.2.8. SALE 3. Supplier evaluates the order

This communicative event also affects the client order so, with regards to the class diagram derivation, it also extends the class CLIENTORDER. Note that the reference field Order refers to a client order (its domain is Order). Apart from the three new attributes that correspond to the three data fields of the message structure (e.g. the data field Decision leads to adding an attribute named decision), an service named Sale3_evaluate is added to the class (so as to set the value of these attributes when this communicative event occurs)

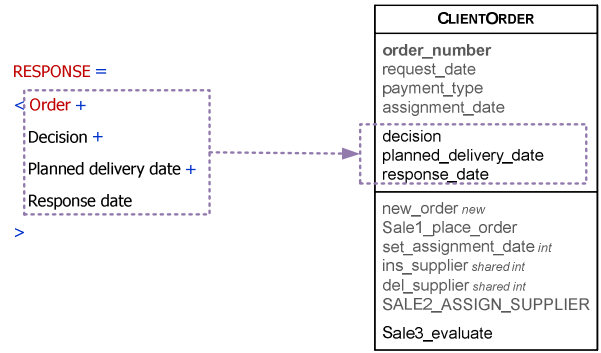

Figure 19.   Class diagram view of SALE 3

### 4.2.9. LOGI 10. Transport manager hires truck driver

The communicative event LOGI 10 results in the creation of a truck driver record (which is a new business object that appears for the first time in the derivation process). Thus, the processing of the message structure leads to the derivation of a new class that is named TRUCKDRIVER. Each data field implies adding an attribute to the new class (e.g. the data field Telephone leads to the derivation of the attribute telephone).

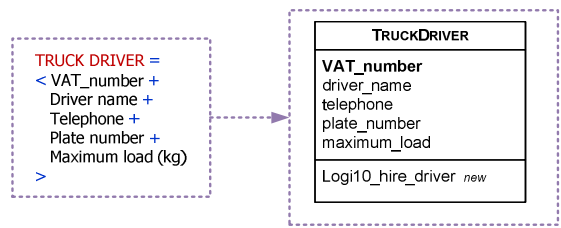

Figure 20.   Class diagram view of LOGI 10

Given the contextual restriction about the identification of truck driver records "Truck drivers are identified by means of the VAT number" (see page 34), the attribute VAT_number is designated as class identifier. A creation service named Logi10_hire_driver is added to the class in order to create new instances and to set the values of the attributes.

### 4.2.10. SALE 4. Transport manager arranges logistics

This communicative event affects the client order by adding logistic information to it. Therefore, it extends the class CLIENTORDER by adding a new attribute named logistic_comments (which is derived from the data field Logistic comments).

---

[21] Section 6 includes a discussion on a possible improvement to the derivation guidelines that allows avoiding the derivation of this transaction.

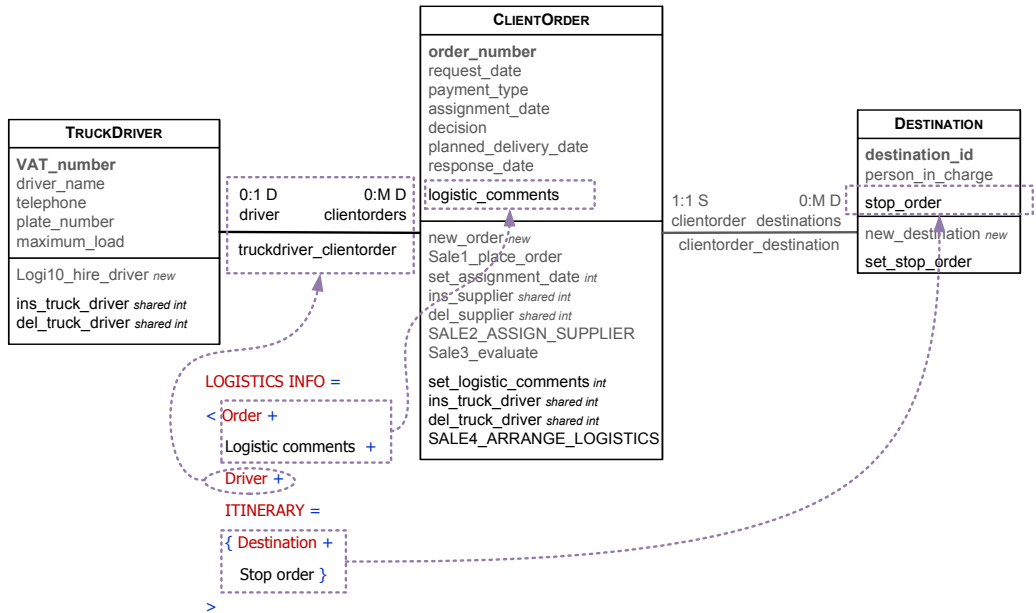

Figure 21. Class diagram view of SALE 4

Also, the reference field Driver (whose domain is Driver so it refers to a truck driver record) leads to adding a structural relationship between CLIENTORDER and TRUCKDRIVER. Given the cardinalities of the structural relationship, two shared services are added (ins_truck_driver and del_truck_driver) both to CLIENTORDER and TRUCKDRIVER.

Furthermore, a transaction named SALE4_ARRANGE_LOGISTICS is added to class ClientOrder in order to ensure that the services set_logistic_comments and ins_truck_driver are executed atomically.

Additionally, the iteration substructure ITINERARY includes a reference field named Destination that refers to an order destination (its domain is Destination). Therefore, the corresponding class (i.e. DESTINATION) is extended with an attribute that is named stop_order after the data field Stop order. Additionally, a service named set_logistic_comments is added to the class DESTINATION so as to set the value of this new attribute.

### 4.2.11. RISK 4. Insurance department clerk contracts insurance policy

Communicative event RISK 4 results in the creation if an insurance policy record. This new business object appears for the first time during the processing of this communicative event. A new class named INSURANCEPOLICY is added to the class diagram and, for each data field in the message structure, this class is added an attribute (e.g. the data field Company leads to the derivation of the attribute COMPANY_NAME). Given the contextual restriction "The company and the policy number together allow identifying an insurance policy" (see page 36), the attributes company_name and policy_number are designated as class identifiers.

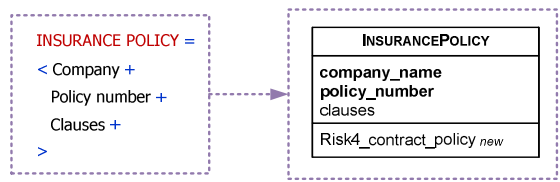

Figure 22. Class diagram view of RISK 4

A creation service named R<small>ISK4_CONTRACT_POLICY</small> is added in order to create new instances and to set the values of the attributes.

## 4.2.12. S<small>ALE</small> 5. Insurance department clerk specifies clauses

This communicative event affects the client order because it assigns the order the insurance policy that will insure the shipment against damages, theft, etc. Also, in some cases, the insurance department clerk adds some extra insurance clauses that clarify or restrict the conditions of the insurance policy.

The reference field <span style="color:red">Order</span> indicates that the client order is affected and, therefore, the class C<small>LIENT</small>O<small>RDER</small> is extended. The data field Extra clauses leads to the adding a new attribute named extra_clauses. A service named set_extra_clauses is added in order to set the value of this new attribute.

The reference field Policy (since its domain is <span style="color:blue">Insurance policy</span>; i.e. an insurance policy) implies adding an structural relationship between the classes C<small>LIENT</small>O<small>RDER</small> and I<small>NSURANCE</small>P<small>OLICY</small>. In absence of further structural restrictions, the only cardinality that can be inferred from the requirements model is the maximum cardinality 1 on the side of I<small>NSURANCE</small>P<small>OLICY</small>, from the fact that the relationship is derived from a reference field. The rest of the cardinalities need to be asked to the user or the analysts should act according to their own criteria. In this case the resulting cardinalities are I<small>NSURANCE</small>P<small>OLICY</small> 0:1 --- 0:M C<small>LIENT</small>O<small>RDER</small>. These cardinalities and the fact that the structural relationship is dynamic imply adding two shared services to both classes; ins_insurance_policy sets the link between instances and del_insurance_policy unsets the link between instances (these services are prescribed by the OO-Method).

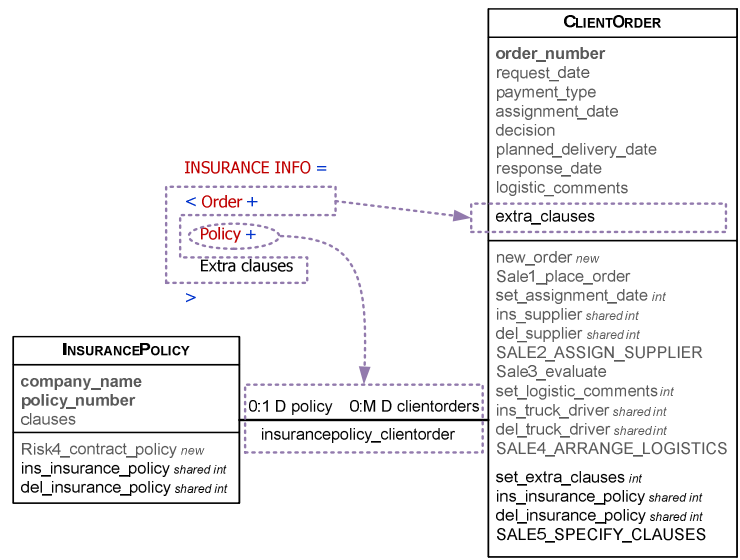

Figure 23. Class diagram view of S<small>ALE</small> 5

Additionally, a transaction named SALE5_SPECIFY_CLAUSES is added to the class in order to ensure the atomic execution of the services set_extra_clauses and ins_insurance_policy.

## 4.2.13. S<small>ALE</small> 6. Supplier notifies the shipping of the goods

This communicative event affects the client order by indicating that the truck has picked the shipment from the supplier warehouse (note that the reference field <span style="color:red">Order</span> confirms this). The data field Shipping timestamp leads to adding a new attribute to the class C<small>LIENT</small>O<small>RDER</small>; it is named shipping_timestamp.

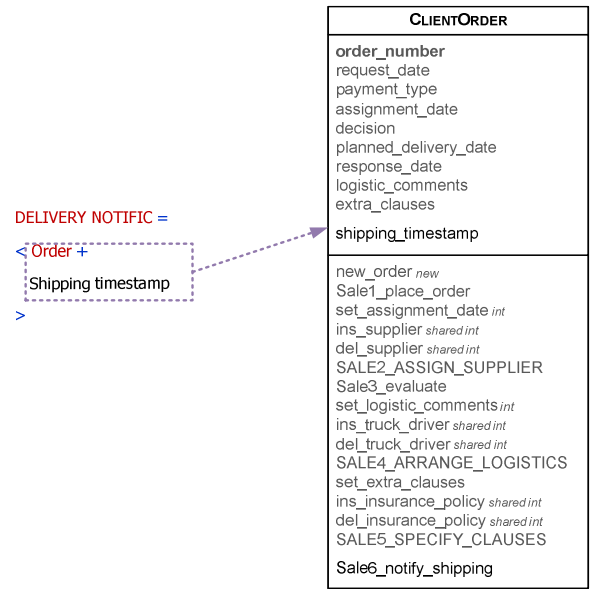

Figure 24.  Class diagram view of SALE 6

A service named Sale6_notify_shipping in order to set the value to the new attribute.

## 4.2.14. Class diagram

We have shown in the previous sections how the communicative events are processed in order to obtain class diagram view. The class diagram views are integrated incrementally. Therefore, at the end of the derivation process, the complete class diagram is obtained (see Figure 25).

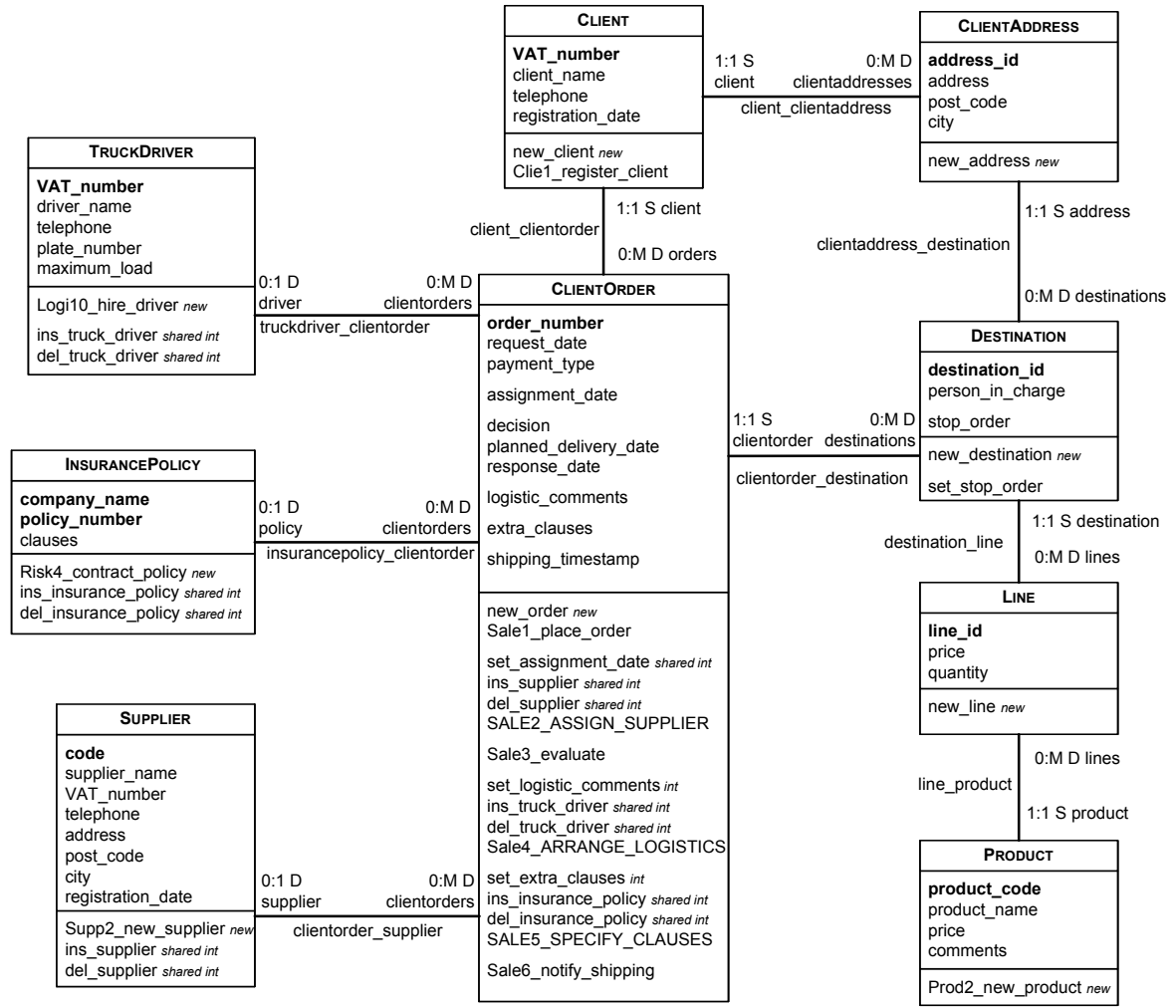

Figure 25.  Class diagram corresponding to SuperStationery Co.

## 4.3. Derivation of the Dynamic Model

Note: This section presents work under revision. It does not provide accurate information but it is left in the document so the readers can picture the derivation of the Dynamic Model in their mind.

The Dynamic Model can be derived from the Communicative Event Diagram. For this purpose, the communicative event diagram is processed in order to create a state transition diagram. The main guideline is the following: communicative events are converted into transitions and precedence relationships are converted into states. Additionally, for each service (i.e. atomic services ascribed to a class) that takes part in a transaction (i.e. molecular services), a transition must also be defined.

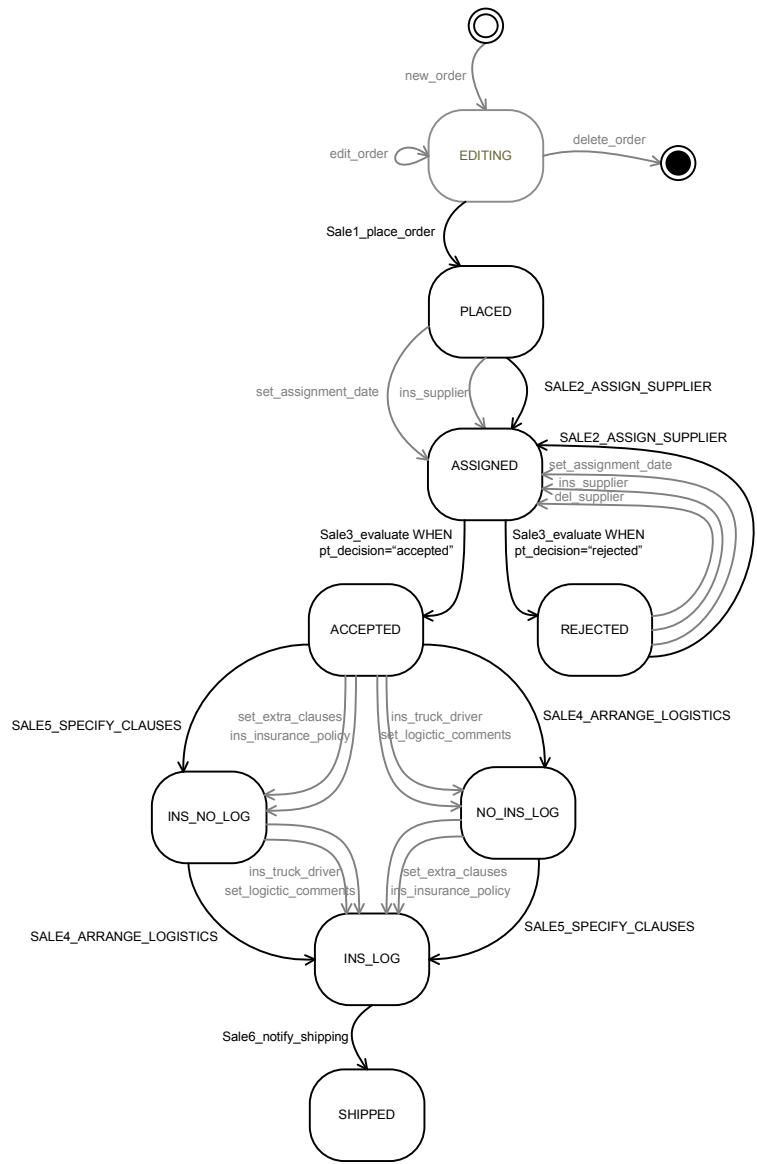

Figure 26.   Dynamic model that corresponds to the SuperStationery Co. case.

Figure 26 shows the state transition diagram that corresponds to the class CLIENTORDER. Transitions that correspond to communicative events appear in black (i.e. black arrows and black service names), whereas transitions that correspond to other services appear in grey (i.e. grey arrows and grey service names).

As a comment, the state transition diagram has been added a state that does not correspond to a precedence relationship; namely EDITING. The client order is the EDITING state from the moment the editing of the order starts, until the moment that the order is completed and the IS reaction is triggered.

# 5. CONSTRUCTION OF THE CONCEPTUAL MODEL IN OLIVANOVA

In the following, a series of snapshots is shown, to illustrate how the class diagram views are created using the *OLIVANOVA* Modeler [CARE Technologies].

## 5.1.1. Supp 2. Salesman registers a supplier

After of the reasoning of Section 4.2.3, it is possible to model the class diagram view in the CASE tool. The following snapshots present the creation of the class Supplier, as well as its attributes and services.

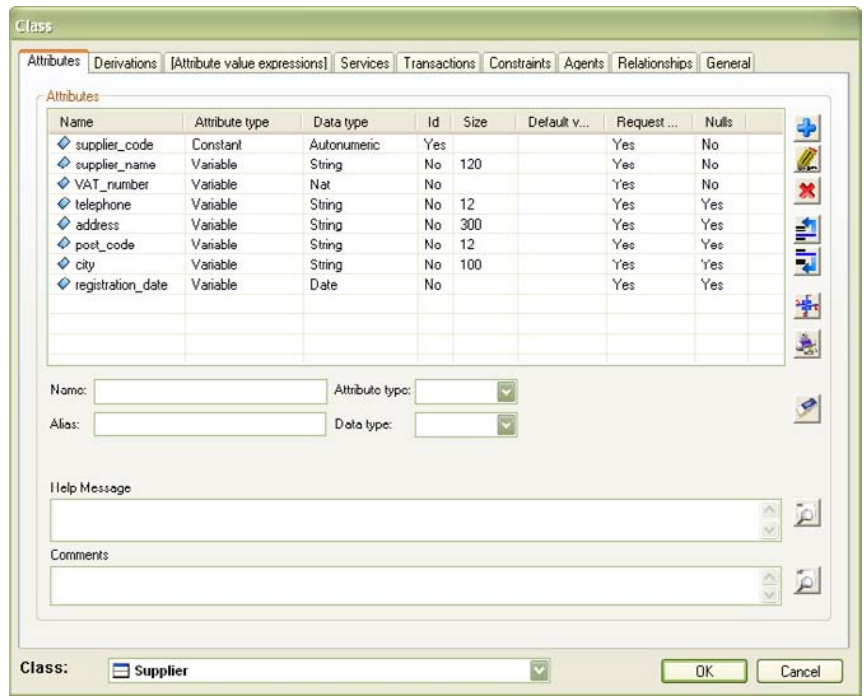

Figure 27.   Attributes of the class Supplier

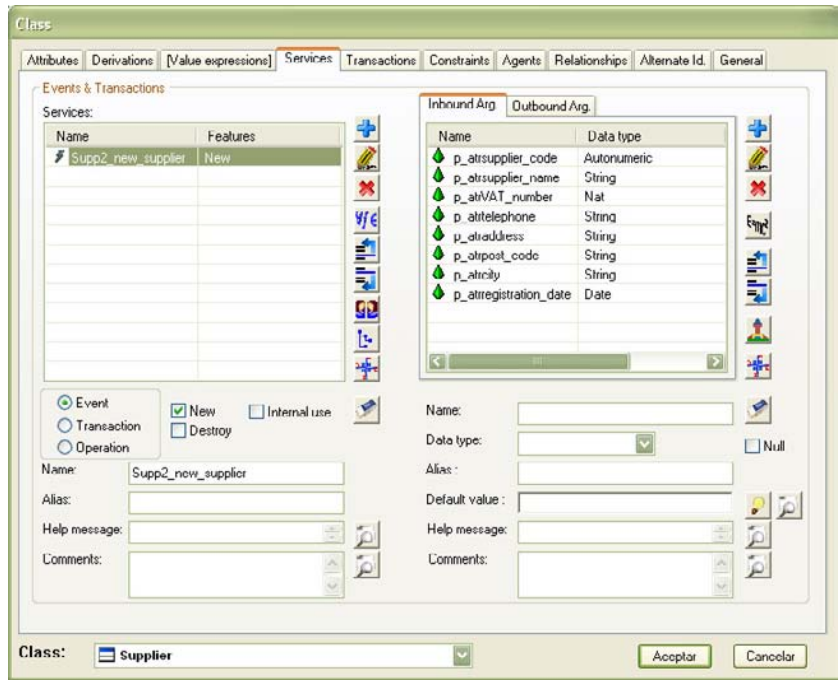

Figure 28.   Addition of a creation service to class SUPPLIER

## 5.1.2.  PROD 2. Company director defines catalogue

After of the reasoning of Section 4.2.4, it is possible to model the class diagram view in the CASE tool. The following snapshots present the creation of the class PRODUCT, as well as its attributes and services.

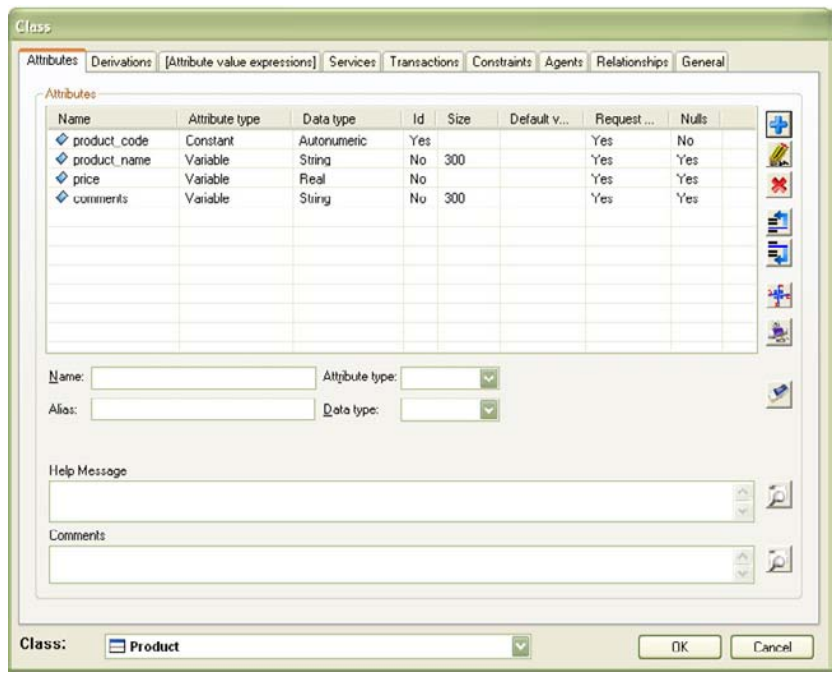

Figure 29.   Attributes of the class PRODUCT

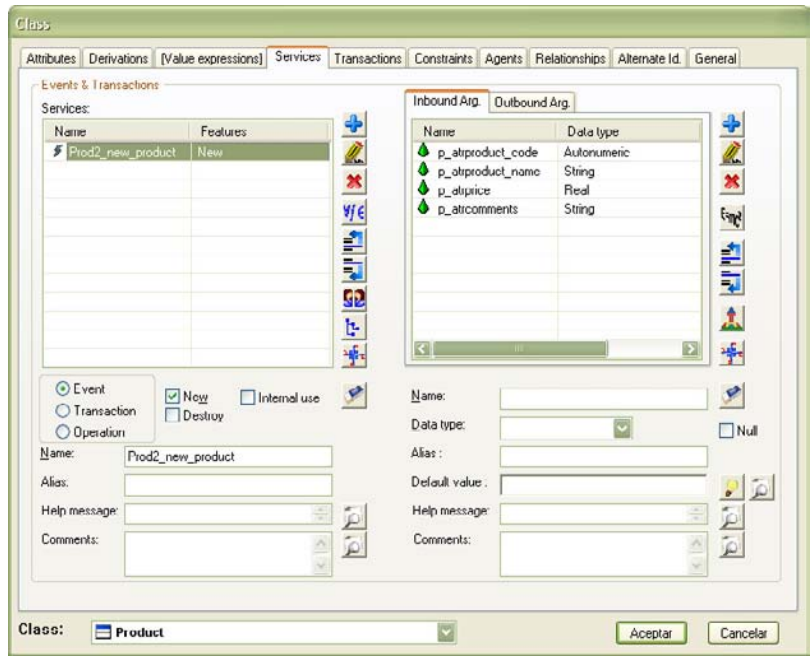

Figure 30.  Addition of a creation service to class PRODUCT

### 5.1.3.  CLIE 1. Salesman registers a client

After of the reasoning of Section 4.2.5, it is possible to model the class diagram view in the CASE tool. The following snapshots present the creation of the class CLIENT and the class CLIENTADDRESS, as well as its attributes and services, and the creation of the structural relationship between these classes.

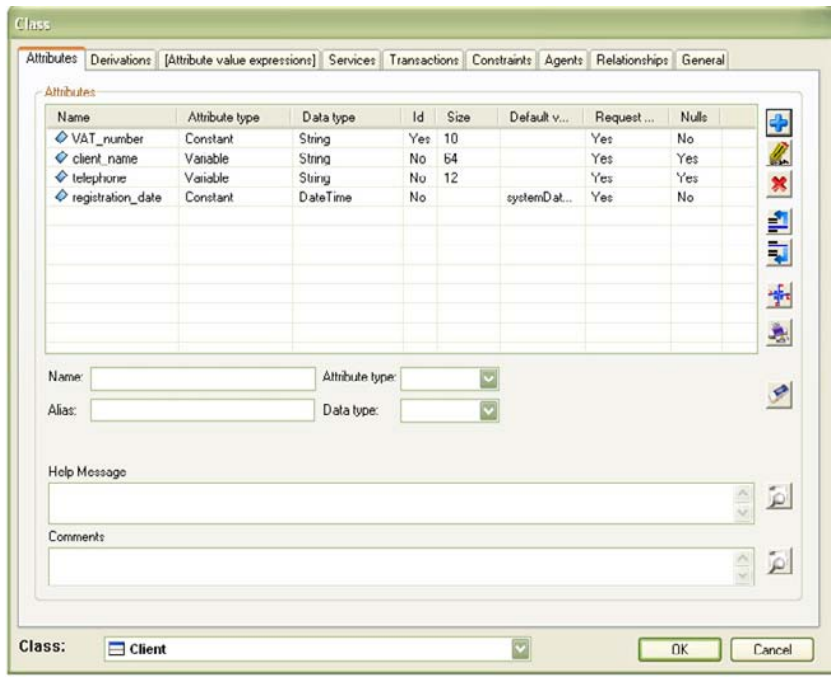

Figure 31.  Attributes of the class CLIENT

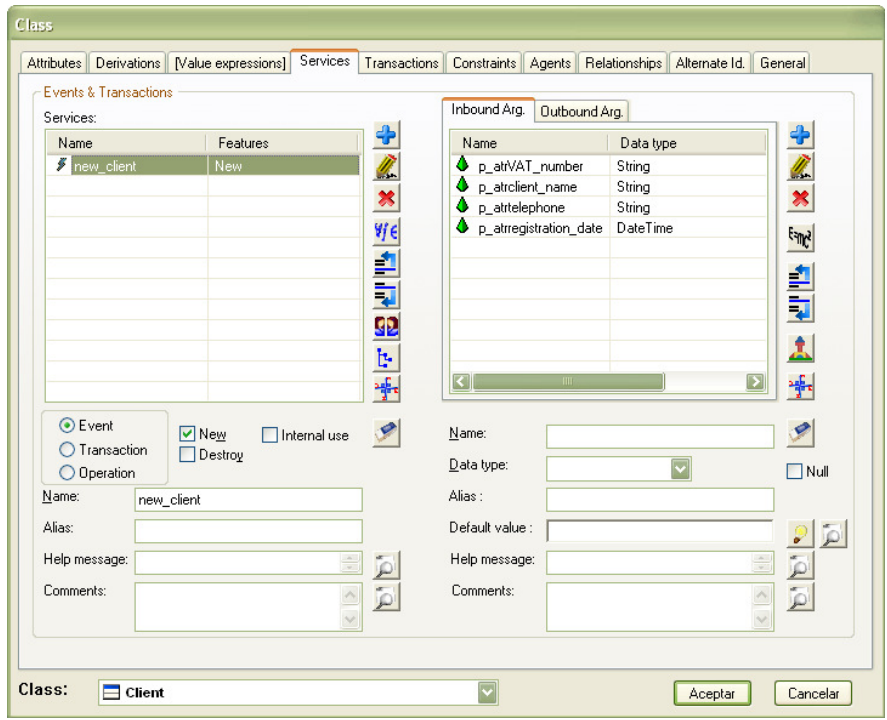

Figure 32.   Addition of the creation service new_client to class Client

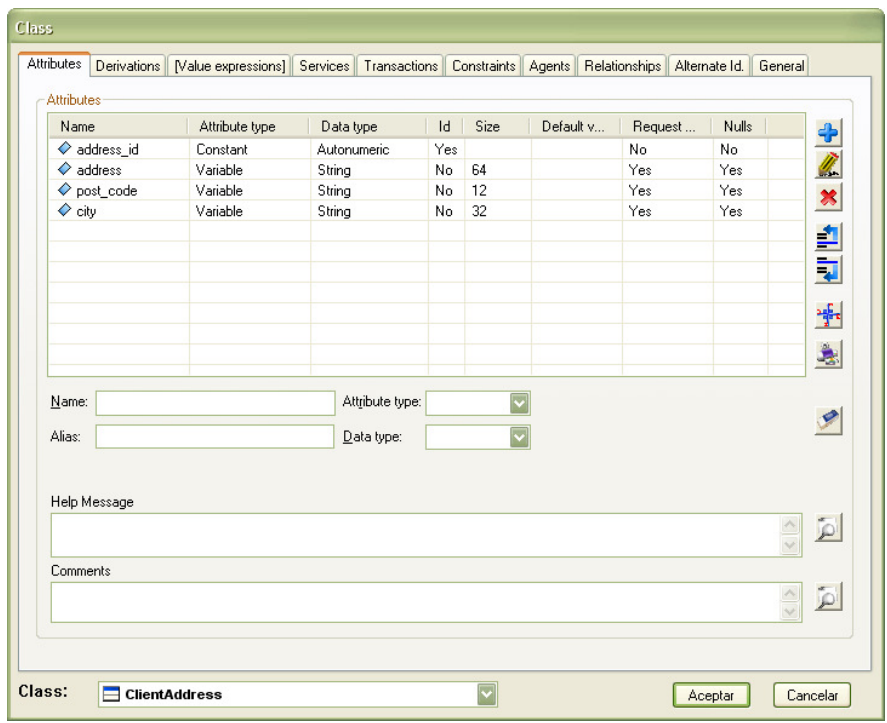

Figure 33.   Attributes of class ClientAddress

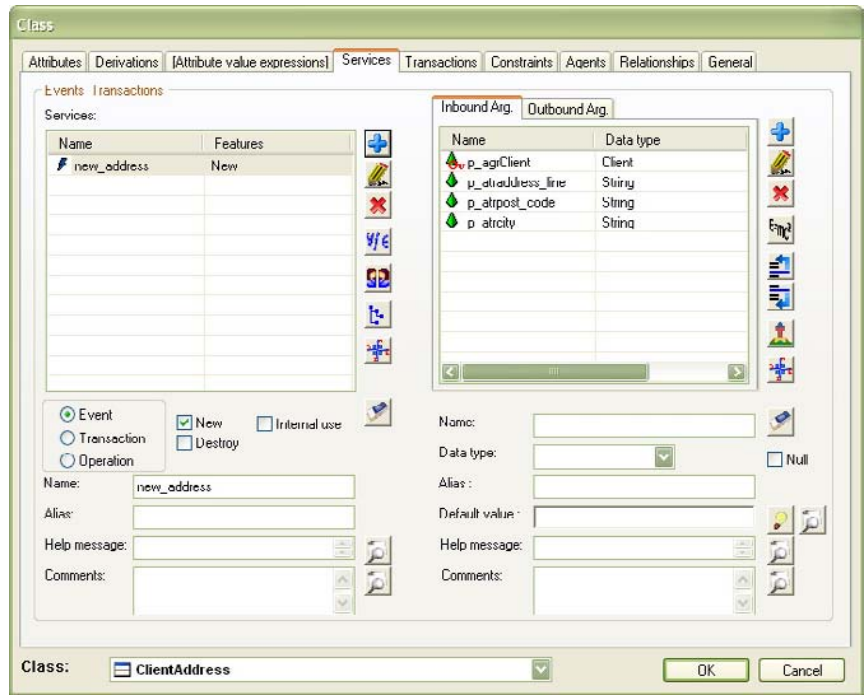

Figure 34. Addition of the creation service new_address to class CLIENTADDRESS

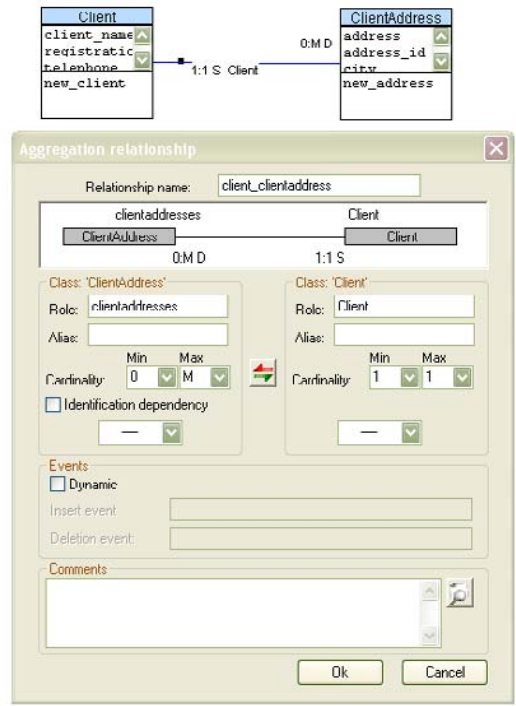

Figure 35. Details of the structural relationship between CLIENT and CLIENTADDRESS

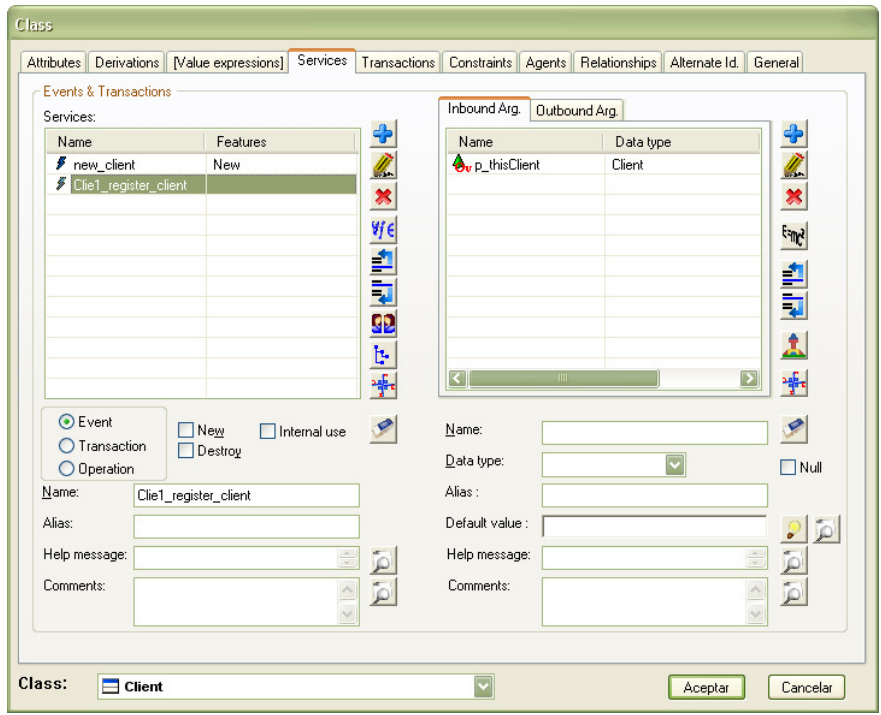

Figure 36.   Addition of an service that corresponds to the *end of edition* of a client record

### 5.1.4. Sale 1. A client places an order

After of the reasoning of Section 4.2.6, it is possible to model the class diagram view in the CASE tool. The following snapshots present the creation of the classes ClientOrder, Destination and Line, as well as its attributes and services, and the creation of the relationships among these classes.

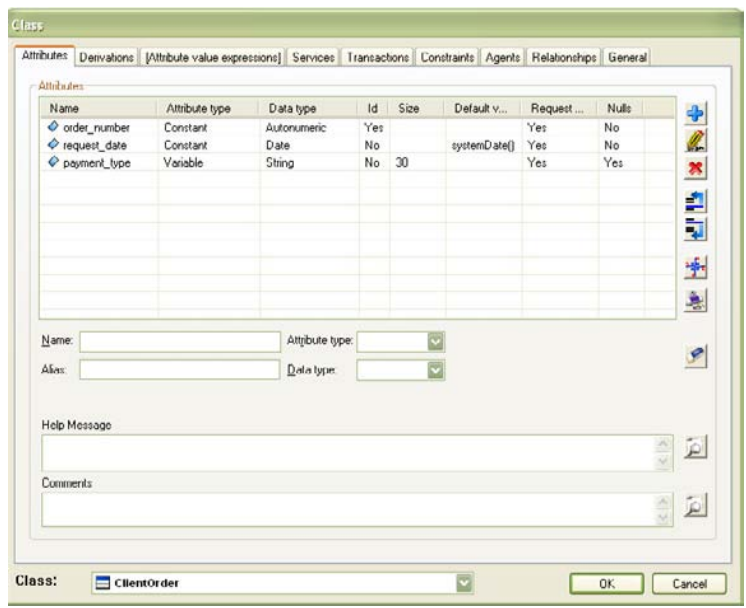

Figure 37.   Attributes of the class ClientOrder

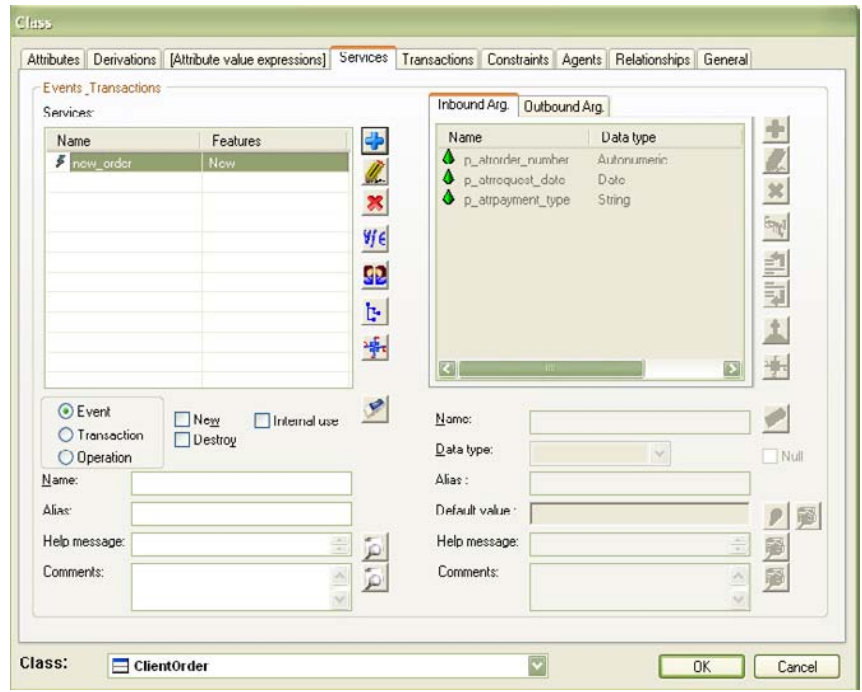

Figure 38.   Addition of a creation service new_order to the class CLIENTORDER

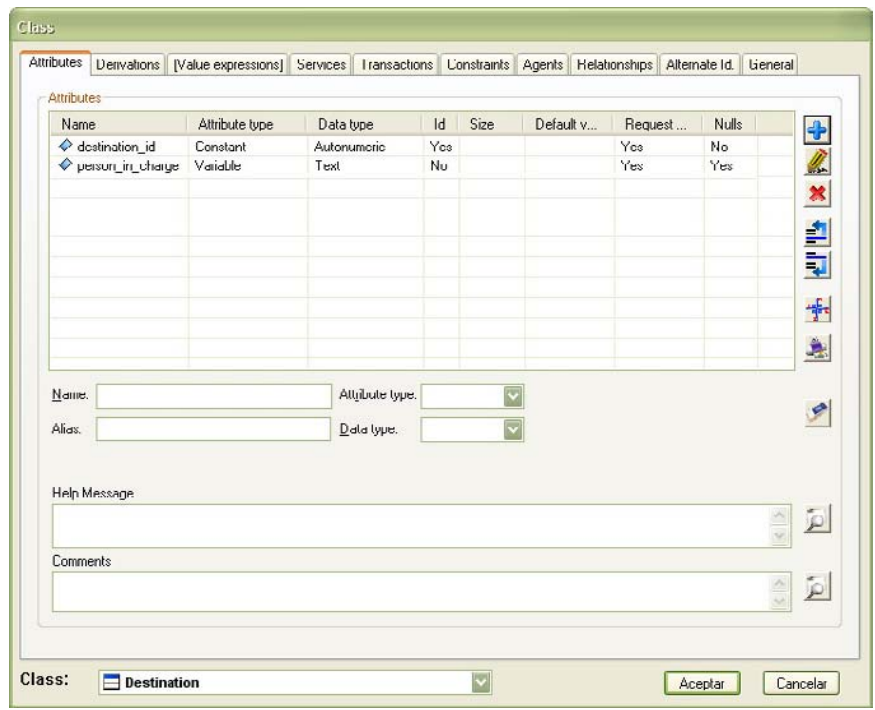

Figure 39.   Attributes of the class DESTINATION

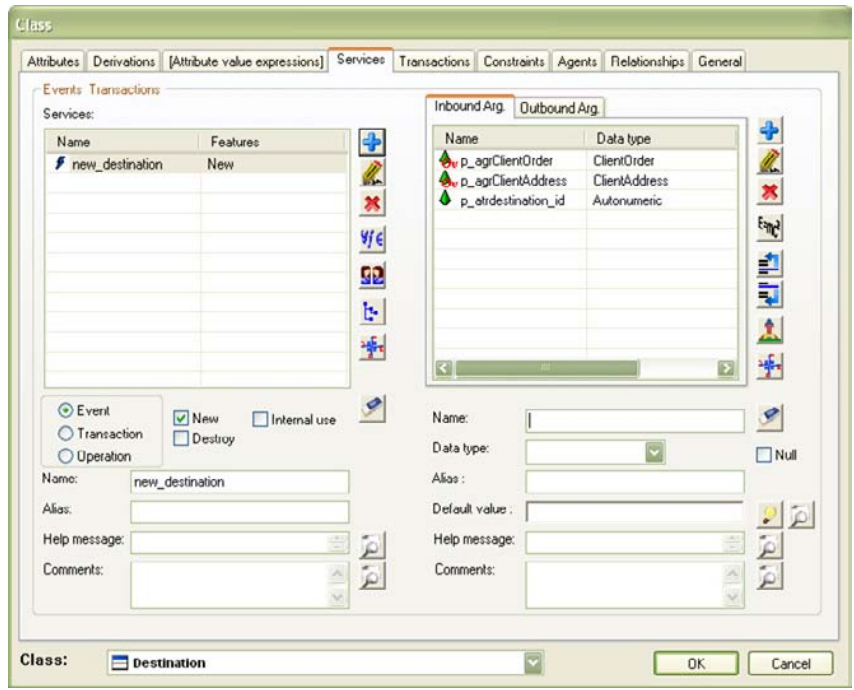

Figure 40.   Addition of a creation service new_destination to class DESTINATION

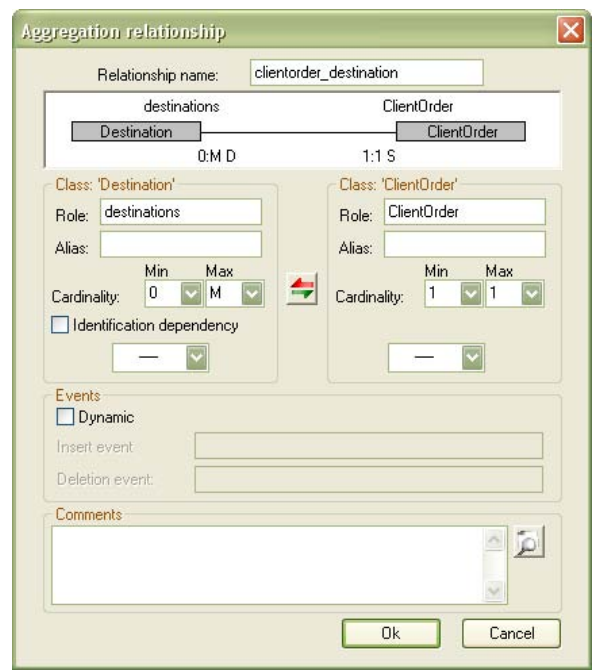

Figure 41.   Details of the structural relationship between DESTINATION and CLIENTORDER

Figure 42. Attributes of the class LINE

Figure 43. Addition of a creation service new_line to the class LINE

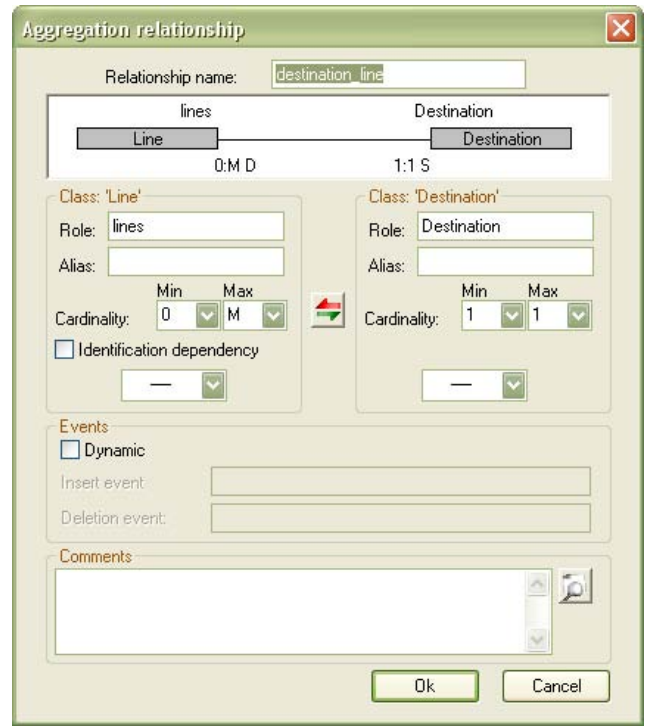

Figure 44.   Details of the structural relationship between classes LINE and DESTINATION

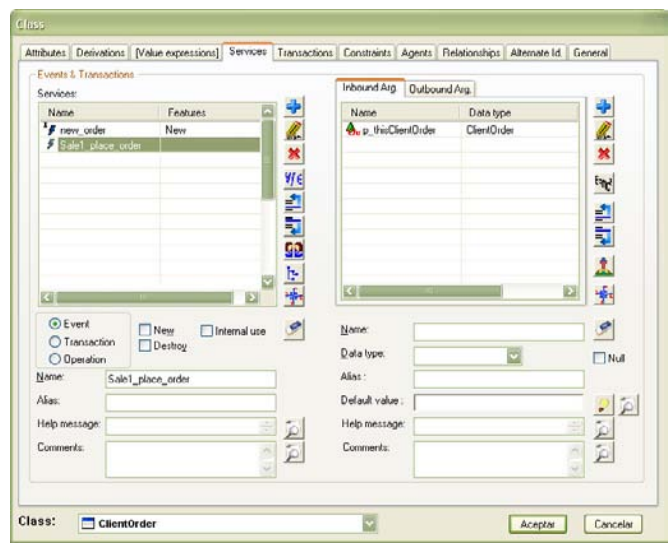

Figure 45.   Addition of an service that corresponds to the *end of edition* of a client order

The figure 46 presents class diagram view at this point. It is possible to depict the creation of the structural relationship between CLIENT and CLIENTORDER, CLIENTADDRESS and DESTINATION and LINE and PRODUCT (these relationships are derived following the steps of the Section 4.2.6).

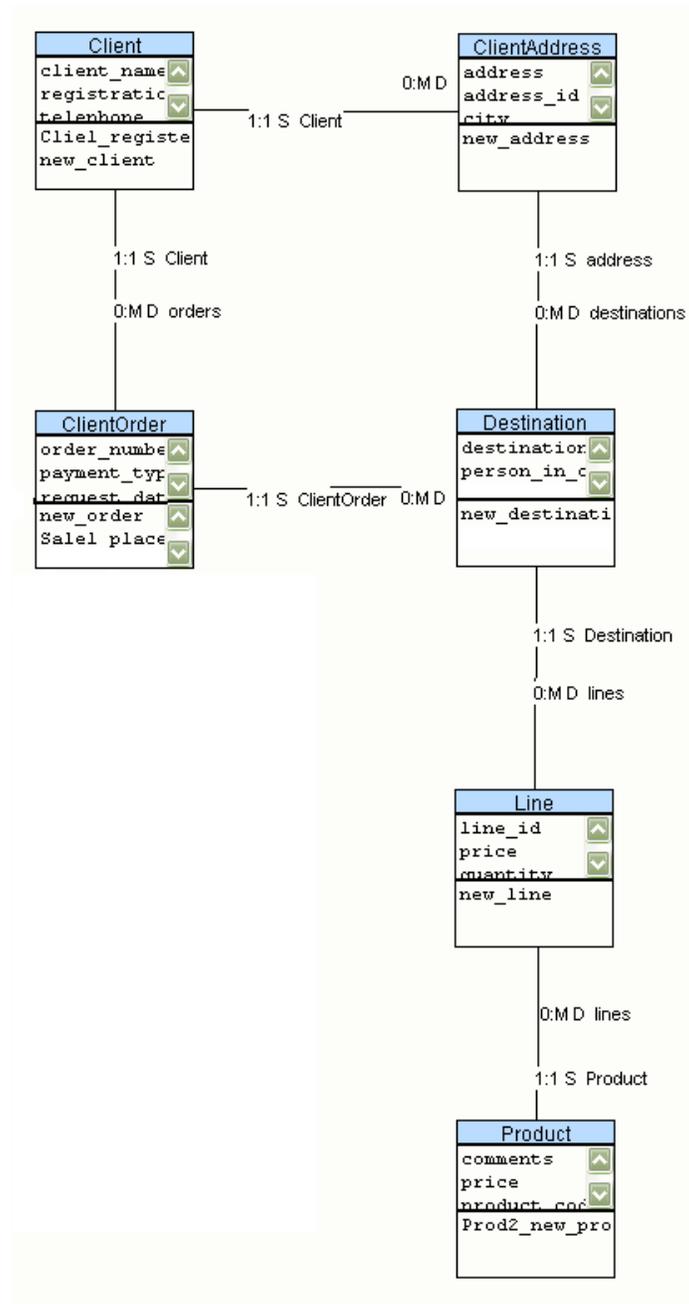

Figure 46.   The class diagram so far: clients and their orders

### 5.1.5.  SALE 2. Sales Manager assigns supplier

After of the reasoning of Section 4.2.7, it is possible to model the class diagram view in the CASE tool. The following snapshots present the creation of the attribute assignment_date, and the creation of the services set_asignment_date, ins_supplier, del_supplier and the transaction SALE2_ASSIGN_SUPPLIER of the class CLIENTORDER, as well as the creation of the structural relationship between the class CLIENTORDER and SUPPLIER.

Figure 47.   Addition of a new attribute assignment_date to the class CLIENTORDER

Figure 48.   Details of the structural relationship between the classes CLIENTORDER and SUPPLIER

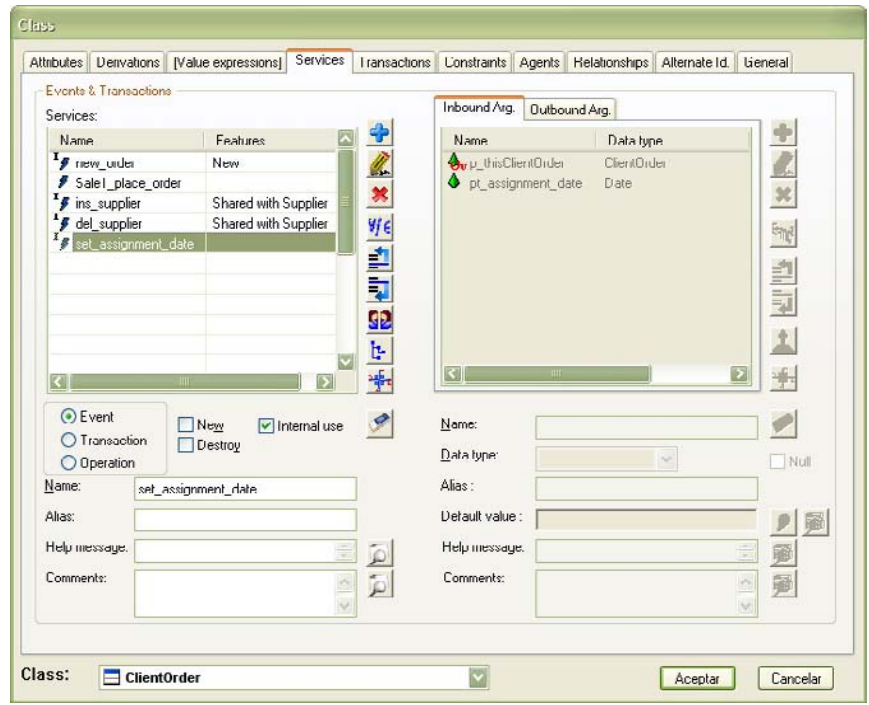

Figure 49.   Addition of an service set_assignment_date that gives value to the attribute assignment_date

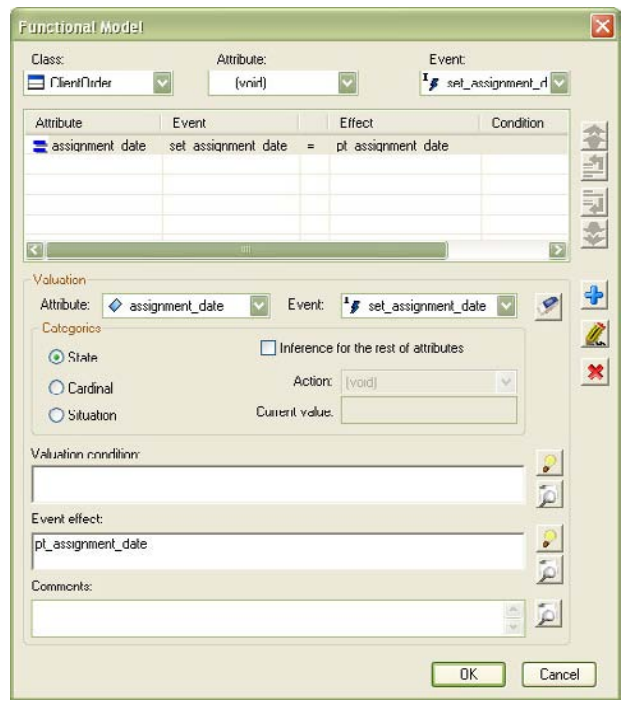

Figure 50.   Valuation rule that sets the value of the attribute assignment_date

Then the shared services ins_supplier and del_supplier are set as internal (see Figure 51); the service set_assignment_date is also set as internal. A transaction named S2_ASSIGN_SUPPLIER is created, so as to ensure the atomic execution of services ins_supplier and set_assignment_date. If a previous assignment of a supplier had been done (see the loopback between the communicative events SALE 3 and Sᴀʟᴇ 2 in the communicative event diagram, page 13), this transaction also deletes the previous assignment. See the transaction formula in Figure 52.

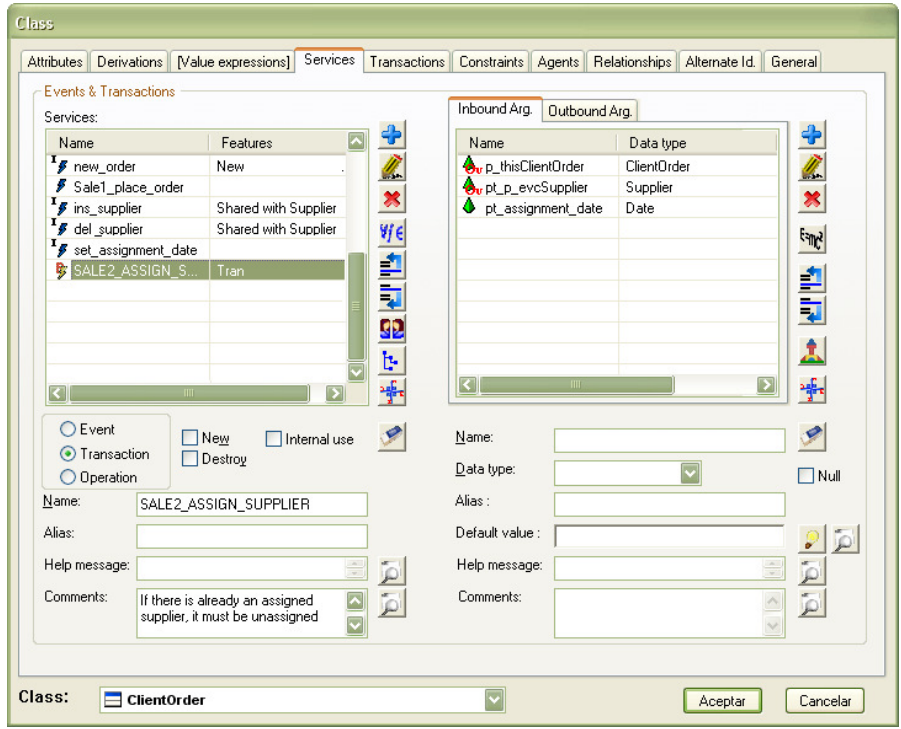

Figure 51. Addition of the transaction SALE2_ASSIGN_SUPPLIER

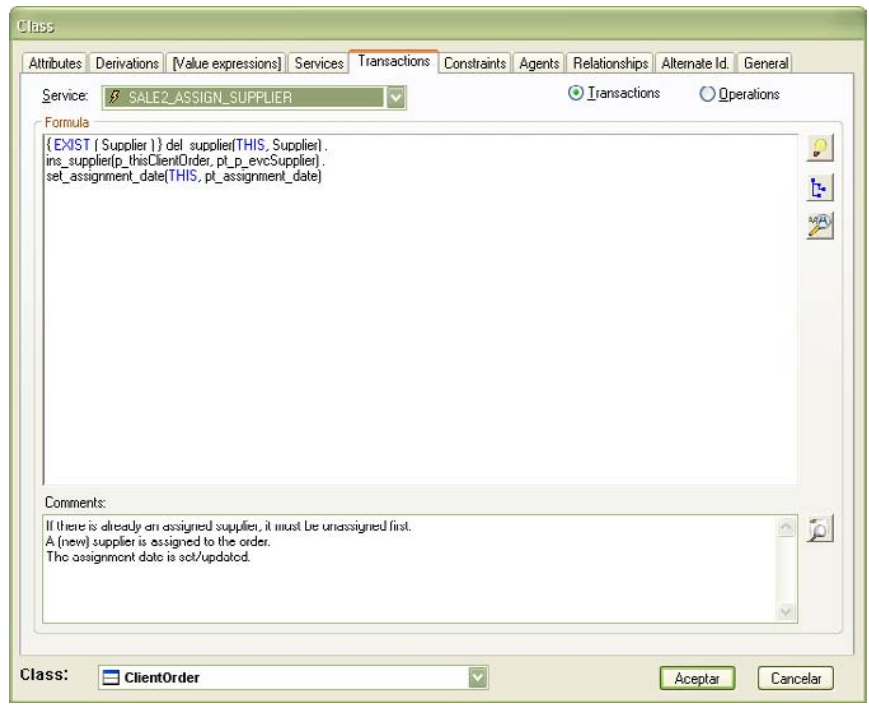

Figure 52. Transaction formula that corresponds to SALE2_ASSIGN_SUPPLIER

### 5.1.6. SALE 3. Supplier evaluates the order

After of the reasoning of Section 4.2.8, it is possible to model the class diagram view in the CASE tool. The following snapshots present the creation of the attribute decision, planned_delivery_date, response_date and the creation of the service sale3_evaluate.

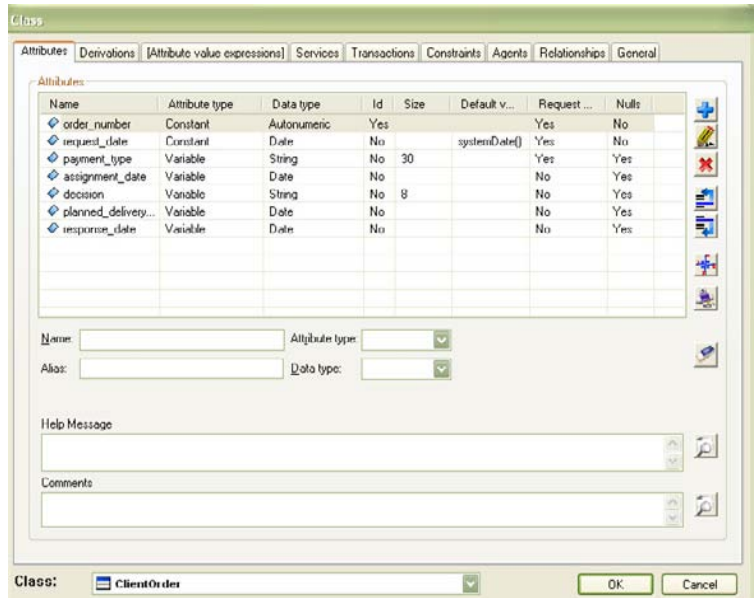

Figure 53.  Addition of attributes decision, planned_delivery_date and response_date to class CLIENTORDER

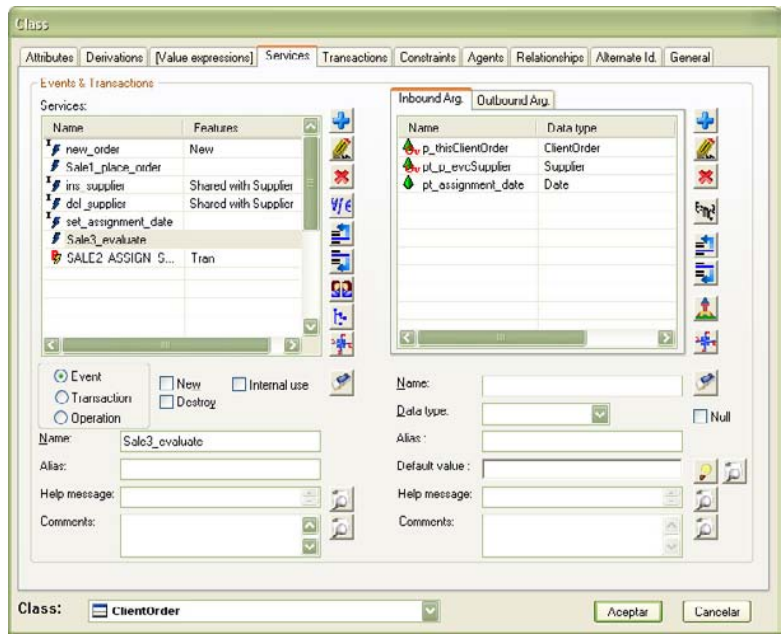

Figure 54.  Addition of service Sale3_evaluate to class CLIENTORDER

Figure 55. Valuation rules of the service S3_evaluate

### 5.1.7. LOGI 10. Transport manager hires truck driver

After of the reasoning of Section 4.2.9, it is possible to model the class diagram view in the CASE tool. The following snapshots present the creation of the class TRUCKDRIVER as well as its attributes and services.

Figure 56. Attributes of the class TRUCKDRIVER

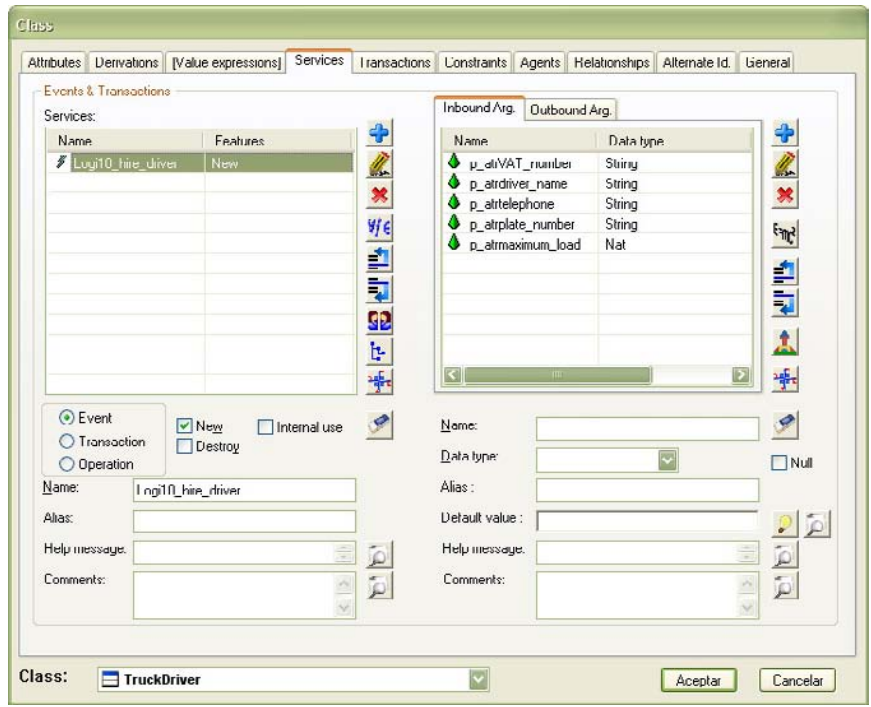

Figure 57. Addition of a creation service to class TRUCKDRIVER

## 5.1.8. SALE 4. Transport manager arranges logistics

After of the reasoning of Section 4.2.10, it is possible to model the class diagram view in the CASE tool. The following snapshots present the creation of the attribute logistic_comments, and the creation of the services set_logistic_comments, ins_truck_driver, del_truck_driver and the transaction SALE4_ARRANGE_LOGISTICS of the class CLIENTORDER, as well as the creation of the structural relationship between CLIENTORDER and TRUCKDRIVER. Besides it presents the creation of the attribute stop_order and the service set_stop_order of the class DESTINATION.

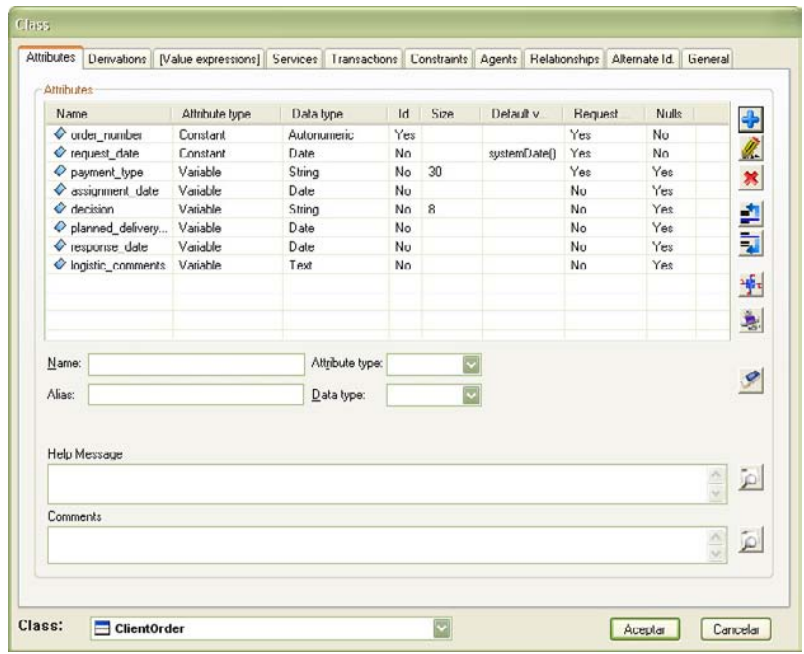

Figure 58. Addition of attribute logistic_comments to class CLIENTORDER

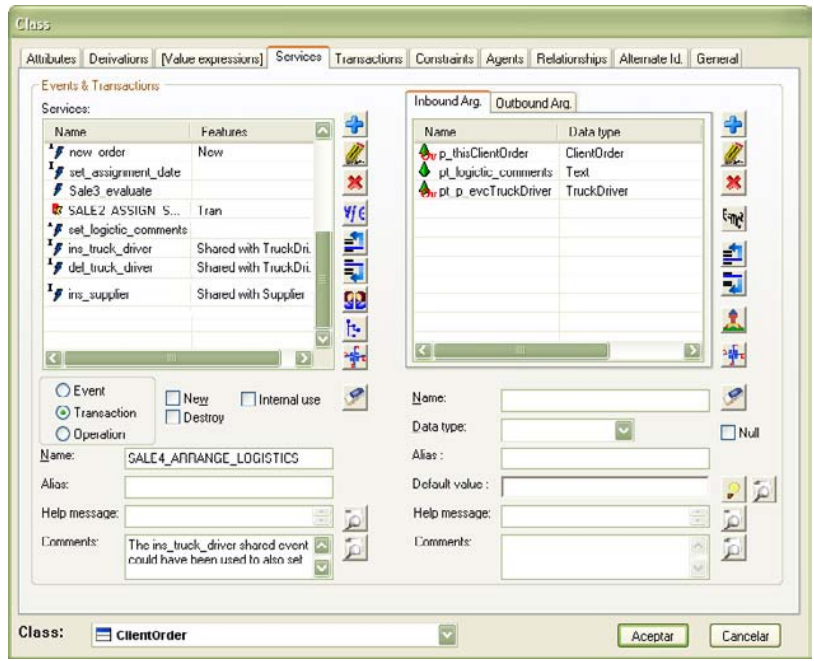

Figure 59.   Addition of services set_logistic_comments, ins_truck_driver, del_truck_driver to class CLIENTORDER

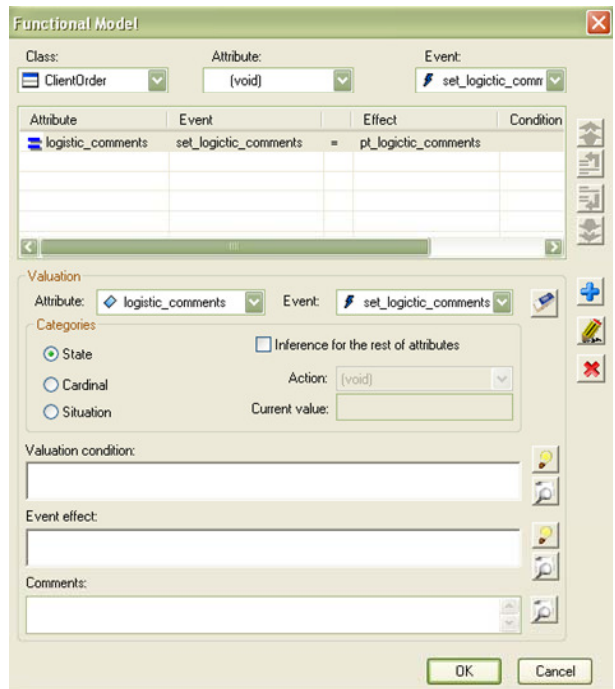

Figure 60.   Valuation rules of the service set_logistic_comment

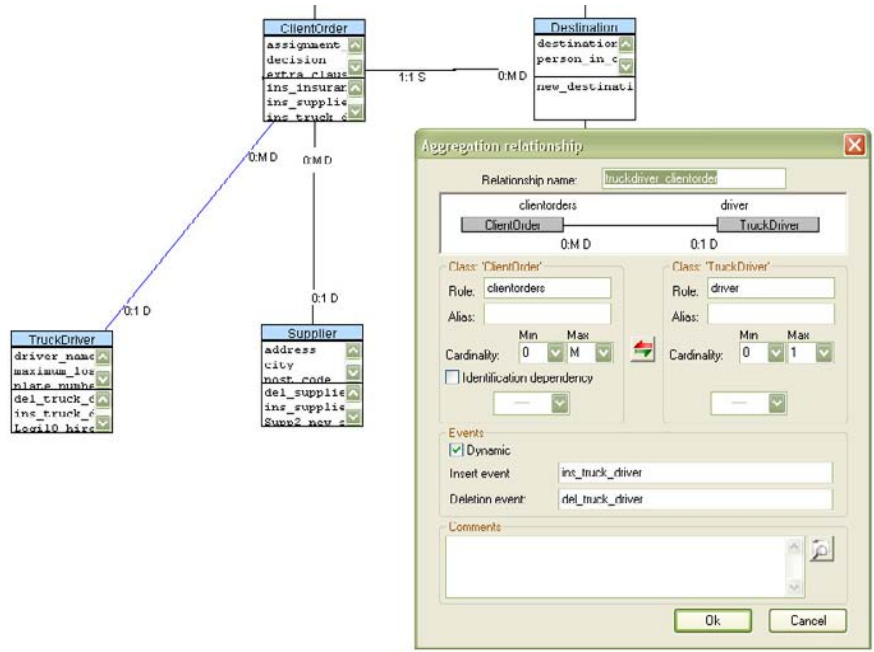

Figure 61.    Details of the structural relationship between CLIENTORDER AND TRUCKDRIVER

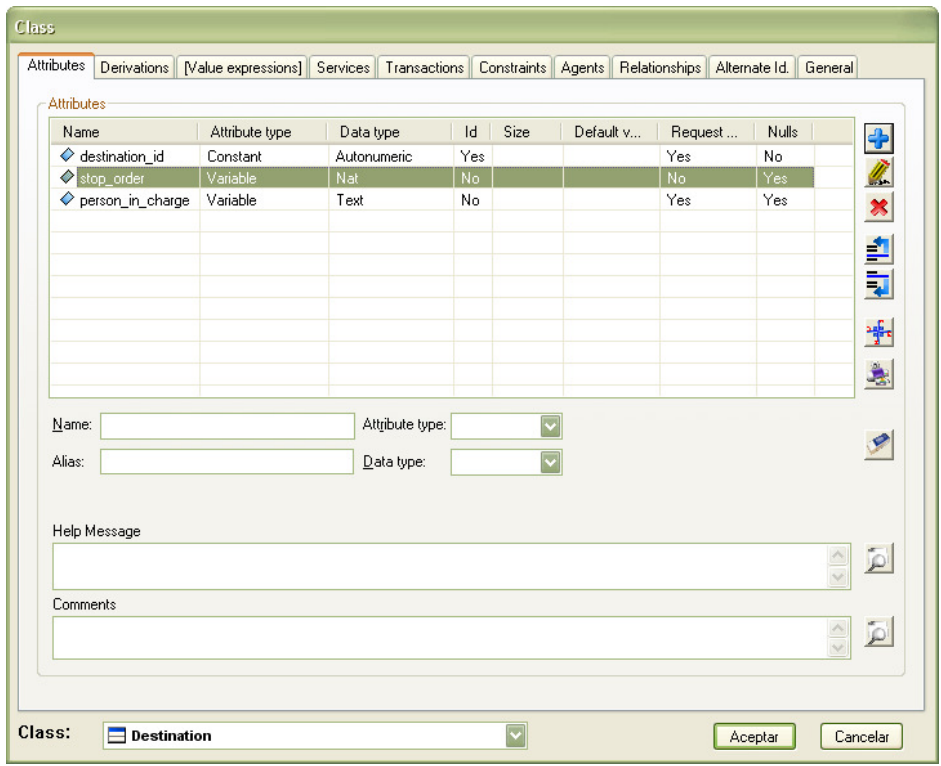

Figure 62.    Addition of attribute stop_order to the class DESTINATION

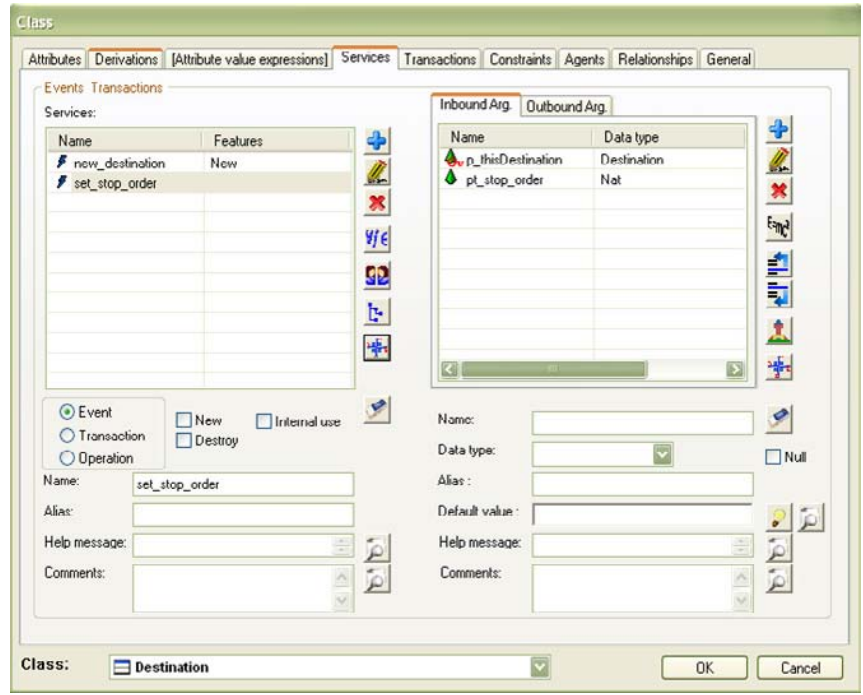

Figure 63. Addition of the service **set_stop_order** to the class DESTINATION

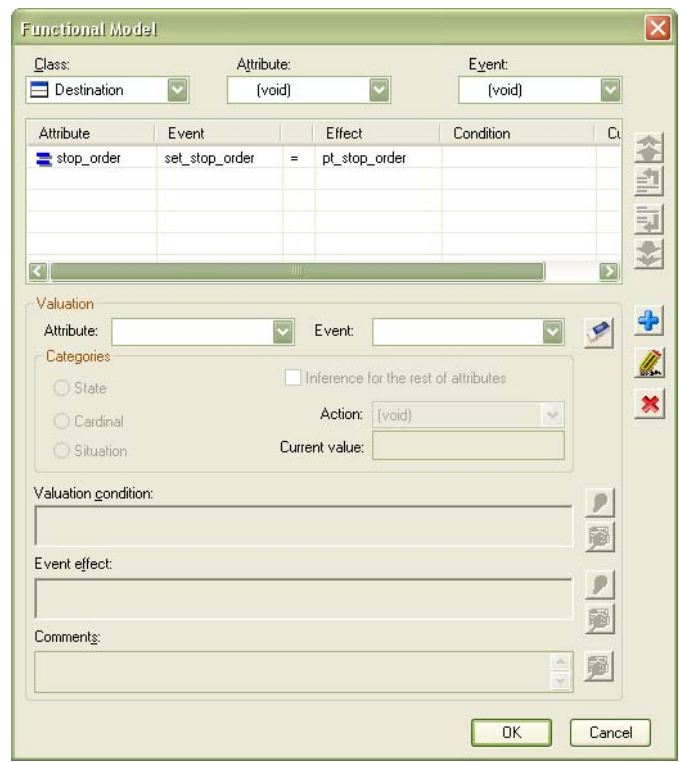

Figure 64. Valuation rule of the service set_stop_order

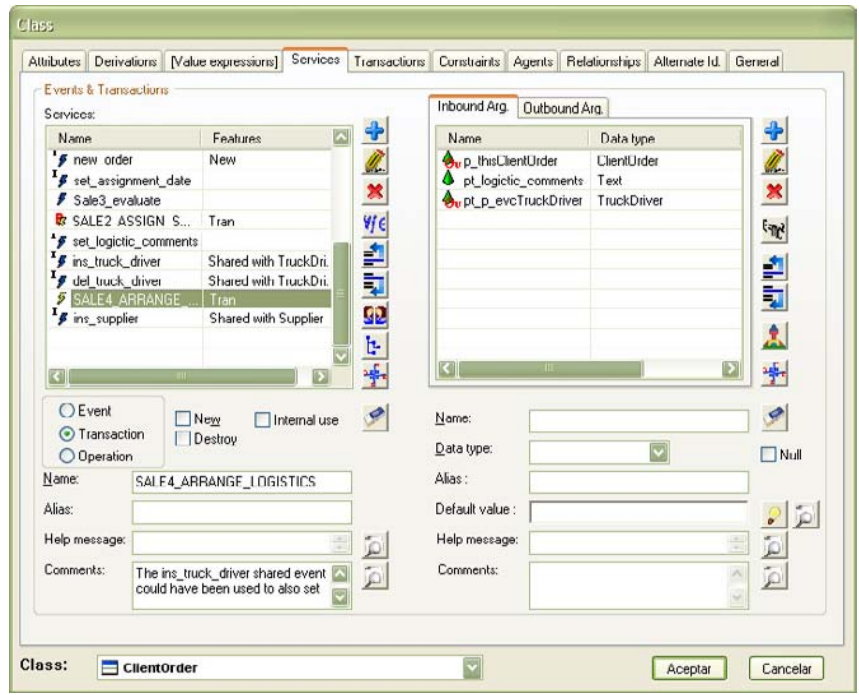

Figure 65.   Addition the transaction SALE4_ARRANGE_LOGISTICS to the class ᴄʟɪᴇɴᴛᴏʀᴅᴇʀ

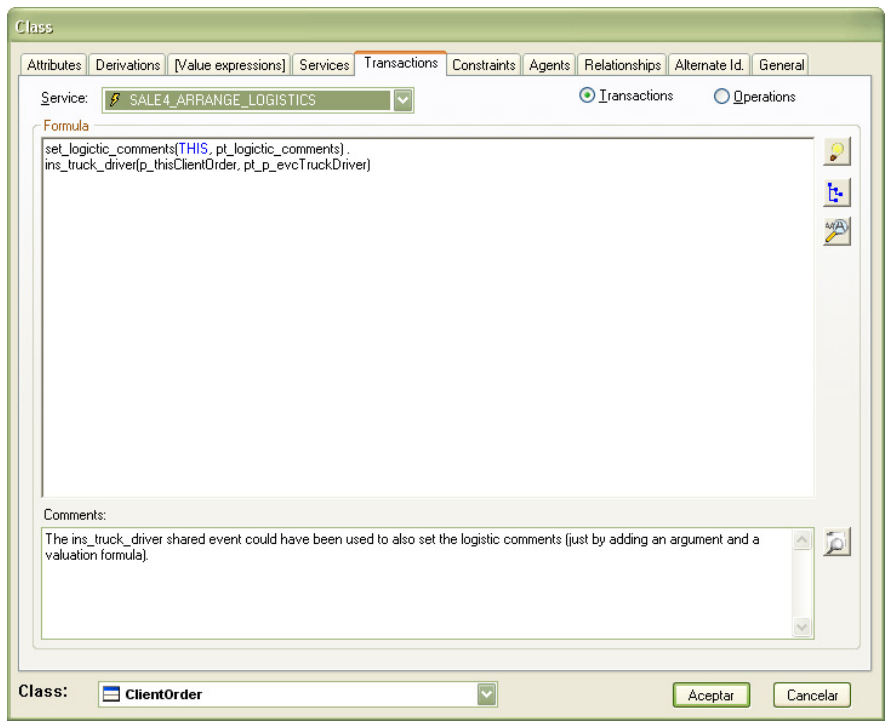

Figure 66.   Transaction formula that corresponds to SALE4_ARRANGE_LOGISTICS

## 5.1.9.  Rɪsᴋ 4. Insurance department clerk contracts insurance policy

After of the reasoning of Section 4.2.11 , it is possible to model the class diagram view in the CASE tool. The following snapshots present the creation of the class ɪɴsᴜʀᴀɴᴄᴇᴘᴏʟɪᴄʏ as well as its attributes and services.

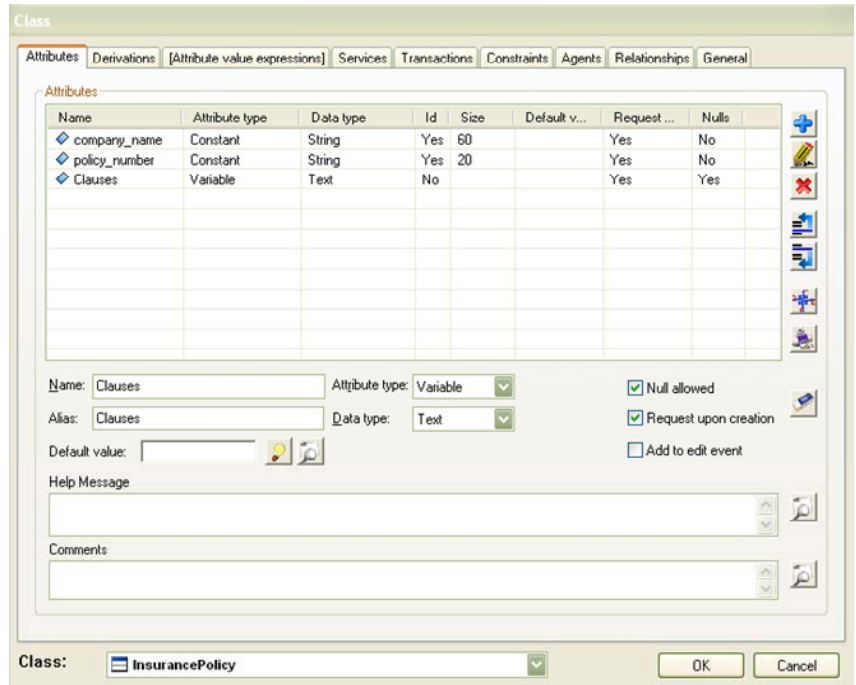

Figure 67.   Attributes of the class INSURANCEPOLICY

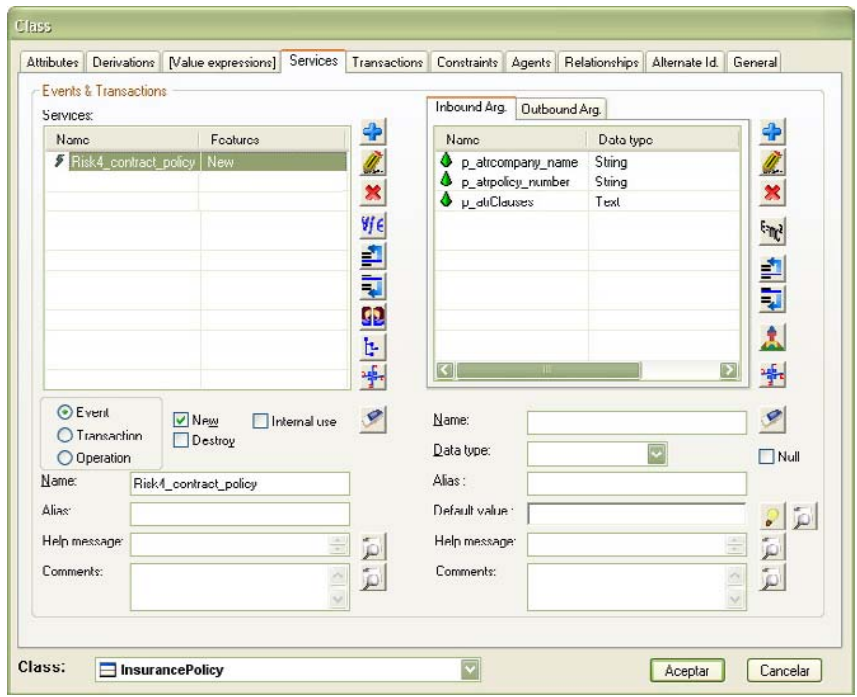

Figure 68.   Addition of a creation service to the class INSURANCEPOLICY

## 5.1.10.SALE 5. Insurance department clerk specifies clauses

After of the reasoning of Section 4.2.12, it is possible to model the class diagram view in the CASE tool. The following snapshots present the creation of the attribute extra_clauses, and the creation of the services set_extra_clauses, ins_insurance_policy, del_insurance_policy and the transaction

SALE5_SPECIFY_CLAUSES of the class CLIENTORDER, as well as the creation of the structural relationship between CLIENTORDER and INCURANCEPOLICY.

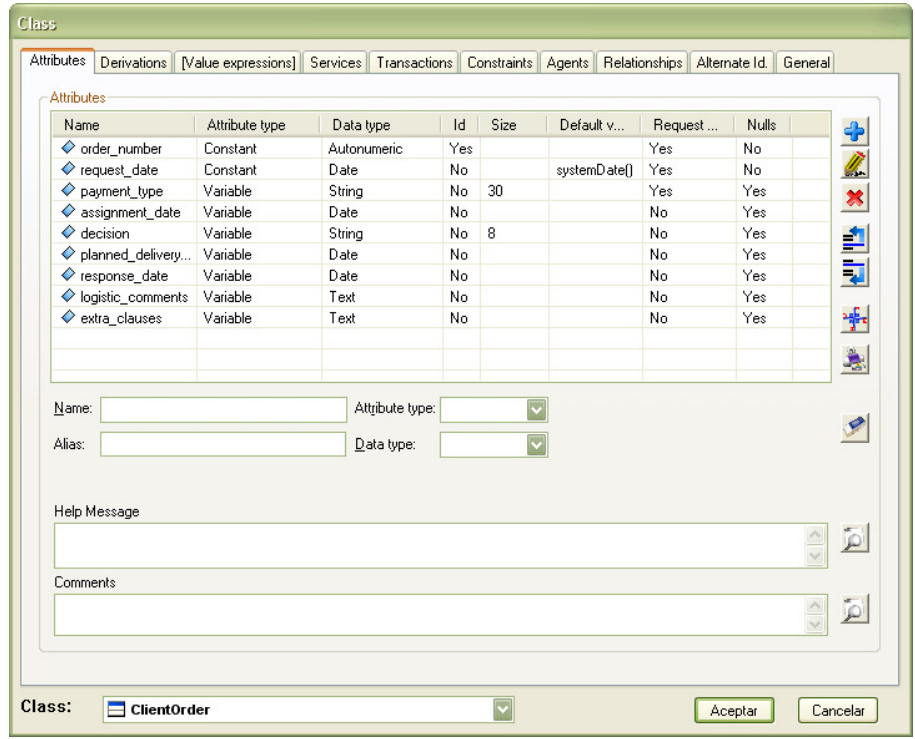

Figure 69.   Addition of attribute extra_clauses to the class CLIENTORDER

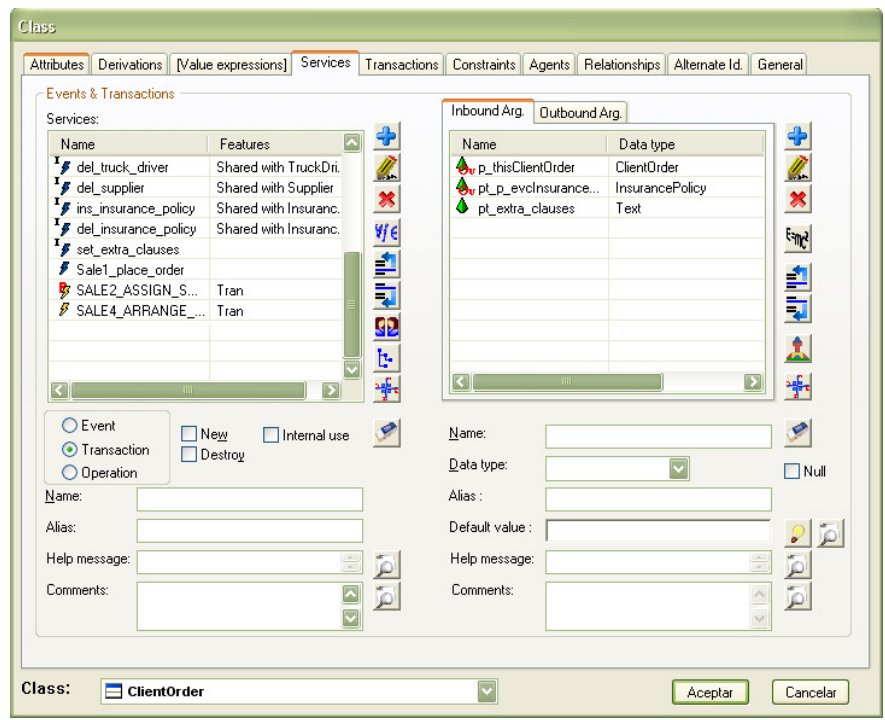

Figure 70.   Addition of the services set_extra_clauses, ins_insurance_policy and del_insurance_policy to the class CLIENTORDER

Figure 71. Valuation rules of the service set_extra_clauses

Figure 72. Details of the structural relationship between CLIENTORDER and INSURANCEPOLICY

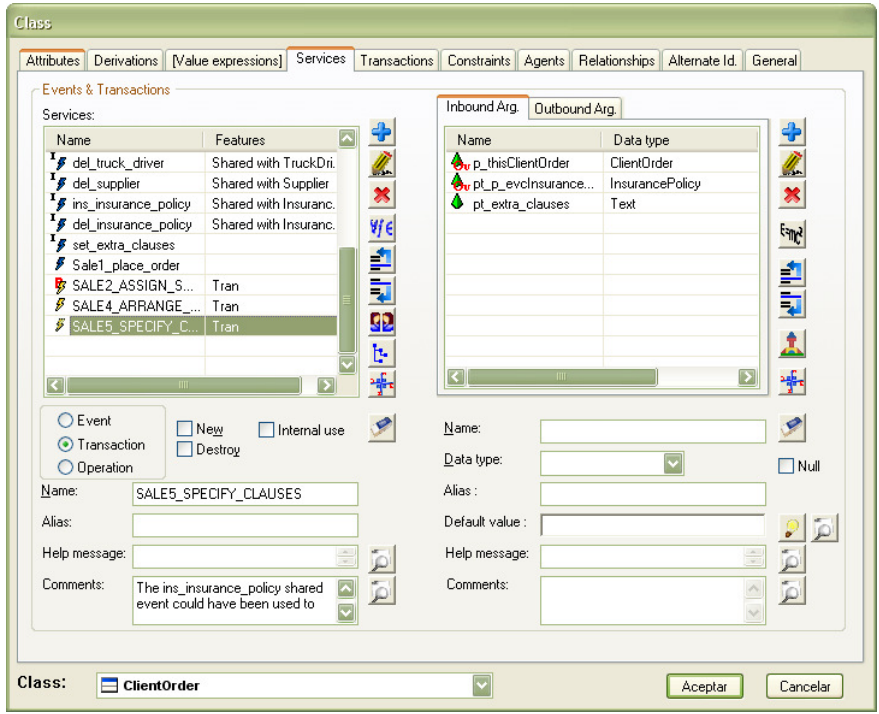

Figure 73. Addition the transaction SALE5_SPECIFY_CLAUSES to the class CLIENTORDER

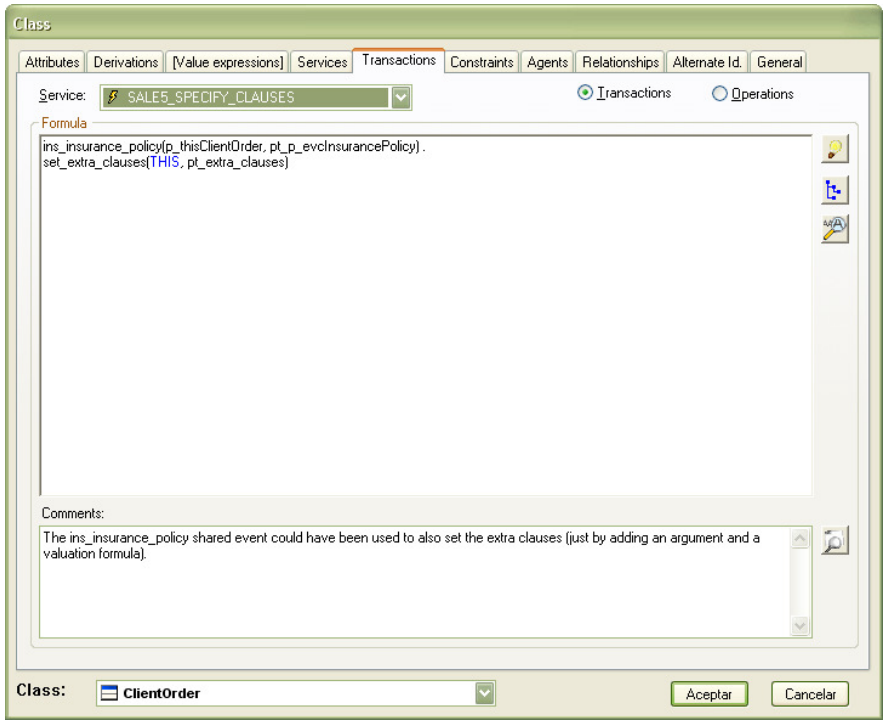

Figure 74. Transaction formula that corresponds to SALE5_SPECIFY_CLAUSES

### 5.1.11.Sale 6. Supplier notifies the shipping of the goods

After the reasoning of Section 4.2.13, it is possible to model the class diagram view in the CASE tool. The following snapshots present the creation of the attribute shipping_timestamp, and the creation of the services sale6_notify_shipping of the class Client Order.

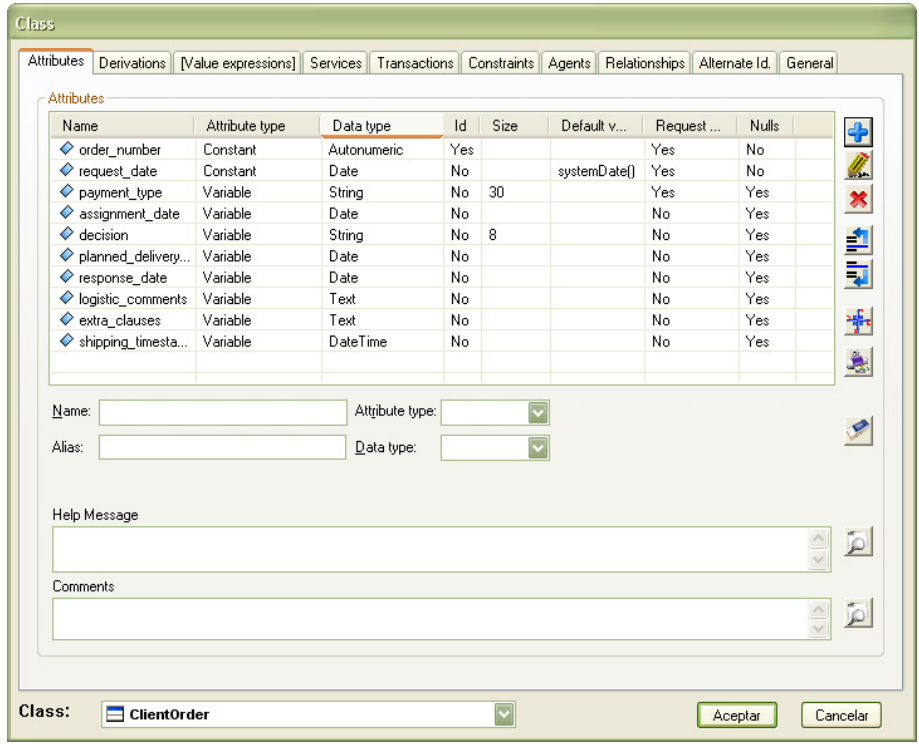

Figure 75. Addition of attribute shipping_timestamp to the class Client Order

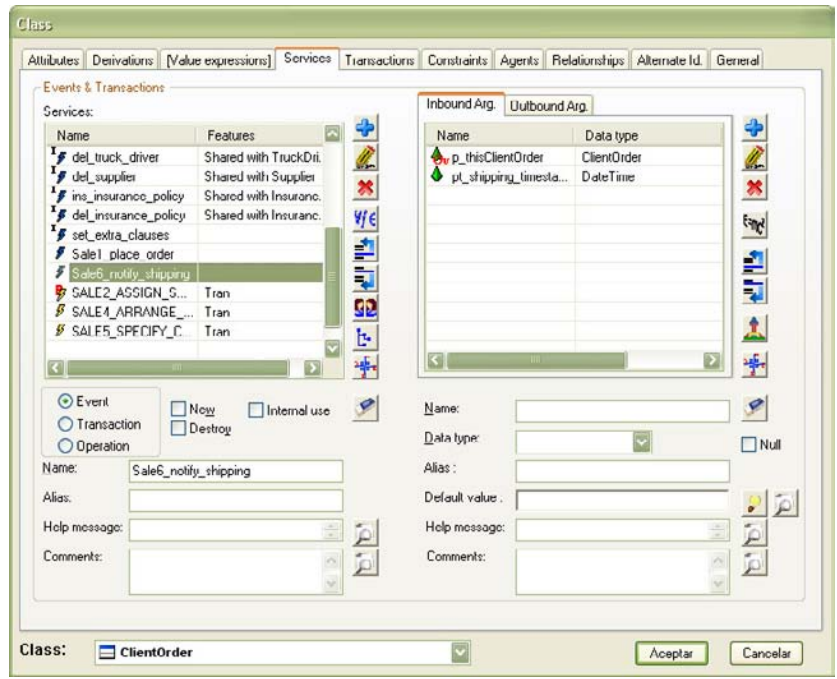

Figure 76. Addition of the service sale6_notify_shipping to the class Client Order

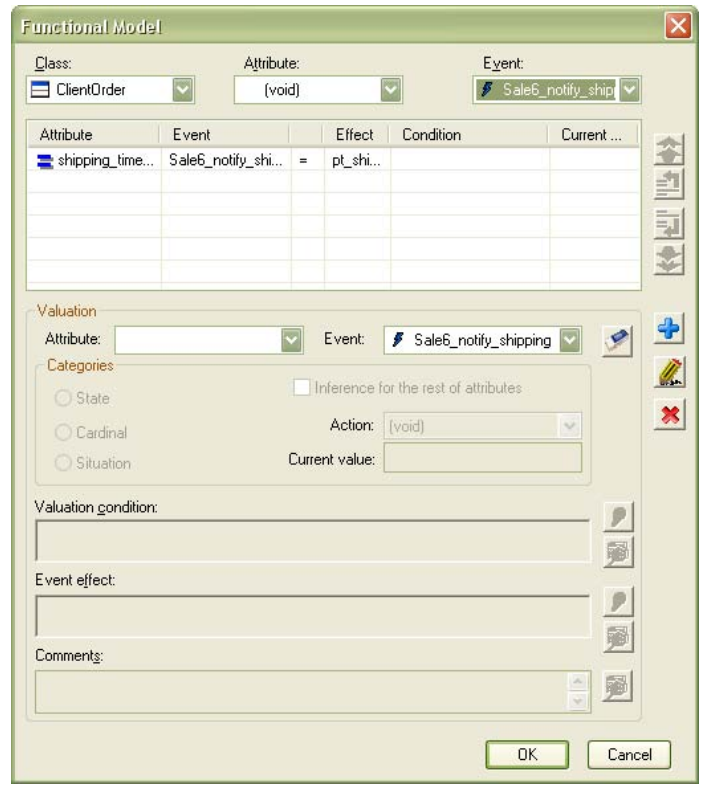

Figure 77. Valuation rules of the service sale6_notify_shipping

## 5.1.12. Revising creation services

*OLIVA***NOVA** Modeler adds by default several parameters to the creation services:

i)    A parameter with a simple data type for each class attribute for which "request upon creation" has been marked.

ii)   An object-valued parameter for each class that is accessible via a structural relationship with 0:1 or 1:1 cardinality (in the role end of the referenced class).

Due to ii), after defining structural relationships among classes, some creation services may have been added object-valued parameters that are not according to the requirements specification anymore. Thus, the creation services need to be revised and, if necessary, the creation service is marked as "internal" and a creation transaction is added.

In the SuperStationery Co. model, the creation service of the class ClientOrder, named new_order, has been added the following attributes: p_agrSupplier, p_agrTruckDriver and p_agrInsurancePolicy. These three attributes do not correspond to the communicative event that corresponds to placing the order (Sale 1), but to later communicative events (Sale 2, Logi 10 and Risk 4 respectively). Therefore, the creation service is marked as internal ("Internal use" is checked) and a creation transaction is added.

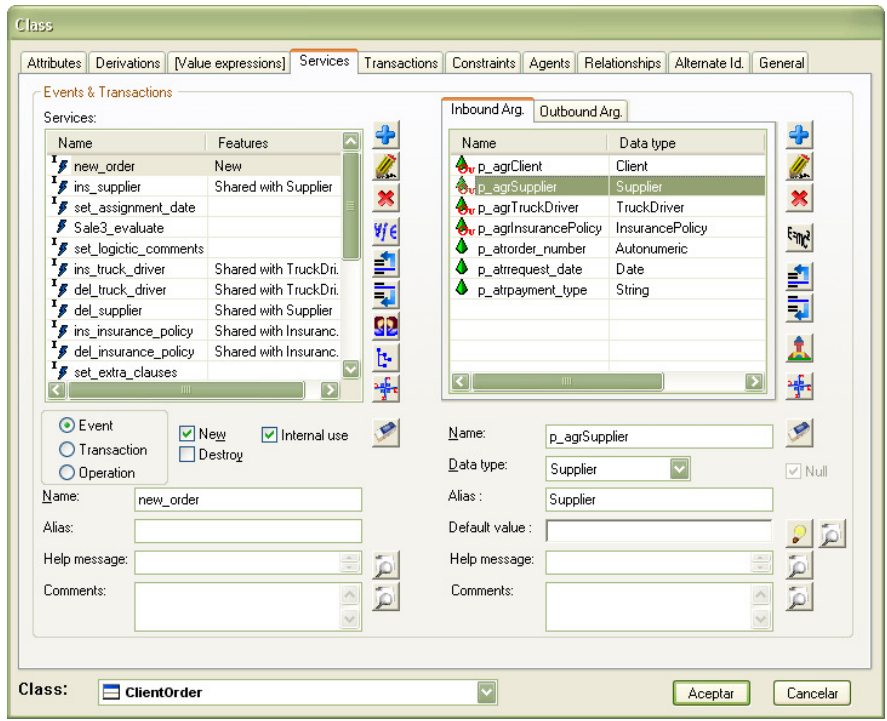

Figure 78.   Addition of the attributes p_agrSupplier, p_agrTruckDriver and p_agrInsurancePolicy to the creation service new_order

The creation transaction is named CREATE_ORDER and it simply executes the creation service, passing a Null value for the above-mentioned parameters.

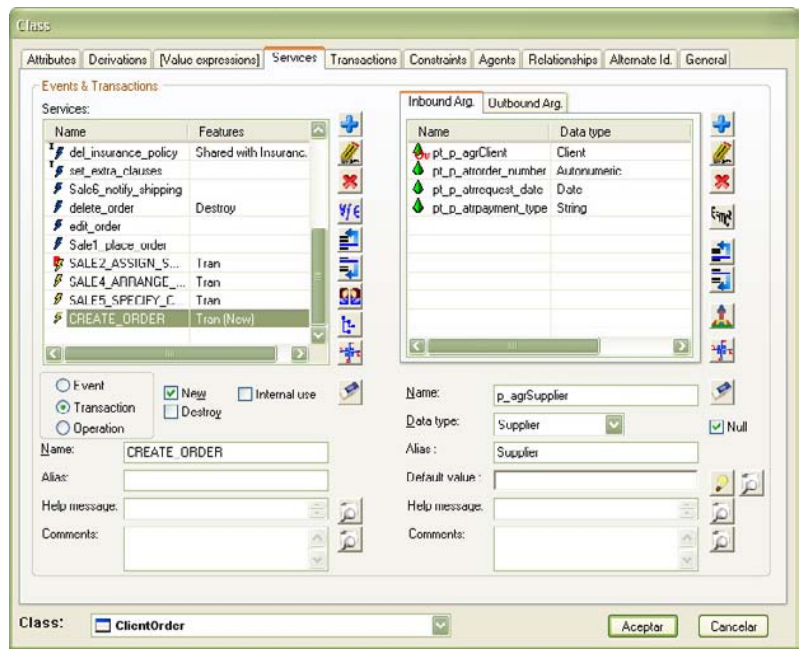

Figure 79.   Addition of the transaction CREATE_ORDER to the class CLIENTORDER

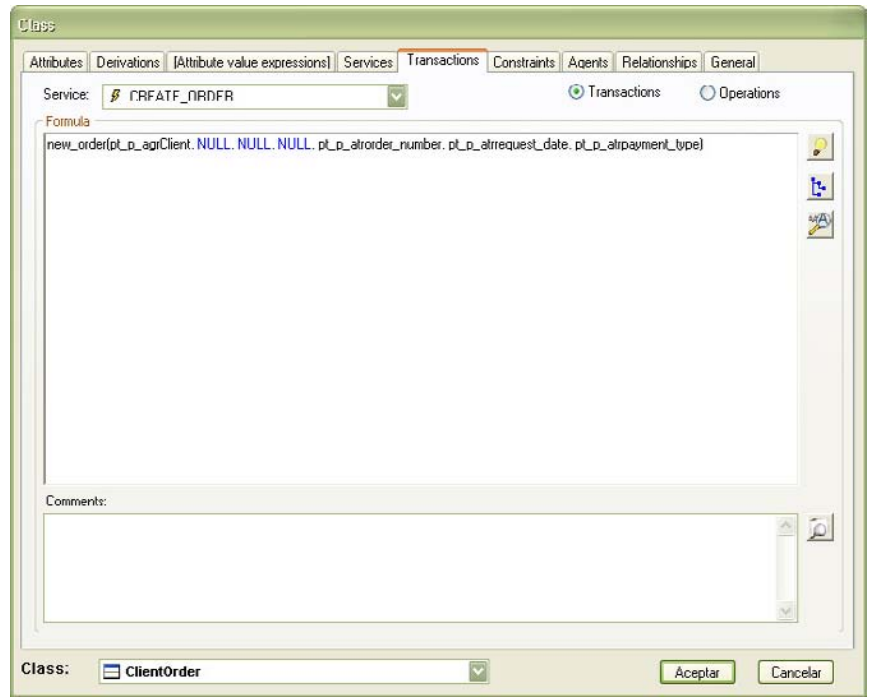

Figure 80.   Transaction formula that corresponds to CREATE_ORDER

## 5.1.13. State Transition Diagram

> Note: This section presents work under revision. It does not provide accurate information but it is left in the document so the readers can picture the derivation of the Dynamic Model in their mind.

The state transition diagram of the SuperStationery Co. case is the following. It is created according to the diagram shown in Figure 10.

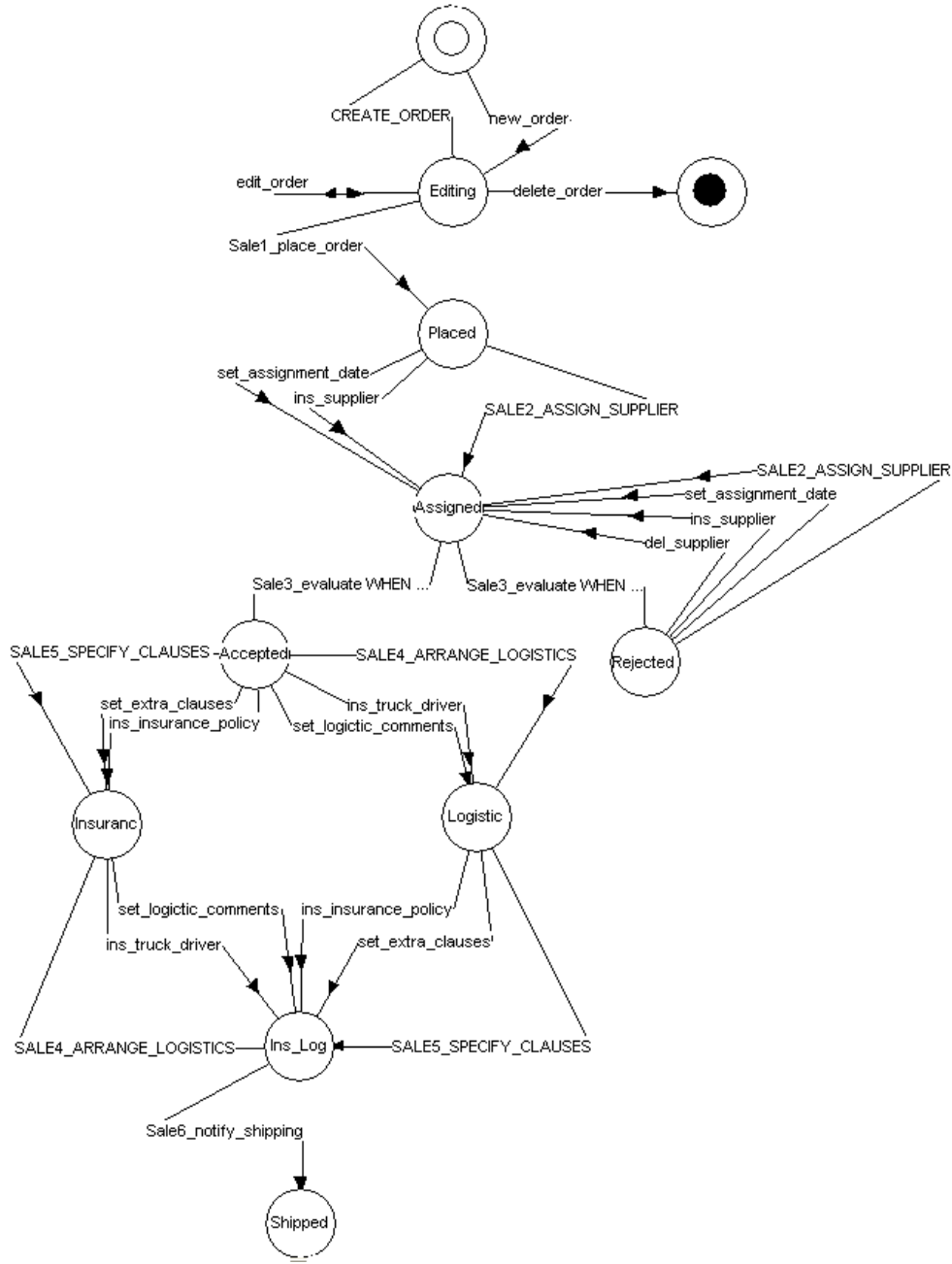

Figure 81.   Dynamic model of class ORDER

# 6. DISCUSSION

The experience with this lab demo has allowed the researchers to ascertain that the approach is feasible in practice. An advantage of following a systematic procedure to create conceptual models is that it allows to ground conceptual modelling decisions in the information from the requirements model (diagrammatic elements, textual statements, etc.). The derivation technique, in its current state, mainly focuses on the construction of the Object Model (a class diagram). The derivation of the Dynamic Model is only drafted and, although it offers some guidelines for the construction of state-transition diagrams, it still needs a workout. With regards to the Object Model derivation, most modelling decisions could be reasoned from (and traced back to) the requirements model; however, other decisions needed to be made during the construction of the conceptual model. We have found several reasons for this:

- In some cases, the requirements model is incomplete and, although it could, it does not provide all the necessary information to ground each and every conceptual modelling decision (e.g. a missing structural restriction). One possible solution is to prescribe that the requirements model must be complete before applying the derivation technique. However, it is not always feasible tom complete the requirements model. In fact, there is a trade-off between fulfilling requirements model completeness and the available resources (e.g time, money). The time to terminate a modelling activity is thus not when the model is perfect (which will never happen) but when it has reached a state where further modelling is less beneficial than applying the model in its current state [Lindland, Sindre et al. 1994]. Thus, the derivation technique should still be possible to be applied in the face of incompleteness (or even invalidity).

- In other cases, the information that is needed to make a given conceptual modelling decision is not provided in the requirements model because it refers to a design aspect (e.g. an initialisation formula, derived information related to business objects such as total amounts). There are several options: (i) either to include this information in the model, being aware of its design nature, (ii) to provide conceptual modelling patterns that the analyst can choose between, (iii) and to let the analyst design his/her own solution. A in-depth analysis of these situations will help to improve the derivation technique.

- In other cases, the derivation technique lacked the proper guidelines to tackle the situation. For instance, the property by which a structural relationship is designated as static/dynamic is not addressed by the derivation technique. The result is that the analysts have to decide by themselves. This is not a problem *per se*. However, we consider that it is worth to investigate whether this conceptual modelling decision can be grounded in the information specified in the requirements model.

The lab demo has also allowed the researchers to realize that there is still space for improvement. In the following, we highlight some issues that require further investigation.

## Not all possible updates of the business objects are considered

Aside from those updates that fulfil the unity criteria [González, España et al. 2009] and therefore deserve the status of communicative event (e.g. the registry of a client; see CLIE 1) there may exist some changes in the state of business objects that are nonetheless important for the IS (e.g. updating the telephone number of a client, deleting a client record). These state changes mainly correspond to basic *update* and *delete* operations (i.e. the U and D operations of the typical CRUD acronym). From the point of view of Communication Analysis, these events do not entail much analytical complexity and it is advised not to include them in the communicative event diagram. For the sake of simplicity, we opt for modelling only *create* operations and those *update* and *delete* operations that are really meaningful for the organisation and affect business process understanding. This guideline has proved to be valuable in practice when complex organisational work practice if tackled.

In any case, although the above-mentioned events are of minor importance in analysis time, they constitute operations that must be supported by the IS. In real development projects that apply Communication Analysis, the IS is designed to include these operations even though they are not included in a communicative event diagram. They are often specified as textual requirements[22].

Therefore, we face the challenge of formalising the specification of basic CRUD operations in a way that they do not burden the analyst with additional workload (it needs to be an agile modelling practice, supported by the requirements structure in a way that these operations are easy and fast to be modelled). Moreover, the specification of CRUD operations needs to be systematically processed during the conceptual model derivation. In short, some primitives need to be added to the Requirements Model and the derivation technique needs to be extended in order to take these new primitives into account. We plan to build upon a previous work that provides a viable notation for CRUD operations [España 2005] to improve Communication Analysis and the derivation technique.

## Service derivation can be further optimised

In some situations, a transaction that groups two atomic services that correspond to the reaction of the same communicative event could be avoided by merging the two services into a single one. It is the case of communicative events that update the state of an existing business object by linking it to another business object (e.g. when a client order is assigned to a supplier, the assignment date is recorded and the order form is linked to a supplier record; see Section 4.2.7). When processing such communicative events during conceptual model derivation, new attributes are added to an existing class (e.g. assignment_date is added to class CLIENTORDER), an atomic service to set the value of the new attributes is added to the class as well (e.g. set_assignment_date), a structural relationship between the affected classes is added to the class diagram (e.g. clientorder_supplier), and the shared services that correspond to the relationship are added to both classes (ins_supplier and del_supplier). Also, in order to ensure the atomic execution of both the service that sets the value of attributes and the service that sets the link between instances (e.g. set_assignment_date and ins_supplier, respectively), a transaction is added to the main class (e.g. SALE2_ASSIGN_SUPPLIER is added to CLIENTORDER).

However, in some cases it would be possible to avoid the creation of a transaction by taking advantage of the insertion shared service and using it to also set the value of the new attributes. This implies adding valuation rules to the shared service. However, valuation rules and transaction formulas are not yet addressed by the derivation technique so new derivation guidelines are needed.

## There is no guidance for the Functional Model

The OO-Method Functional Model offers a very expressive pseudocode to express the reaction of the IS. The technique should be improved to guide the modelling of valuation rules (rules that specify the behaviour of atomic services) and transaction formulas (specifications of the behaviour of transactions). Most valuation rules are expected to be easily derived because they mainly set the value of class attributes using the values of inbound arguments. Transaction formulas will probably entail more complexity since their expressiveness is much higher.

## Not all complex business objects need an end of editing service

An *end of editing* service is an atomic service that is added to a class in order to indicate that all the information that constitutes the message related to a communicative event has been already entered and that the IS can now react to the communicative event. For instance, such a service is needed for the client order so as to indicate when the information about the order (including the destinations and the order lines) has been completely entered; until that moment, the order is still being edited and the IS cannot yet react (e.g. the order should not be communicated to the Sales Manager or else he would get confused by the incomplete information).

---

[22] Especially when the software implementation is outsourced, every requirement needs to be explicitly stated.

On the contrary, the client record does not need an end of editing service because, although it is a complex business object, it does not require further IS reaction aside from having the information recorded. Adding such a service increases the workload of the salesman because it complicates the interaction with the interface of the IS: it requires at least one additional, unnecessary mouse click and users dislike this.

So we face the challenge of investigating which type of business objects require an end of editing service and which do not. We may come up with some derivation guidelines that allow to make a modelling decision using heuristics such as "if the reaction of the IS to a communicative event that creates a new complex business object includes a linked communicative interaction to another actor then an end of editing service is needed to explicitly trigger the IS reaction."[23] In other cases, the users or the analyst should make the decision and, in order to account for this decision in the requirements model, a new primitive should be added to the Requirements Model. This new type of requirements to be added to the Communication Analysis requirements structure can be of the form "Explicitly confirm end of editing"; it would only apply to a communicative event that involves creating a new complex business object (or, more specifically, to a complex substructure that represents a complex business object). Its value is simply a Boolean value (a Yes/No or a checkbox).

Also, the corresponding derivation guidelines should be added to the derivation technique in order to process this new type of requirements. The new type of requirement specifies whether an end of editing service needs to be derived or not. Note that this type of requirements belongs to *L4. Usage environment* requirements level of Communication Analysis [España, González et al. 2009]. Therefore, it specifies an aspect of design.

### Attribute default values can be derived from the Requirements Model

Message structures can be used in analysis time and in design time. Depending on the stage, different properties of the message fields can be defined. This way, when switching from IS analysis to design, the message structures can be extended in order to include initialisation formulas [González, Ruiz et al. 2011] (see Figure 82).

| FIELD | OP | DOMAIN | INITIALISATION | EXAMPLE VALUE |
|---|---|---|---|---|
| ORDER = | | | | |
| < Order number + | g | number | | 10352 |
| Request date + | i | date | today() | 31-08-2009 |
| Payment type + | i | text | | Cash |
| Client + | i | Client | | 56746163-R, John Papiro Jr. |
| DESTINATIONS = | | | | |
| { DESTINATION = | | | | |
| < Address + | i | Client address | | Blvd. Blue mountain, 35-14A, 2363 Toontown |
| Person in charge + | i | text | | Brayden Hitchcock |
| LINES = | | | | |
| { LINE = | | | | |
| < Product + | i | Product | | ST39455, Rounded scissors (cebra) box-100 |
| Price + | i | money | Product.Price | 25,40 € |
| Quantity > | i | number | | 35 |
| } | | | | |
| > | | | | |
| } | | | | |
| > | | | | |

Figure 82.   A message structure that includes the Initialisation property (it corresponds to Sale 1)

---

[23] This guideline is just a draft and it needs a further investigation.

As explained in Section 4, data fields lead to the derivation of class attributes and reference fields lead to the derivation of structural relationships. For data fields, this property can be processed in order to determine the formula that defines the default value for the attributes. For reference fields, this property can be processed in order to define a transaction that initialises the link.

## The type of structural relationship is not derived

We refer as structural relationships to semantic relationships between classes as defined by Pastor and Molina [2007]. Although the OO-Method defines several types of structural relationships (namely, association, aggregation and composition), only associations are derived for the moment.

Actually, this is not a serious issue because the type of structural relationship actually has no implication in the automatic code generation (it does not change the software behaviour). However, we could take advantage of their different notation and semantic meaning in order to improve class diagram understanding.

## Heuristic guidelines could be offered to improve the derivation of the attribute type

A class attribute can be of the following types:

- *Constant*. Its value cannot be changed after the initialisation
- *Variable*. Its value can be changed after the initialisation, provided that the appropriate service is defined.
- *Derived*. Its value depends on the value of other attributes and, therefore, it is determined by a derivation formula (it is not entered by the user).

We face the challenge of investigating whether the Requirements Model can provide enough information to recommend whether an attribute should be constant or variable. For instance, the fact that two data fields that appear in two different communicative events actually refer to the same piece of information of a business object (and, therefore, to the same class attribute) is a sign that the attribute should be variable (it may be necessary to update the value of that attribute after its initialisation).

## Domain ontologies could give support for a more accurate derivation

This research challenge is very wide. Just as an example, the determination of the data type of class attributes (including its size, in the case of the String type) could be supported by a domain ontology. If an attribute refers to the name of fish species, then an ontology if the fishing industry could provide guidance for determining the proper size of this attribute.

## The derivation of properties for structural relationships could be improved

No explicit guidelines are provided for the dynamic/static property of structural relationships. Analysts need to ask the users or act upon their own criteria. However, we should investigate whether the requirements model provides enough information so as to provide guidelines.

## There is no provision for agent derivation

Organisational actors in the Requirements Model could be processed to derive *agent classes*, which are a special type of classes of the Object Model that are used by the *OLIVANOVA* Model Compiler to generate software users and to grant them permissions (rights to access pieces of information and trigger software functionality).

Primary actors provide the new information, whereas support actors are responsible for communicating this information to the Information System. Therefore, support actors are candidates to derive agent classes. This way, an agent class named Salesman would be derived during the processing of Sᴀʟᴇ1 and it would be granted rights to access client order information and to trigger the services new_order and Sale1_place_order.